\documentclass{article}
\usepackage{graphicx}
\usepackage{amsmath}
\usepackage{amsfonts}
\usepackage{amssymb}
\usepackage{enumerate}
\usepackage{lscape}
\usepackage{longtable}
\usepackage{rotating}
\usepackage{multirow}
\usepackage{color}
\usepackage{url}
\usepackage{subfigure}
\usepackage{rotating}
\usepackage{comment}
\usepackage[utf8]{inputenc}
\usepackage{acro}
\usepackage{color}
\usepackage{booktabs}
\usepackage{float}
\usepackage{hyperref}
\usepackage{geometry}  

\title{Survey of Mathematical Models of and Numerical Methods for Fluid Dynamics Water Engineering }
\author{Anshu Kumar, Kemi Olimba, Vyacheslav Kungurtsev and Fabio V. Difonzo}
\date{July 2024}

\newcommand{\R}{\mathbb{R}}

\begin{document}

\maketitle

\begin{abstract}
    Computational fluid dynamics (CFD) has become a cornerstone of modern water engineering, providing quantitative tools for the analysis, prediction, and management of complex hydraulic systems across a wide range of spatial and temporal scales. This survey reviews the mathematical models and numerical methods that underpin CFD applications in water engineering, from depth-averaged formulations such as the shallow water equations to fully three-dimensional Navier–Stokes models, as well as selected alternative modeling approaches. We examine the historical development of these models, their mathematical structure, and the numerical discretization and solution strategies commonly employed in practice, including finite difference, finite volume, and finite element methods. Beyond core solver technology, the survey addresses practical modeling issues such as source-term treatment, wetting and drying, turbulence modeling, free-surface representation, and computational efficiency. The growing role of data integration is also discussed, encompassing data assimilation, uncertainty quantification, and emerging machine-learning-assisted approaches that complement physics-based solvers. To illustrate the impact of modeling and numerical choices on real-world applications, representative case studies from large-scale water management systems are reviewed. By integrating theory, numerical techniques, and applied perspectives, this survey provides a unified reference for researchers and practitioners seeking to understand both the foundational principles and contemporary challenges of CFD in water engineering.
\end{abstract}

\section{Introduction}

Computational fluid dynamics (CFD) plays a central role in contemporary water engineering, providing numerical tools for the analysis and prediction of river flows, coastal and estuarine dynamics, groundwater systems, and hydraulic infrastructure. Depending on the spatial and temporal scales of interest, CFD models for water applications range from depth-averaged formulations such as the shallow water equations (SWE) to fully three-dimensional Navier--Stokes models incorporating turbulence closures and free-surface tracking. Over the past several decades, a substantial body of literature has developed around both the governing equations and the numerical techniques required to solve them efficiently and robustly in practical settings.

Several survey and review studies have contributed to organizing this literature. General reviews of computational hydraulics in the context of water resources engineering summarize the governing equations, numerical discretization frameworks—such as finite difference, finite element, and finite volume methods—and their practical implementations across open-channel and environmental flows includes \cite{sarker2021short}. Dedicated reviews of shallow water equation modeling examine the derivation of the SWE, their mathematical structure, and common numerical treatments, including source-term discretization, handling of wetting and drying, and solution strategies suitable for hydraulic applications includes \cite{delis2021shallow,ortloff2024hydraulic}. These surveys are complemented by classical and modern CFD texts that provide the numerical and physical foundations underpinning many of the methods used in hydraulic modeling\cite{versteeg2007introduction,toro2024computational}. 

In parallel, broader investigations in CFD methodology review emerging numerical advances that increasingly influence water-related applications. Recent surveys highlight the growing role of data-driven and machine-learning-assisted modeling in CFD, where surrogate models, hybrid physics–data approaches, and reduced-order techniques are used to accelerate simulations or enhance predictive capabilities \cite{wang2024recent}. While such approaches are not yet replacements for physics-based solvers in water engineering, they are increasingly viewed as complementary tools within modern CFD workflows.

While these studies provide valuable insights into specific aspects of CFD for water engineering, they often emphasize particular components—such as governing equations, numerical schemes, or application domains—depending on their focus. Building on this body of work, the present survey aims to bring these complementary perspectives together by offering an integrated discussion of CFD modeling for water engineering\cite{bates2005computational}. The discussion addresses: (1) the historical evolution of CFD approaches in hydraulic research, (2) descriptions of water flow models ranging from depth-averaged formulations to fully three-dimensional systems, as well as a few alternative models (3) numerical analysis and solution strategies employed in modern CFD practice, (4) additional practical considerations as far as faithful modeling of fluid flow for water engineering purposes and (5) representative case studies illustrating how CFD models are applied in real-world water management contexts. Specifically, we examine notable applications such as California’s CALSIM planning model, Australia’s SOURCE model for the Murray–Darling Basin, Canada’s Green Kenue Water Resource Model developed by the National Research Council, and India’s FloodHub platform by the Central Water Commission, highlighting how different modeling choices influence performance and predictive outcomes.

Beyond governing equations and numerical solvers, the interaction between CFD models and observational data has become increasingly important in water engineering. Topics such as data assimilation, system identification, and uncertainty quantification \cite{keesman2011system,smith2024uncertainty} are essential for calibrating and validating simulations, ensuring that predictions reflect observed hydraulic behavior under real operating conditions. Through selected case studies across diverse water management systems, this survey illustrates how modeling and numerical choices affect accuracy, robustness, and interpretability, providing practical insights for researchers and practitioners alike.

By integrating theory, numerical techniques, and real-world applications within a unified framework, this discussion offers a structured pathway for readers to engage with both the foundational principles and practical considerations that define modern CFD in water engineering.

\section{History of Computational Fluid Dynamics}

The history of Computational Fluid Dynamics (CFD) spans over more than two centuries of developments in mathematics, physics, and computational technology. What began as analytical descriptions of fluid motion has evolved into a critical tool for engineering, aerospace, meteorology, and many other domains. This section merges early developments in CFD modeling with the broader historical narrative of CFD, providing a continuous and consistent overview without altering the core facts or citations.

From the 1700s to the 1900s, the foundations of fluid dynamics were laid by physicists such as \textit{Leonhard Euler} and \textit{Daniel Bernoulli}, whose equations described the behavior of ideal fluids. These were later expanded by \textit{Claude-Louis Navier} and \textit{George Gabriel Stokes}, resulting in the Navier--Stokes equations that govern modern fluid dynamics. In 1904, \textit{Ludwig Prandtl} introduced boundary layer theory, providing essential insights for aerodynamic studies and informing many computational methods.

Early attempts at numerical solutions date back to Lewis Fry Richardson, whose efforts were limited by the absence of computing power. The true beginning of practical CFD came after World War II with the advent of electronic computers such as ENIAC, which enabled the first computational simulations of fluid flow.

The beginnings of computational methods in fluid dynamics emerged in the 1940s and 1960s, driven by advancements in numerical analysis and the introduction of digital computing. Foundational work by \textit{John von Neumann} and others set the stage for numerical approaches to solving fluid dynamics problems \cite{anderson1997history}. In 1942, \textit{Richard Courant}, \textit{Kurt Friedrichs}, and \textit{Hans Lewy} introduced the Courant--Friedrichs--Lewy (CFL) condition, which became essential for ensuring stability in finite-difference methods. By 1947, \textit{John von Neumann} and \textit{Robert D. Richtmyer} at Los Alamos National Laboratory conducted pivotal work on numerical stability, further advancing the field \cite{anderson1997history}. 

With the rise of digital computers in the 1950s and 1960s, more complex numerical solutions of the Navier--Stokes equations became feasible. Organizations such as NASA and Los Alamos National Laboratory (LANL) played major roles in the development of finite-difference and finite-element methods (FDM and FEM). Between 1955 and 1957, LANL and the RAND Corporation applied finite-difference methods to weather forecasting and missile trajectory calculations. In 1959, NASA adopted FDM for supersonic and hypersonic flight studies, and in 1963, LANL extended FDM techniques for shock wave simulations. The 1960s also saw the introduction of finite-volume methods (FVM) by \textit{Brian Spalding} in 1967, along with the development of the SIMPLE algorithm, which became a cornerstone of modern CFD \cite{BrianSpalding}. In 1965, \textit{Raymond Clough} formalized the finite-element method (FEM) for structural analysis, and by 1968, NASA was exploring FEM for spacecraft structural dynamics.

The 1970s and 1980s marked the period when CFD became practical and commercially viable. Improvements in computing hardware allowed NASA and Boeing to simulate airflow over aircraft, reducing reliance on physical prototypes. In 1970, research on FDM and FEM expanded across fluid dynamics applications. At NASA Ames, \textit{Thomas Pulliam} advanced FDM for transonic and supersonic flows in 1971, while NASA Langley Research Center implemented FEM for aerothermal reentry simulations in 1972, contributing to the Space Shuttle program \cite{anderson1997history}. The publication of \textit{The Finite-Element Method} by \textit{O.C.~Zienkiewicz} and \textit{R.L.~Taylor} in 1975 became a foundational resource. By 1980, \textit{Brian Spalding}'s work led to the founding of Computational Fluid Dynamics Services (CFDS) in the UK, which developed commercial CFD software \cite{anderson1997history}. NASA Ames further integrated FDM and FEM with turbulence models for complex aerodynamic simulations. In 1982, NASA expanded its use of FDM and FEM to include fluid--structure interaction, a critical development for multiphysics CFD tools. The commercialization of CFD accelerated in 1987 with the release of packages such as PHOENICS and ANSYS Fluent, greatly increasing accessibility.

The 1990s and 2000s saw major advancements in algorithms and software, driven by high-performance computing (HPC). Widely used codes such as ANSYS Fluent and Star-CD matured during this period. In 1990, LANL used HPC to advance FDM in large-scale simulations, including hypersonic flows. NASA further advanced FEM and FDM applications in 1992 for spacecraft reentry, while the development of Large Eddy Simulation (LES) enhanced turbulence modeling. By 1996, HPC made large-scale three-dimensional FEM and FDM simulations possible, increasing accuracy in turbulent and shock-dominated flows \cite{anderson1997history}. Concurrently, the 1990s and 2000s saw software such as OpenFOAM emerge, with its open-source release in 2004 encouraging community-driven innovation. Automotive industries, including Tesla and Ford, integrated CFD extensively between 2006 and 2010 for design optimization.

Modern CFD is characterized by high-fidelity simulations and widespread use of high-performance computing, including GPU acceleration. Techniques such as Direct Numerical Simulation (DNS) are increasingly used for research. Multiphysics simulations—coupling CFD with structural mechanics, heat transfer, or electromagnetics—have become common. Lattice Boltzmann Methods (LBM) have gained popularity for flows involving complex geometries \cite{chen1998lattice}, while hybrid RANS--LES strategies such as Detached Eddy Simulation (DES) offer a balance between accuracy and computational cost \cite{ferziger2002computational}. 

Contemporary trends include the integration of artificial intelligence and machine learning into CFD workflows. Advances in computing power, the availability of real-time data from the Internet of Things (IoT), and the development of cloud-based solvers have enabled real-time and on-demand CFD. Digital twins-virtual representations fed with real-time data are emerging as transformative technology for predictive maintenance and live system monitoring. Reduced-order modeling, GPU computing, and data-driven methods continue to push CFD towards real-time capabilities \cite{brenner2019perspective}. As digital transformation advances, the fusion of CFD with AI, cloud computing, and real-time analytics is poised to define the future of fluid dynamics applications.

\begin{figure}[H]
    \centering
    \includegraphics[width=0.5\linewidth]{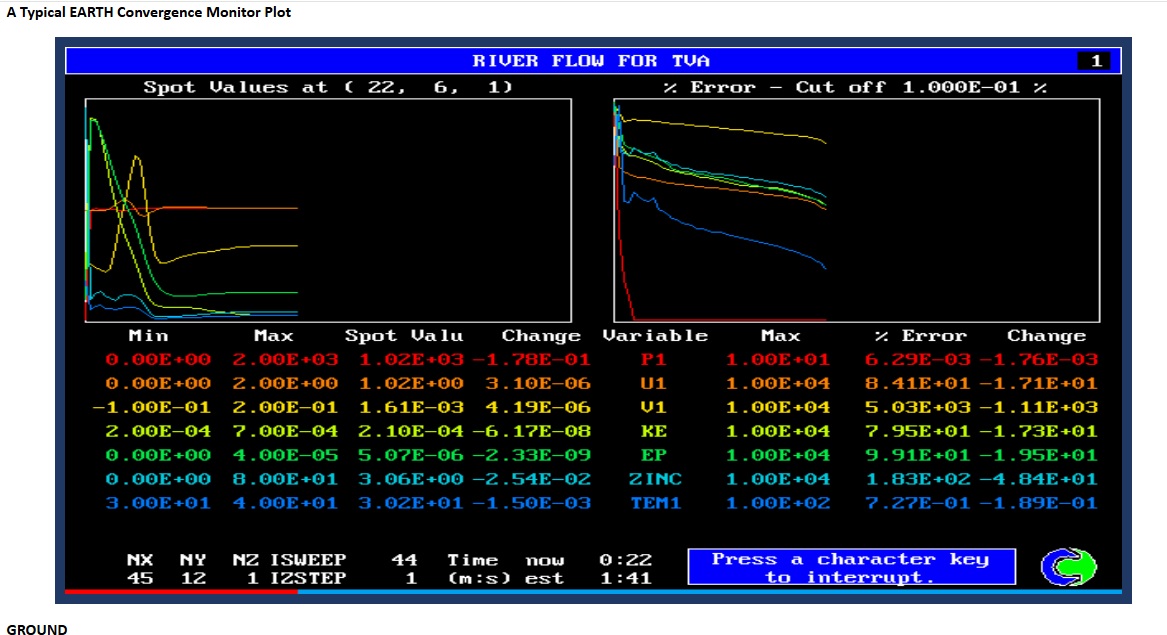}
    \caption{Phoenics Display in the late 1980s.}
    \label{fig:phoenics}
\end{figure}

\begin{figure}[H]
    \centering
    \includegraphics[width=0.75\linewidth]{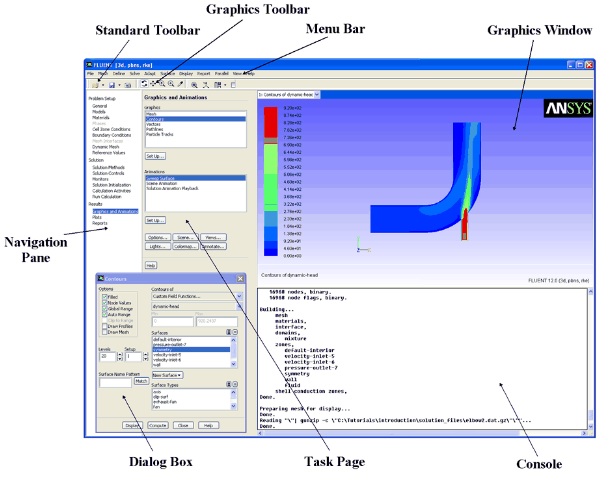}
    \caption{Ansys Fluent Version 12.0 GUI.}
    \label{fig:ansys}
\end{figure}

\begin{figure}[H]
    \centering
    \includegraphics[width=0.75\linewidth]{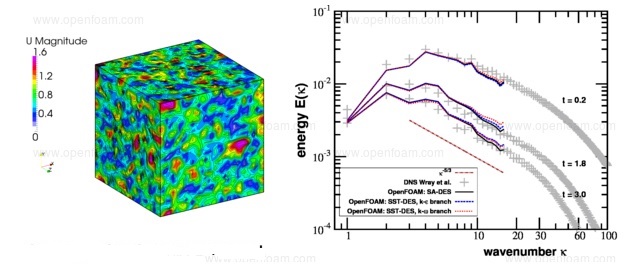}
    \caption{OpenFoam Version 3.0 - Decaying Isotropic Turbulence.}
    \label{fig:openfoam}
\end{figure}

\begin{figure}[H]
    \centering
    \includegraphics[width=0.75\linewidth]{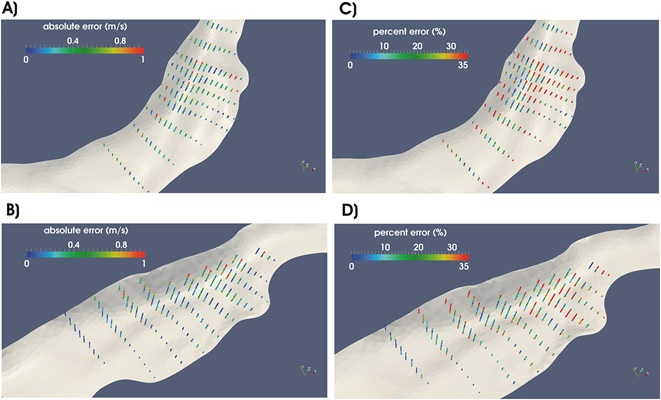}
    \caption{Display of a Detached Eddy Simulation Model of a River.}
    \label{fig:des}
\end{figure}

\section{ Foundations of CFD}

Computational Fluid Dynamics (CFD) is based on the numerical approximation of the governing equations of fluid motion, chiefly the Navier-Stokes equations expressing conservation of mass, momentum, and energy. The continuous problem is discretized in space and time using finite difference, finite volume, or finite element methods, leading to large systems of algebraic equations whose accuracy and robustness depend on consistency, stability, and convergence properties \cite{anderson1995}. Additional challenges arise from turbulence, often addressed through modeling strategies such as RANS or LES, which balance physical fidelity and computational cost \cite{moin1998direct,leveque2002finite}.

\subsection{Fundamentals and The Navier-Stokes Equations}

The foundation of fluid dynamics, and therefore CFD, lies in the Navier-Stokes equations. These equations describe the motion of viscous Newtonian fluids. They are derived from the fundamental principles of mass conservation (continuity) and momentum conservation.

\begin{enumerate}
    \item \textbf{Conservation of Mass (Continuity):} This is a zero divergence condition on the momentum of the fluid. In differential form for an incompressible fluid:
        \[
            \nabla \cdot \mathbf{u} = 0
        \]
        where $\mathbf{u}$ is the velocity vector.
        The continuity equation arises from the conservation of mass, implying that the rate of change of mass within a volume must equal the net flux across its boundary.
    \item \textbf{Conservation of Momentum (Newton's Second Law):}  The momentum equation is derived by applying Newton's second law to an infinitesimal fluid element and incorporating the effects of viscous stresses and pressure forces.
 Considering forces due to pressure, viscosity, and external body forces (such as gravity):
        \[
            \rho \left( \frac{\partial \mathbf{u}}{\partial t} + \mathbf{u} \cdot \nabla \mathbf{u} \right) = -\nabla p + \mu \nabla^2 \mathbf{u} + \mathbf{f}
        \]
        where $\rho$ is the fluid density, $p$ is the pressure, $\mu$ is the dynamic viscosity and $\mathbf{f}$ represents external forces.
\end{enumerate}

The Navier-Stokes equations (specifically the momentum equation) are a set of non-linear partial differential equations. Analytical solutions are only possible for very simple cases. Direct Numerical Simulation (DNS), which resolves all scales of turbulence by directly solving the Navier-Stokes equations, is computationally extremely expensive and currently only feasible for very low Reynolds number flows and simplified geometries. Therefore, various approximations and modeling strategies are employed in practical CFD applications.




\paragraph{State-of-the-Art and Future Directions}

Exascale computing, digital twins, and cloud-based CFD are driving further advancements. Future research will focus on improved accuracy and efficiency, enhanced multiphysics coupling, and increased use of AI/ML.





The Navier-Stokes equations are notoriously challenging to solve due to their non-linear nature, particularly the $(\mathbf{u} \cdot \nabla) \mathbf{u}$ term, which represents convective acceleration. In turbulent flows, the wide range of interacting spatial and temporal scales further complicates both analytical and numerical solutions.
Solving the full Navier-Stokes equations computationally over a large space-time domain with fine spatial and temporal resolution is extremely demanding. The required computational resources grow rapidly due to the need to resolve small-scale features and the stiffness introduced by high Reynolds number flows. This makes direct numerical simulation (DNS) infeasible for many practical applications, especially in three dimensions. As a result, in practice, simplified models such as the Shallow Water Equations and the Boussinesq approximation are often employed in computational fluid dynamics (CFD). These models capture essential large-scale flow features while reducing computational complexity, allowing for more efficient simulations in atmospheric, oceanographic, and engineering contexts.

\subsection{Summary of Numerical Methods for Solving Fluid Dynamics}

Fluid dynamics equations, including the Navier-Stokes and shallow water equations, are typically nonlinear and often defined over complex domains. Analytical solutions are rarely available except for highly simplified cases. 
As a result, numerical methods are essential for approximating the evolution of fluid variables over time and space. 
These methods allow the discretization of the governing equations on a computational grid or mesh, enabling simulations of realistic physical scenarios, including flows with varying bathymetry, free surfaces, and boundary interactions. 
Numerical approaches also provide flexibility in handling different initial and boundary conditions, making them indispensable tools for both research and engineering applications.

\paragraph{Finite Difference Method (FDM)}
The finite difference method approximates derivatives by finite differences on a structured grid. 
For example, the first-order spatial derivative of a variable $u$ can be approximated as:
\begin{equation}
\frac{\partial u}{\partial x} \approx \frac{u_{i+1} - u_i}{\Delta x},
\end{equation}
where $u_i$ and $u_{i+1}$ are values at adjacent grid points. 
FDM is simple, intuitive, and computationally efficient, making it popular for one-dimensional and simple multi-dimensional problems. However, it struggles with complex geometries and unstructured meshes. 
It is widely used in academic research and in software for structured-grid simulations~\cite{ferziger2002computationalFDM}.

\paragraph{Finite Volume Method (FVM)}
The finite volume method integrates the governing equations over control volumes, ensuring exact conservation of fluxes across cell boundaries. 
For example, the divergence term in the continuity equation is discretized as:
\begin{equation}
\int_{V} \nabla \cdot \mathbf{u} \, dV = \oint_{\partial V} \mathbf{u} \cdot \mathbf{n} \, dA,
\end{equation}
where $\partial V$ is the surface enclosing the control volume $V$, and $\mathbf{n}$ is the outward normal vector. 
FVM is particularly valued in engineering applications because it naturally conserves fluxes and can handle complex geometries on unstructured grids. It is widely implemented in industrial software packages such as OpenFOAM and ANSYS Fluent~\cite{ferziger2002computationalFVM}.

\paragraph{Finite Element Method (FEM)}
The finite element method uses variational principles to solve the governing equations. 
The fluid domain is divided into finite elements, and the equations are transformed into a weak form. 
For example, the weak form of the momentum equation reads:
\begin{equation}
\int_{\Omega} \mathbf{v} \cdot \left( \frac{\partial \mathbf{u}}{\partial t} + (\mathbf{u} \cdot \nabla) \mathbf{u} \right) \, d\Omega = -\int_{\Omega} \mathbf{v} \cdot \nabla p \, d\Omega + \int_{\Omega} \mathbf{v} \cdot \nu \nabla^2 \mathbf{u} \, d\Omega,
\end{equation}
where $\mathbf{v}$ is a test function. 
FEM is highly flexible and suitable for arbitrary geometries and boundary conditions, making it widely used in both research and industrial simulations. Software such as COMSOL, FEniCS, or ANSYS commonly implement FEM for complex CFD problems~\cite{ferziger2002computationalFEM}.

\paragraph{Other Methods}\medskip
In addition to the dominant finite difference, finite volume, and finite element approaches, several alternative numerical methods have been explored in CFD and hydraulic modeling, each with specific strengths and limitations.

\textbf{Spectral methods} approximate solutions using global basis functions such as Fourier series or Chebyshev polynomials. For smooth problems in simple or periodic domains, they offer very high accuracy and rapid (often exponential) convergence. These properties make spectral methods particularly effective for idealized flows and direct numerical simulation (DNS) of turbulence. However, their reliance on global basis functions makes them less suitable for complex geometries, localized discontinuities, or sharp gradients, where issues such as the Gibbs phenomenon degrade performance \cite{canuto2006spectral, boyd2001chebyshev}.

\textbf{Viscosity solutions} provide a rigorous framework for interpreting solutions of fully nonlinear partial differential equations, especially in the presence of shocks or discontinuities. The approach introduces a vanishing viscous regularization that yields smooth approximate solutions, which converge to a physically meaningful weak solution as the regularization parameter tends to zero. While viscosity solutions are primarily a theoretical construct rather than a standalone numerical method, they play a central role in the analysis of stability and convergence of numerical schemes for hyperbolic conservation laws and fluid flow models \cite{crandall1983viscosity, evans2022partial}.

\textbf{Lattice Boltzmann methods (LBM)} offer a mesoscopic alternative to conventional discretizations by evolving particle distribution functions whose macroscopic moments recover the shallow water variables. For SWE applications, shallow-water–adapted lattice models enable the simulation of wave propagation, hydraulic jumps, and inundation processes through local streaming and collision steps. The method’s explicit formulation and locality make it well suited for parallel computing, although incorporating complex boundary conditions and maintaining stability at high Reynolds numbers remain active areas of research.

\subsection{Duties and Activities of 
National Water Authorities}


National water authorities play a critical role in managing water resources sustainably. They are responsible for overseeing the allocation and distribution of water resources to meet the needs of agriculture, industry, and domestic use, ensuring a balance between demand and supply while considering long-term sustainability. Policies and regulations are enforced to control water extraction, maintain water quality, and prevent pollution. For example, standards for wastewater discharge are monitored to protect ecosystems and safeguard public health.

Continuous monitoring of water bodies is conducted to ensure compliance with environmental standards. Data on water quality, flow rates, and ecosystem health are collected and analyzed to support informed decision-making. In addition, authorities develop infrastructure and strategies to mitigate the impacts of floods and droughts. This includes constructing reservoirs, flood barriers, and implementing early warning systems to enhance resilience against extreme weather events.

Public engagement and education are also essential components of their work. Authorities promote awareness about water conservation and responsible usage while supporting research into innovative water management technologies. By integrating these efforts, national water authorities play a vital role in ensuring the sustainable and equitable management of water resources for current and future generations.


 \section{Water Transport Equations }

The mathematical description of water transport lies at the core of computational fluid dynamics in water engineering. Across rivers, canals, coastal zones, and engineered hydraulic systems, the governing equations capture how mass and momentum evolve under the influence of gravity, pressure gradients, terrain, and external forces. In practice, a hierarchy of models is used, each derived from fundamental conservation laws but tailored to different flow regimes, spatial and temporal scales, and levels of physical fidelity.
\\
This section presents the principal differential equations employed to model water transport. We examine simplifications appropriate for large-scale or predominantly inertial flows—namely the Euler equations, which neglect viscosity, and the shallow water equations, which exploit scale separation between horizontal and vertical dimensions. Such reduced models retain the essential physics while enabling efficient numerical simulation in practical applications ranging from flood forecasting to river hydraulics and coastal flow prediction.
\\
For each model class, we outline the underlying assumptions, the derivation from conservation principles, and the implications these choices have for solution behavior. We also connect these equations to the numerical methods discussed later, highlighting how different formulations-compressible versus incompressible, conservative versus non-conservative-shape the choice of discretization strategies and influence stability and accuracy. 

\subsection{Basic Formulas}

\subsubsection{Flow Equation}
\label{swe:ce}

The volumetric flow rate, denoted by \(Q\), describes the quantity of fluid passing through a given cross-section per unit time. It is directly related to the cross-sectional area of flow, \(A\), and the average flow velocity, \(v\), and is expressed mathematically as:

\begin{equation}
Q = A \cdot v
\end{equation}

This relationship indicates that the flow rate increases proportionally with either an increase in the area through which the fluid moves or an increase in the velocity of the fluid.

The variables used in the equation and their corresponding descriptions are summarized in the table below~\ref{tab:flow_variables}.

\begin{table}[ht]
\centering
\caption{Description of variables used in the flow rate equation}
\label{tab:flow_variables}
\begin{tabular}{lll}
\hline
\textbf{Symbol} & \textbf{Variable description} & \textbf{Typical units} \\
\hline
\(Q\) & Volumetric flow rate & \(m^3/s\) \\
\(A\) & Cross-sectional area of the flow & \(m^2\) \\
\(v\) & Average flow velocity & \(m/s\) \\
\hline
\end{tabular}
\end{table}

This formulation is widely applied in fluid mechanics and hydraulic engineering to characterize flow behavior in pipes, channels, and other conveyance systems.

\subsubsection{Manning's Equation}
\label{swe:me}

Manning’s equation \cite{HCManning} is widely used in hydraulic engineering to estimate the discharge in open channels and to determine the average flow velocity in rivers and streams. The equation relates the channel geometry, surface roughness, and slope to the volumetric flow rate and is expressed as:

\begin{equation}
Q = \frac{1}{n} A R^{2/3} S^{1/2}
\end{equation}

In this expression, the discharge \(Q\) depends on the cross-sectional area of flow \(A\), the hydraulic radius \(R\), the slope of the energy grade line \(S\), and Manning’s roughness coefficient \(n\). The hydraulic radius is defined as the ratio of the cross-sectional area of flow to the wetted perimeter, and it provides a measure of the efficiency of the channel section in conveying flow. The roughness coefficient accounts for the resistance offered by the channel bed and banks, while the energy slope represents the driving force of the flow.

The variables used in Manning’s equation and their corresponding descriptions are summarized in Table~\ref{tab:manning_variables}.

\begin{table}[ht]
\centering
\caption{Variables used in Manning’s equation}
\label{tab:manning_variables}
\begin{tabular}{llc}
\hline
\textbf{Symbol} & \textbf{Description} & \textbf{Unit} \\
\hline
\(Q\) & Flow rate (discharge) & \(m^{3}\,s^{-1}\) \\
\(A\) & Cross-sectional area of flow & \(m^{2}\) \\
\(R\) & Hydraulic radius (\(A\) divided by wetted perimeter) & \(m\) \\
\(S\) & Slope of the energy grade line & -- \\
\(n\) & Manning’s roughness coefficient & -- \\
\hline
\end{tabular}
\end{table}

\subsubsection{Chezy Formula}

The average velocity of flow in an open channel, such as a river, can be estimated using the Chezy formula. This empirical relationship links the mean flow velocity to the hydraulic characteristics of the channel and the slope of the energy grade line. Chezy’s formula is expressed as:

\[
v = C \sqrt{R S}
\]

where the parameters appearing in the equation represent the physical and hydraulic properties of the channel, as summarized in Table~\ref{tab:chezy_variables}.

\begin{table}[ht]
\centering
\caption{Description of variables used in Chezy’s formula}
\label{tab:chezy_variables}
\begin{tabular}{ll}
\hline
\textbf{Variable} & \textbf{Description} \\
\hline
$v$ & Mean velocity of flow ($m/s$) \\
$C$ & Chezy coefficient ($m/s$) \\
$R$ & Hydraulic radius ($m$), defined as the ratio of the cross-sectional flow area to the wetted perimeter \\
$S$ & Slope of the energy grade line or channel slope (dimensionless) \\
\hline
\end{tabular}
\end{table}

The Chezy coefficient encapsulates the effects of channel roughness and flow resistance, while the hydraulic radius and channel slope together describe the geometric and energetic conditions governing the flow. This formulation is widely applied in hydraulic engineering for estimating flow velocities in natural and artificial open channels.

\subsection{Source and Action of the Primary Forces Prevalent in CFD}

There are 5 main forces that affect the dynamics of a fluid and depending on which force is assumed or determined to be negligible, there is an ideal equation that should be used to model the dynamics of a fluid. Below is a list of the 5 forces

\paragraph{Gravity Force}

It is the force exerted by the Earth on a fluid due to its mass. It plays a fundamental role in driving fluid motion in open channels, influencing hydrostatic pressure in fluids at rest, such as water stored behind a dam and contributing to phenomena like buoyancy. The gravitational force acts uniformly on the fluid's mass, pulling it downward, and is commonly represented as a body force per unit mass, denoted by \( g \). In static fluids, gravity balances the pressure forces, while in dynamic situations, it opposes upward fluid motion, as observed in the rise of air bubbles. Moreover, gravity interacts with viscous forces to shape flow patterns in various fluid systems. It induces pressure differences within fluid columns, which in turn give rise to effects such as buoyancy and natural convection. In many fluid mechanics problems, the gravitational force is counteracted by pressure gradients to maintain equilibrium. Consequently, gravity is incorporated as a fundamental term in most governing equations of fluid dynamics, reflecting its pervasive influence on fluid behavior.

\paragraph{Pressure force}
Fluid pressure force refers to the force exerted by a fluid due to its pressure. It plays a crucial role in driving fluid flow from high to low pressure regions and supporting fluid columns, known as hydrostatic pressure. This pressure significantly impacts engineering designs, such as dam walls and retaining structures, which require reinforcements like counterforts on the inside or buttresses on the outside to counteract its effects. Fluid pressure force also influences fluid acceleration and remains a key factor in both static and dynamic fluid conditions. It interacts with other forces by balancing gravity in static fluids, counteracting viscosity, and being affected by fluid compressibility and turbulence. The differences in pressure serve as primary drivers of fluid flow, especially in incompressible fluids. For example, fluids in pipes move from regions of high pressure to low pressure, while in aerodynamic applications, pressure forces contribute to lift and drag on objects. In river systems, it plays a role in sediment transport, erosion, and the shaping of riverbanks. Moreover, fluid pressure force is a fundamental component in various equations, including the Navier-Stokes, Bernoulli’s, and Euler’s equations, which govern fluid mechanics.

\paragraph{Viscosity force}
Viscosity refers to the resistance to fluid flow caused by internal friction within the fluid. It plays a crucial role in resisting fluid motion, converting kinetic energy into heat, and influencing the formation of boundary layers. In fluid dynamics, viscosity interacts with other forces by opposing pressure and gravity, determining whether a flow regime is laminar or turbulent, and contributing to energy losses in fluid systems. Its effects include energy dissipation in the form of heat and the development of boundary layers, where fluid velocity transitions from zero at a solid boundary to its free-stream value. High-viscosity fluids tend to exhibit slow, smooth movement characteristic of laminar flow, whereas low-viscosity fluids are more prone to turbulent behavior. The impact of viscosity is mathematically represented in the Navier-Stokes equations, where viscosity forces appear as terms involving the viscosity coefficient and velocity gradient

\paragraph{Turbulence flow}
 Turbulence Force refers to the irregular fluctuations in fluid velocity and pressure caused by chaotic and unpredictable motion within the flow. It plays a crucial role in increasing energy dissipation, enhancing mixing, and influencing both heat and mass transfer in fluid systems. Turbulence arises from the complex interplay of pressure, viscosity, and inertial forces, and it interacts with all other forces in a fluid in intricate ways. Its effects are far-reaching—turbulent flows significantly enhance momentum and energy transfer, create complex flow structures such as eddies and vortices, and increase drag on objects due to higher friction losses compared to laminar flow. In fluid dynamics, turbulence is often modeled through modifications of the Navier-Stokes equations, incorporating additional terms like Reynolds stresses. Computational approaches, such as the widely used $k-\varepsilon$ model in computational fluid dynamics (CFD), are employed to account for the enhanced momentum and energy transport characteristic of turbulent flows.

\paragraph{Compressibility Force}
It arises from variations in fluid density due to changes in pressure and becomes particularly significant in high-speed flows, such as those encountered in supersonic or hypersonic regimes. This force plays a vital role in phenomena like sound propagation and the formation of shock waves—sudden, steep changes in pressure, temperature, and density. It interacts notably with pressure and gravity forces, especially under conditions of high velocity or pressure. Compressibility effects are essential in high-speed aerodynamics, where assuming a constant density (as in incompressible flow) is no longer valid. These effects introduce additional complexity into the governing equations of fluid motion. In mathematical modeling, compressibility is incorporated into the continuity and momentum equations by allowing fluid density $\rho$ to vary. The speed of sound
$c$ becomes a key parameter in such cases, influencing the behavior of compressibility waves, including both sound and shock waves.

\subsubsection{Interconnections and Examples}

Fluid motion in natural and engineered systems results from the combined action of several fundamental forces whose relative importance varies with the flow environment. In pressurized pipe systems, flow is primarily driven by pressure gradients, while viscous forces oppose motion by dissipating energy. Gravitational effects may influence the direction and magnitude of flow in inclined or vertical pipes, and turbulence significantly increases head losses beyond those predicted by laminar flow theory. Compressibility effects are generally negligible in pipe flow unless velocities approach the speed of sound, at which point density variations become non-trivial.

In river systems, gravity is the dominant driving force, pulling water downslope and establishing pressure differences that sustain flow. Viscosity governs resistance along the channel bed and banks, while turbulence enhances momentum transfer and mixing across the flow depth. Under typical river conditions, water can be treated as incompressible without loss of accuracy.

Ocean wave dynamics arise from the interaction of gravity and pressure forces acting at the free surface. These forces generate and sustain wave motion, while viscous effects act to dissipate energy and dampen wave amplitude over time. Turbulence plays a critical role in wave breaking and energy dissipation near the surface. Although compressibility is usually negligible for surface waves, it becomes increasingly important in extreme events such as tsunami propagation, where pressure disturbances travel over large distances.

It is essential to recognize that these forces are rarely independent. Turbulence, for example, emerges from the complex interaction between pressure gradients, viscous resistance, and inertial effects. Consequently, the dominance of any single force depends strongly on the specific flow conditions, geometry, and scale of the system under consideration.

\subsubsection{Relationships and Interactions}

The interaction between gravity and pressure is evident in hydrostatics, where gravity establishes a vertical pressure distribution in stationary fluids. In moving fluids, gravity contributes to pressure gradients that may either assist or oppose the flow, depending on orientation and boundary conditions.

Pressure and viscosity are closely linked in determining flow regimes. In laminar flow, pressure forces are balanced primarily by viscous resistance, resulting in smooth and predictable velocity profiles. In turbulent flow, pressure gradients continue to drive the motion, but resistance increases substantially due to the presence of eddies and fluctuating velocity components.

Viscosity plays a critical role in the onset and suppression of turbulence. Fluids with low viscosity are more susceptible to turbulent instabilities, whereas high-viscosity fluids tend to dampen velocity fluctuations. Once turbulence develops, it effectively increases the apparent viscosity of the flow, often necessitating additional modeling approaches to represent its influence accurately.

Compressibility further complicates these interactions. In incompressible flow assumptions, pressure variations do not affect density, greatly simplifying the governing equations. In compressible flows, however, pressure changes lead to significant density variations, introducing strong coupling between pressure, viscosity, turbulence, and energy transfer processes.

\begin{table}[ht]
\centering
\renewcommand{\arraystretch}{1.2}
\caption{Key variables commonly appearing in the governing equations of fluid flow}
\begin{tabular}{p{2cm} p{3cm} p{4cm} p{5.5cm}}
\hline
\textbf{Symbol} & \textbf{Variable} & \textbf{Physical Meaning} & \textbf{Role in Fluid Flow} \\
\hline
$p$ & Pressure & Normal force per unit area exerted by the fluid & Drives flow and establishes pressure gradients \\
$\rho$ & Density & Mass per unit volume of the fluid & Governs inertia and compressibility effects \\
$\mu$ & Dynamic viscosity & Measure of internal resistance to deformation & Controls viscous dissipation and flow resistance \\
$\mathbf{g}$ & Gravitational acceleration & Body force acting on the fluid mass & Drives free-surface and open-channel flows \\
$\mathbf{u}$ & Velocity vector & Rate and direction of fluid motion & Describes flow kinematics \\
$\nabla p$ & Pressure gradient & Spatial change in pressure & Primary driving force in many flow systems \\
$\tau$ & Shear stress & Force due to viscosity acting tangentially & Transfers momentum within the fluid \\
-- & Turbulence & Chaotic velocity fluctuations & Enhances mixing and increases energy losses \\
\hline
\end{tabular}
\end{table}

\subsubsection{Summary}

These forces are interconnected and collectively determine the behavior of fluid flows. Depending on the flow conditions (e.g., speed, viscosity, external forces), different forces may dominate, leading to different flow regimes (e.g., laminar vs. turbulent, compressible vs. incompressible). Understanding their relationships is key to solving fluid dynamics problems in both theoretical and practical contexts.

\section{\textbf {Steady River Flow }}

\subsection{\textbf{Shallow Water Equations }}
\label{subsec:swe}



The \textbf{One-Dimensional (1D) Shallow Water Equations} are fundamental for modeling fluid flow in rivers, channels, and floodplains. These equations are derived from the Navier-Stokes equations under the assumption of hydrostatic pressure and negligible vertical velocity, a framework known as the \textbf{Saint-Venant Theory}. The \textbf{continuity equation}, which represents mass conservation, is given by:

\[
\frac{\partial A}{\partial t} + \frac{\partial Q}{\partial x} = 0,
\]

where \( A \) is the cross-sectional area of the flow (\( \mathrm{m^2} \)), \( Q = A u \) is the discharge (\( \mathrm{m^3/s} \)), \( u \) is the depth-averaged velocity (\( \mathrm{m/s} \)), \( t \) is time (\( \mathrm{s} \)), and \( x \) is the horizontal coordinate (\( \mathrm{m} \)). The \textbf{momentum equation}, which describes the conservation of momentum, is expressed as:

\[
\frac{\partial Q}{\partial t} + \frac{\partial}{\partial x} \left( \frac{Q^2}{A} + g A h \right) = g A \left( S_0 - S_f \right),
\]

where \( h \) is the water depth (\( \mathrm{m} \)), \( S_0 \) is the bed slope (\( \mathrm{m/m} \)), and \( S_f \) is the friction slope (\( \mathrm{m/m} \)). These equations are widely used for river flow and flood modeling due to their simplicity and efficiency, although they are limited in capturing rapid changes in flow dynamics \cite{castro2019shallow}.

For flows where vertical accelerations and non-hydrostatic pressure effects are significant, the \textbf{non-hydrostatic shallow water equations} are employed. These equations extend the Saint-Venant framework by including additional terms to account for vertical dynamics. The continuity equation in this context is:

\[
\frac{\partial h}{\partial t} + \frac{\partial u h}{\partial x} = 0,
\]

and the momentum equation becomes:

\[
\frac{\partial u}{\partial t} + u \frac{\partial u}{\partial x} = -g \frac{\partial h}{\partial x} + \frac{\partial}{\partial x} \left( \frac{p_{\text{nh}}}{\rho} \right),
\]

where \( p_{\text{nh}} \) is the non-hydrostatic pressure correction (\( \mathrm{Pa} \)) and \( \rho \) is the fluid density (\( \mathrm{kg/m^3} \)). These equations are more computationally intensive but provide greater accuracy for modeling surface waves and other complex flow phenomena \cite{castro2019shallow}.

In some applications, \textbf{cross-sectional averaged equations} are used to simplify calculations. The continuity equation in this form is:

\[
\frac{\partial \bar{h}}{\partial t} + \frac{\partial}{\partial x} \left( \bar{u} \bar{h} \right) = 0,
\]

where \( \bar{h} \) is the cross-sectional average depth (\( \mathrm{m} \)) and \( \bar{u} \) is the cross-sectional average velocity (\( \mathrm{m/s} \)). The corresponding momentum equation is:

\[
\frac{\partial (\bar{u} \bar{h})}{\partial t} + \frac{\partial}{\partial x} \left( \bar{u}^2 \bar{h} + g \bar{h}^2 \right) = g \bar{h} \left( S_0 - S_f \right).
\]

These equations are particularly useful in hydraulic engineering, where local effects can be neglected for large-scale simulations.

For modeling unsteady flows in ideal fluids, the \textbf{Serre equations} (also known as the Green-Naghdi equations) are employed. These equations include dispersive terms to account for nonlinear wave propagation. The continuity equation is:

\[
\frac{\partial h}{\partial t} + \frac{\partial}{\partial x} \left( h u \right) = 0,
\]

and the momentum equation is:

\[
\frac{\partial u}{\partial t} + u \frac{\partial u}{\partial x} + g \frac{\partial h}{\partial x} = -\frac{h}{3} \frac{\partial^3 u}{\partial x^3}.
\]

The dispersive term \( -\frac{h}{3} \frac{\partial^3 u}{\partial x^3} \) makes these equations particularly suitable for modeling tsunamis and other wave-dominated flows \cite{castro2019shallow}.

Moving to \textbf{Two-Dimensional (2D) Shallow Water Equations}, these are used to describe flow in horizontal planes, making them applicable to a wide range of problems in hydrology, oceanography, and meteorology. The \textbf{Saint-Venant equations} in two dimensions are derived from the conservation of mass and momentum. The continuity equation is:

\[
\frac{\partial h}{\partial t} + \frac{\partial (hu)}{\partial x} + \frac{\partial (hv)}{\partial y} = 0,
\]

where \( h \) is the water depth, and \( u \) and \( v \) are the velocity components in the \( x \)- and \( y \)-directions, respectively. The momentum equations are:

\[
\frac{\partial (hu)}{\partial t} + \frac{\partial (hu^2 + \frac{1}{2}gh^2)}{\partial x} + \frac{\partial (huv)}{\partial y} = -gh \frac{\partial z_b}{\partial x} - \tau_x,
\]
\[
\frac{\partial (hv)}{\partial t} + \frac{\partial (huv)}{\partial x} + \frac{\partial (hv^2 + \frac{1}{2}gh^2)}{\partial y} = -gh \frac{\partial z_b}{\partial y} - \tau_y,
\]

where \( g \) is the gravitational acceleration, \( z_b \) is the bed elevation, and \( \tau_x \) and \( \tau_y \) are friction terms in the \( x \)- and \( y \)-directions. These equations assume shallow water, hydrostatic pressure distribution, and negligible vertical accelerations.

For \textbf{2D steady potential flow}, the governing equation is derived from the Laplace equation for the velocity potential:

\[
\nabla^2 \phi = 0,
\]

where \( \phi \) is the velocity potential and \( \nabla^2 \) is the Laplacian operator. This approach is accurate for irrotational, incompressible flows but is unsuitable for turbulent or rotational flows \cite{castro2019shallow}.

The \textbf{2D Serre equations} extend the Saint-Venant equations to include non-hydrostatic pressure and dispersive effects, making them ideal for modeling wave propagation. The continuity equation is:

\[
\frac{\partial h}{\partial t} + \frac{\partial (hu)}{\partial x} + \frac{\partial (hv)}{\partial y} = 0,
\]

and the momentum equation is:

\[
\frac{\partial u}{\partial t} + u\frac{\partial u}{\partial x} + v\frac{\partial u}{\partial y} + \frac{g}{h} \frac{\partial h}{\partial x} = -\frac{\partial}{\partial x}\left(\frac{h^2}{2}\frac{\partial^2 u}{\partial x^2}\right).
\]

These equations are computationally demanding but provide high accuracy for problems like tsunami modeling \cite{castro2019shallow}.

Numerical methods such as the \textbf{Finite Difference Method (FDM)}, \textbf{Finite Element Method (FEM)}, and \textbf{Finite Volume Method (FVM)} are commonly used to solve the 2D shallow water equations. The FDM discretizes the equations using a grid of points, replacing partial derivatives with finite differences. For example, the time derivative of water depth \( h \) is approximated as:

\[
\frac{\partial h}{\partial t} \approx \frac{h^{n+1}_{i,j} - h^n_{i,j}}{\Delta t}.
\]

The FDM is simple to implement for regular grids but is limited by stability constraints such as the Courant-Friedrichs-Lewy (CFL) condition. The FEM, on the other hand, discretizes the domain into triangular or quadrilateral elements and solves the equations using weighted residuals. This method is flexible for complex geometries but is computationally intensive. The FVM integrates the equations over control volumes, ensuring conservation of mass and momentum. It is particularly suitable for shock capturing and high-resolution fluid dynamics \cite{castro2019shallow}.

The \textbf{Gaussian Quadrature Galerkin Method} enhances the accuracy of FEM by approximating integrals using Gaussian quadrature:

\[
\int_\Omega f(x) dx \approx \sum_{i=1}^n w_i f(x_i),
\]

where \( w_i \) are weights and \( x_i \) are quadrature points. This method is particularly useful for nonlinear problems but requires careful selection of quadrature points \cite{khan2014modeling2DGQ}.

In \textbf{Three-Dimensional (3D) Shallow Water Equations}, vertical variations and turbulent mixing are incorporated to model more complex flows. The \textbf{3D diffusion equation}, which governs the transport of scalar quantities such as temperature or concentration, is given by:

\[
\frac{\partial \phi}{\partial t} = \nabla \cdot (D \nabla \phi),
\]

where \( \phi \) is the scalar quantity and \( D \) is the diffusion coefficient. This equation is discretized using finite difference approximations and is widely used in oceanography and meteorology \cite{maliska2023fundamentals}.

For \textbf{3D interpolation}, trilinear interpolation is commonly used to approximate field variables between discrete points. The scalar field \( \phi(x, y, z) \) is expressed as:

\[
\phi(x, y, z) \approx \sum_{i=0}^1 \sum_{j=0}^1 \sum_{k=0}^1 \phi_{ijk} \cdot (1 - \xi)^i \xi^{1-i} (1 - \eta)^j \eta^{1-j} (1 - \zeta)^k \zeta^{1-k},
\]

where \( \xi, \eta, \zeta \) are normalized coordinates within a grid cell. This method is widely used in FEM and CFD applications.

The \textbf{3D advection-diffusion equation} models the transport of scalar quantities due to both advection and diffusion:

\[
\frac{\partial \phi}{\partial t} + \mathbf{u} \cdot \nabla \phi = \nabla \cdot (D \nabla \phi),
\]

where \( \mathbf{u} = (u, v, w) \) is the velocity vector. This equation is used for pollutant dispersion and heat transport simulations.

For \textbf{3D Hamiltonian incompressible flows}, the governing equations are derived from the incompressibility condition \( \nabla \cdot \mathbf{u} = 0 \) and the Hamiltonian formulation:

\[
\mathbf{u} = \nabla \phi + \nabla \times \psi,
\]

where \( \phi \) is the scalar potential and \( \psi \) is the vector stream function. These equations are used for theoretical analysis of fluid flows and vortex dynamics \cite{shivamoggi2023introduction3DHIF}.

Finally, the \textbf{3D irrotational flow} equations are based on the assumption of zero vorticity \( \nabla \times \mathbf{u} = 0 \). The velocity potential \( \phi \) satisfies the Laplace equation:

\[
\nabla^2 \phi = 0.
\]

These equations are used for idealized flow models, such as potential flow over bodies \cite{shivamoggi2023introduction3DIF}.

\paragraph{Shallow Water Flow Equations in Rectangular Channels}

The one-dimensional (1D) shallow water equations in conservative form are widely used to model the flow of water in rivers, floodplains, and other shallow systems. These equations are expressed as:

\[
\frac{\partial}{\partial t} \begin{bmatrix} h \\ hu \end{bmatrix} + \frac{\partial}{\partial x} \begin{bmatrix} hu \\ hu^2 + \frac{1}{2} g h^2 \end{bmatrix} = \begin{bmatrix} 0 \\ -g h \frac{\partial z_b}{\partial x} \end{bmatrix},
\]

where \( h \) represents the water depth, \( u \) is the flow velocity, \( z_b \) is the bed elevation, and \( g \) is the gravitational acceleration. These equations describe the conservation of mass and momentum in shallow water systems, making them essential for flood modeling and hydraulic engineering.

One of the challenges in solving the shallow water equations is handling dry bed conditions, where the water depth \( h \) approaches zero. This can lead to numerical instability due to the hyperbolic nature of the equations. To address this, a dry bed treatment is applied by ensuring that the water depth remains above a small positive threshold \( \epsilon \). This is achieved by setting:

\[
h = \max(h, \epsilon).
\]

Additionally, to prevent non-physical negative depths, the flux is modified to preserve the positivity of \( h \):

\[
\hat{F} = \max(0, \hat{F}).
\]

This approach ensures numerical stability and accuracy when simulating wetting and drying processes in shallow water systems.

A specific example of wetting and drying can be illustrated using a parabolic bowl with bathymetry \( z_b(x) \) defined as:

\[
z_b(x) = z_0 - \alpha x^2,
\]

where \( z_0 \) is the initial elevation and \( \alpha \) is a curvature parameter. The governing equations for this system are:

\[
\frac{\partial h}{\partial t} + \frac{\partial (hu)}{\partial x} = 0,
\]
\[
\frac{\partial (hu)}{\partial t} + \frac{\partial}{\partial x} \left( hu^2 + \frac{1}{2} g h^2 \right) = -g h \frac{\partial z_b}{\partial x}.
\]

To handle wetting and drying in this scenario, the water depth is corrected using:

\[
h = \max(h + z_b - z_0, 0).
\]

The flux is evaluated only for wet regions where \( h > \epsilon \), and the numerical scheme is modified to ensure smooth transitions between wet and dry states. This approach is critical for accurately simulating the dynamics of water flow in systems with variable topography.

The Saint-Venant equations, a specific form of the shallow water equations, are widely used in hydrology and civil engineering to describe the flow of shallow water. These equations are derived from the principles of shallow water theory and are expressed as:

\[
\frac{\partial h}{\partial t} + \frac{\partial (hu)}{\partial x} = 0,
\]
\[
\frac{\partial (hu)}{\partial t} + \frac{\partial \left( hu^2 + \frac{1}{2} g h^2 \right)}{\partial x} = -g h \frac{\partial h}{\partial x},
\]

where \( h \) is the water depth, \( u \) is the flow velocity, and \( g \) is the acceleration due to gravity. These equations provide a mathematical framework for analyzing and predicting water flow in open channels and floodplains.

The Reynolds Transport Theorem offers a general framework for converting between a control volume and the system perspective in fluid dynamics. It is expressed as:

\[
\frac{d}{dt} \int_{V(t)} \rho \phi \, dV = \int_{V(t)} \frac{\partial (\rho \phi)}{\partial t} \, dV + \int_{S(t)} \rho \phi \mathbf{u} \cdot \mathbf{n} \, dS,
\]

where \( \phi \) represents a property such as mass, momentum, or energy, \( \rho \) is the fluid density, \( \mathbf{u} \) is the velocity vector, \( \mathbf{n} \) is the outward-pointing unit normal vector, \( V(t) \) is the control volume, and \( S(t) \) is the control surface. This theorem is fundamental for deriving conservation laws in fluid mechanics and is widely used in the analysis of fluid systems.

In conclusion, the shallow water equations, Saint-Venant equations, and Reynolds Transport Theorem provide a robust mathematical foundation for modeling and analyzing fluid flow in various applications. Techniques such as dry bed treatment and wetting-drying algorithms ensure numerical stability and accuracy, making these tools indispensable for hydraulic engineering and environmental modeling.

\subsection{Analytical Solutions from SWASHES}

A crucial step in the development and validation of numerical solvers for the Shallow Water Equations (SWEs) is testing them against known analytical solutions. The work of Delestre et al.~\cite{delestre2013swashes} addresses this need by providing a comprehensive and open-source compilation known as \textbf{SWASHES (Shallow Water Analytic Solutions for Hydraulic and Environmental Studies)}. This library gathers a broad range of exact and semi-analytical solutions to the one- and two-dimensional SWEs under diverse physical scenarios, offering a consistent foundation for verifying numerical schemes.

SWASHES contains solutions for both steady and unsteady configurations and accounts for a variety of source terms, including bed slope, friction modeled through Manning--Strickler or Darcy--Weisbach laws, rainfall inputs, and diffusive processes. Its test cases span subcritical, supercritical, and transcritical regimes and include flows with wetting and drying fronts. Particularly important are the scenarios involving shocks, rarefaction waves, and moving bathymetry interfaces, which allow researchers to assess the capability of numerical schemes to resolve discontinuities and nonlinear wave interactions. The database also incorporates classical benchmarks such as dam-break problems, Thacker’s oscillating basin, and analytical solutions for parabolic geometries.

The SWASHES framework is implemented in an extensible C++ code base, enabling researchers to reproduce analytical solutions or integrate new ones. This modularity, combined with the physical diversity of the test problems, makes SWASHES especially valuable for verifying finite volume and finite element solvers. Its emphasis on reproducibility and standardized benchmarks has established it as a widely used reference in computational hydraulics and environmental modeling, serving both as a validation tool and as a means of comparing the performance of different numerical methods.

\subsection{Nonlinear Shallow Water Equations}

The shallow water equations (SWE) describe the evolution of a free-surface fluid layer under the hydrostatic assumption and depth-averaged dynamics. Their nonlinear character arises from the coupling between flow depth, velocity, and momentum fluxes, making them a fundamental example of quasilinear hyperbolic systems.

For a fluid layer of height $h(x,y,t)$ and depth-averaged velocity $(u(x,y,t), v(x,y,t))$, the two-dimensional nonlinear SWE can be written as
\begin{align}
\frac{\partial h}{\partial t} + \frac{\partial (hu)}{\partial x} + \frac{\partial (hv)}{\partial y} &= 0, \\
\frac{\partial (hu)}{\partial t} + \frac{\partial}{\partial x}\!\left( hu^2 + \tfrac{1}{2} g h^2 \right)
+ \frac{\partial (huv)}{\partial y} &= - g h\,\frac{\partial b}{\partial x}, \\
\frac{\partial (hv)}{\partial t} + \frac{\partial (huv)}{\partial x}
+ \frac{\partial}{\partial y}\!\left( hv^2 + \tfrac{1}{2} g h^2 \right) &= - g h\,\frac{\partial b}{\partial y},
\end{align}
where $b(x,y)$ denotes the bathymetry and $g$ is gravitational acceleration. Even with a flat bottom, the governing equations remain nonlinear because the momentum variables appear quadratically in terms such as $hu^2$, $hv^2$, and $huv$, and the pressure contribution $\tfrac{1}{2}gh^2$ depends nonlinearly on the water depth.

Mathematically, the SWE are quasilinear hyperbolic equations whose characteristic speeds depend on the solution itself. In one spatial dimension the eigenvalues take the form
\[
\lambda_{1,3} = u \pm \sqrt{gh},
\]
indicating that wave propagation depends directly on the local depth and velocity. This dependence causes solutions to develop steep gradients and, in many cases, discontinuities such as hydraulic jumps and bores. Because smooth solutions may cease to exist in finite time, weak solutions and entropy conditions are required to describe the physically correct behavior.

Analysis of the nonlinear features typically involves studying characteristic fields, deriving energy estimates to understand stability, and examining how variations in $h$ and $u$ influence wave steepening. These tools explain why nonlinear interactions in the SWE lead to complex flow structures and why numerical methods must account for shocks and rapidly varying gradients.

For a rigorous treatment of the nonlinear structure of the shear shallow water model and related hyperbolic systems, standard references include the mathematical fluid mechanics framework of Chorin and Marsden \cite{chorin1990mathematical} and the finite-volume analysis of hyperbolic conservation laws developed by LeVeque \cite{leveque2002finite}.

\section{Turbulent River Flow }

Turbulent flow in river systems is characterized by chaotic, irregular fluid motion that significantly impacts flow velocity, sediment transport, and energy dissipation. Several mathematical models are used to analyze and predict turbulence, each tailored to specific scenarios and computational resources. These models include the Reynolds-Averaged Navier-Stokes (RANS) equations, Large Eddy Simulation (LES), Direct Numerical Simulation (DNS), and the \( k \)-\(\epsilon\) turbulence model. Each model has its governing equations, solves specific challenges, makes certain assumptions, and comes with its own set of pros and cons, making them suitable for different applications.

The \textbf{Reynolds-Averaged Navier-Stokes (RANS) equations} are widely used for predicting time-averaged turbulent flow characteristics in river systems. The governing equations include the continuity equation and the momentum equation, expressed as:

\[
\frac{\partial \overline{\rho}}{\partial t} + \nabla \cdot (\overline{\rho} \overline{\mathbf{u}}) = 0,
\]
\[
\overline{\rho} \left( \frac{\partial \overline{\mathbf{u}}}{\partial t} + (\overline{\mathbf{u}} \cdot \nabla) \overline{\mathbf{u}} \right) = -\nabla \overline{p} + \mu \nabla^2 \overline{\mathbf{u}} + \nabla \cdot \tau_{ij} + \mathbf{f},
\]

where \(\overline{\rho}\) is the time-averaged density, \(\overline{\mathbf{u}}\) is the time-averaged velocity, \(\overline{p}\) is the time-averaged pressure, \(\mu\) is the dynamic viscosity, \(\tau_{ij}\) is the Reynolds stress tensor representing turbulent stresses, and \(\mathbf{f}\) represents external forces. The RANS equations reduce computational complexity by focusing on averaged properties, making them efficient for large-scale river systems. However, they rely on turbulence models like \( k \)-\(\epsilon\) or \( k \)-\(\omega\), which can introduce inaccuracies. The assumptions include incompressible flow and the representation of turbulence through time-averaged properties. RANS is ideal for steady-state river flow analysis where detailed turbulence resolution is unnecessary.

The \textbf{Large Eddy Simulation (LES)} model resolves large-scale turbulence structures explicitly and captures unsteady and transient phenomena. The governing equations are the filtered Navier-Stokes equations:

\[
\frac{\partial \overline{u}_i}{\partial t} + \frac{\partial (\overline{u}_i \overline{u}_j)}{\partial x_j} = -\frac{1}{\rho} \frac{\partial \overline{p}}{\partial x_i} + \nu \frac{\partial^2 \overline{u}_i}{\partial x_j^2} - \frac{\partial \tau_{ij}}{\partial x_j},
\]

where \(\overline{u}_i\) and \(\overline{u}_j\) are the filtered velocity components, \(\overline{p}\) is the filtered pressure, \(\nu\) is the kinematic viscosity, and \(\tau_{ij}\) is the subgrid-scale stress tensor representing unresolved turbulence. LES assumes that large eddies dominate turbulence and are directly resolved, while small-scale turbulence is modeled using subgrid-scale models. This approach is more accurate than RANS but computationally expensive, requiring fine spatial and temporal resolution. LES is ideal for high-resolution studies of turbulent river flow, such as near hydraulic structures or during flood events.

The \textbf{Direct Numerical Simulation (DNS)} model provides the most accurate representation of turbulence by resolving all scales of motion. The governing equations are the full Navier-Stokes equations:

\[
\frac{\partial \mathbf{u}}{\partial t} + (\mathbf{u} \cdot \nabla) \mathbf{u} = -\frac{1}{\rho} \nabla p + \nu \nabla^2 \mathbf{u},
\]
\[
\nabla \cdot \mathbf{u} = 0,
\]

where \(\mathbf{u}\) is the velocity vector, \(p\) is the pressure, and \(\nu\) is the kinematic viscosity. DNS makes no assumptions about turbulence and directly computes all flow features. While highly accurate, DNS is computationally infeasible for large-scale or practical applications due to its extreme resource requirements. It is ideal for fundamental turbulence studies in controlled or small-scale environments.

The \textbf{\( k \)-\(\epsilon\) turbulence model} is a widely used RANS-based model that focuses on turbulent kinetic energy (\( k \)) and its dissipation rate (\(\epsilon\)). The governing equations are:

\[
\frac{\partial k}{\partial t} + \mathbf{u} \cdot \nabla k = P - \epsilon + \nabla \cdot (\nu_t \nabla k),
\]
\[
\frac{\partial \epsilon}{\partial t} + \mathbf{u} \cdot \nabla \epsilon = C_1 \frac{\epsilon}{k} P - C_2 \epsilon^2 + \nabla \cdot (\nu_t \nabla \epsilon),
\]

where \( P \) is the production of turbulent kinetic energy, \(\nu_t\) is the turbulent viscosity, and \( C_1 \) and \( C_2 \) are model constants. The \( k \)-\(\epsilon\) model assumes isotropic turbulence and steady-state turbulence characteristics. It is simple and robust for engineering applications but has limited accuracy for complex or highly transient flows. This model is ideal for general-purpose turbulence modeling in river systems with moderate complexity.

In summary, the choice of turbulence model depends on the scale of the river system, the level of detail required, and the available computational resources. RANS models are preferred for large-scale systems, LES for detailed and transient studies, and DNS for fundamental research. Simplified models like \( k \)-\(\epsilon\) and \( k \)-\(\omega\) are practical for engineering applications, balancing accuracy and computational cost. Each model addresses specific challenges and assumptions, making them valuable tools for understanding and predicting turbulent flow in river systems.

\subsection{Reynolds Number}

The Reynolds number (\(Re\)) is a dimensionless parameter used to characterize fluid flow regimes and predict whether the flow is laminar, transitional, or turbulent~\cite{GKBReynolds}. It represents the ratio of inertial forces to viscous forces in a fluid. Mathematically, it is defined as
\[
Re = \frac{\rho U L}{\mu},
\]
where \(\rho\) is the fluid density, \(U\) is a characteristic velocity, \(L\) is a characteristic length scale, and \(\mu\) is the dynamic viscosity of the fluid. In terms of kinematic viscosity \(\nu = \mu/\rho\), it can also be expressed as
\[
Re = \frac{U L}{\nu}.
\]

The Reynolds number provides insight into the nature of the flow. Low values (\(Re < 2000\)) indicate that viscous forces dominate, leading to laminar, smooth flow. High values (\(Re > 4000\)) suggest that inertial forces dominate, resulting in turbulent, chaotic flow, while intermediate values (\(2000 < Re < 4000\)) correspond to transitional regimes.

In the context of thin fluid films, several simplifying assumptions are often applied to derive Reynolds-type equations~\cite{SPReynoldsNumber}: the flow is laminar, the fluid is incompressible and Newtonian, the film thickness is much smaller than other characteristic dimensions, and inertial effects are negligible compared to viscous forces. Applying these assumptions to the Navier--Stokes equations and integrating across the film thickness yields the classical Reynolds equation:
\[
\frac{\partial}{\partial x} \left( h^3 \frac{\partial p}{\partial x} \right) + \frac{\partial}{\partial z} \left( h^3 \frac{\partial p}{\partial z} \right) = 6\mu \frac{\partial}{\partial x} (h U) + 12 \mu \frac{\partial h}{\partial t},
\]
where \(p\) is the pressure within the fluid film, \(h(x,z,t)\) is the local film thickness, \(\mu\) is the dynamic viscosity, \(U(x,z)\) is the velocity of the moving surface relative to the stationary surface in the \(x\) direction, and \(t\) is time.

For more general flows, such as in pipes or open channels, the Reynolds number can be expressed in terms of measurable quantities:
\[
Re = \frac{4 \rho Q}{\mu \pi d},
\]
where \(Q\) is the volumetric flow rate and \(d\) is the pipe diameter. This highlights that reducing the characteristic diameter while maintaining the same flow rate increases the Reynolds number, promoting turbulent flow.

The Reynolds number plays a critical role in engineering and environmental applications. In pipe flow, it informs the design by predicting the onset of turbulence. In hydraulics, it helps analyze flow in rivers and channels, guiding strategies to control flooding and reduce erosion. Similarly, in aerodynamics, it is fundamental for evaluating flow behavior over surfaces such as aircraft wings. Understanding the balance between inertial and viscous forces, as captured by \(Re\), is therefore central to both theoretical analysis and practical design in fluid mechanics.

\begin{figure}
    \centering
    \includegraphics[width=0.5\linewidth]{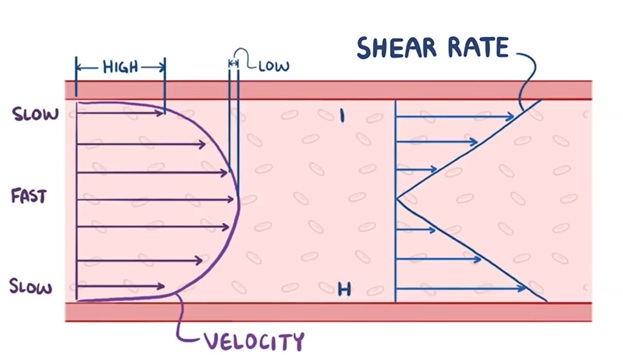}
    \caption{Velocity and shear rate distribution in laminar flow.}
    \label{fig:Vrsrl}
\end{figure}

\begin{figure}
    \centering
    \includegraphics[width=0.75\linewidth]{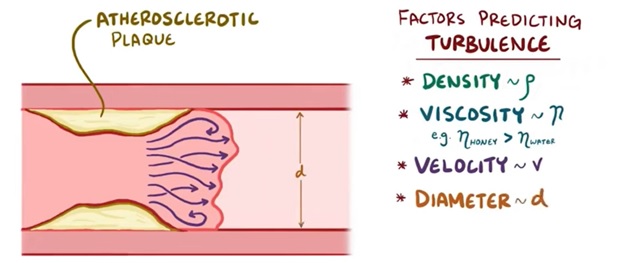}
    \caption{Turbulent Flow Illustration}
    \label{fig:TFI}
\end{figure}

\subsection{Richardson's Equation}
The concept is named after Lewis F. Richardson, whose pioneering work on turbulence, scaling laws, and numerical weather prediction laid much of the foundation for modern geophysical fluid dynamics. His early studies on stratified turbulence and energy cascades remain central themes in contemporary research.
Richardson's equation, and the associated Richardson number (\(Ri\)), are fundamental tools used to assess the stability of stratified flows. Widely applied in fluid dynamics, atmospheric science, and oceanography, they quantify the competition between stabilizing buoyancy forces and destabilizing velocity shear. This balance determines whether a stratified flow remains laminar or transitions toward turbulence.

\subsubsection*{Governing Equation}

A simplified representation of perturbation dynamics in a stratified fluid can be expressed as
\[
\frac{\partial^{2} \phi}{\partial t^{2}} + \left( N^{2} + k^{2} \right)\phi = 0,
\]
where \( \phi \) denotes a small perturbation field, \( k \) is the wavenumber, and 
\[
N^{2} = -\frac{g}{\rho}\frac{\partial \rho}{\partial z}
\]
is the Brunt–Väisälä (buoyancy) frequency, representing the restoring force due to density stratification. Here, \(g\) is gravitational acceleration, \(\rho\) is the fluid density, and \(z\) is the vertical coordinate. A positive \(N^{2}\) indicates stable stratification, meaning vertical displacements experience a restoring buoyant force.

\subsubsection*{Richardson Number}

\paragraph{Definition}

The classical gradient Richardson number is defined as
\[
Ri = \frac{N^{2}}{\left( \frac{\partial U}{\partial z} \right)^{2}},
\]
where \(U(z)\) is the mean horizontal velocity and \(\frac{\partial U}{\partial z}\) is the vertical shear. When \(Ri\) is large, buoyancy dominates and inhibits turbulence; when \(Ri\) is small, shear can overwhelm buoyancy and promote instability. A widely accepted empirical threshold is \(Ri = 0.25\), below which the flow becomes susceptible to Kelvin–Helmholtz instabilities and turbulence.

\paragraph{Physical Interpretation}

The Richardson number quantifies the dynamic interplay between buoyancy, which tends to suppress vertical mixing, and shear, which promotes overturning and turbulent motion. High values (\(Ri \gg 1\)) correspond to strongly stratified flows where vertical motions are inhibited and the flow remains stable. Conversely, low values (\(Ri < 1\)) indicate that velocity shear is sufficiently strong to distort or overturn density layers. When \(Ri\) falls below the critical value of approximately \(0.25\), the flow is unable to resist shear-driven perturbations, and turbulence is likely to develop. This threshold originates from linear stability analysis of shear layers and is widely used in atmospheric and oceanographic modeling. Buoyancy arises from vertical density gradients, such as those created by temperature or salinity differences. In contrast, shear originates from velocity variations between adjacent layers of fluid. The Richardson number thus encapsulates the essential physics of stratified stability by measuring which of these two effects is dominant in a given flow configuration. A flow is considered stably stratified when the Richardson number is sufficiently large, effectively suppressing vertical mixing. Unstable or weakly stratified flows, associated with small Richardson numbers, permit shear-driven turbulence that enhances vertical transport of momentum, heat, and mass. These criteria form the backbone of geophysical stability analysis and are embedded in many modern turbulence parameterizations.

\paragraph{Background and Context}

In geophysical flows, turbulence refers to chaotic, irregular motion characterized by eddies and rapid fluctuations in velocity and pressure. Stratification, caused by density differences, organizes fluids into layers of varying stability. In such settings, the buoyancy and shear terms embedded in the Richardson number describe the conditions under which these layers remain intact or become mixed. This makes Richardson's framework indispensable for understanding atmospheric boundary layers, oceanic thermoclines, and density-driven flows in rivers, estuaries, and reservoirs.

\paragraph{ Derivation and Simplification}

Richardson also proposed relationships that estimate the growth rate of turbulence in stratified fluids. Although simplified and partly empirical, these formulations help connect observed turbulent behavior with underlying physical gradients. The Richardson number emerges naturally from the governing equations of momentum and density transport. Under linearization and assuming small perturbations, the ratio of buoyancy restoring terms to shear production terms simplifies into the form used today. Though idealized, this ratio provides a meaningful first-order indicator of flow stability in many practical scenarios.

\paragraph{Applications }

Richardson's framework has extensive applications. In atmospheric dynamics, it helps determine boundary-layer stability and the likelihood of cloud formation, turbulence, or wave breaking. In oceanography, it guides the study of internal wave propagation, mixing of water masses, and vertical diffusivity. Engineering applications include predicting mixing efficiency in reservoirs, evaluating pollutant dispersion, and analyzing stratified shear layers in industrial flows. Numerical models in computational fluid dynamics (CFD) routinely incorporate Richardson-based criteria to regulate turbulence closure schemes in stratified environments. Generalized versions of the Richardson number appear in rotating, compressible, or strongly nonlinear systems, accounting for Coriolis effects, variable stratification, or complex shear profiles. In CFD models of atmospheric and oceanic flows, variants of the Richardson number often regulate turbulence closures, stability functions, or mixing-length parameterizations in stratified layers.

\paragraph{Limitations}

Despite its usefulness, the Richardson number has limitations. It arises from linear stability theory, and therefore applies most accurately to weak perturbations; fully nonlinear behavior may deviate substantially. It also assumes simple stratification and shear structures, whereas real environments often contain strong heterogeneities and transient dynamics. Moreover, \(Ri\) is an empirical threshold rather than a universal constant; different systems may exhibit turbulence onset at slightly different values.

\subsection{Large Eddy Simulation (LES)}

Large Eddy Simulation (LES) resolves the large, energy-containing turbulent structures explicitly \cite{ferziger2002computationalLESRANS} while the smaller, unresolved scales are modeled through subgrid-scale (SGS) closures \cite{ghaib2019introductionLES}. This approach offers higher fidelity compared to RANS, particularly for unsteady and spatially developing turbulence.

\subsubsection*{Filtered Navier--Stokes Equations}

In LES, the governing equations are obtained by applying a spatial filter to the Navier--Stokes equations:
\[
\frac{\partial \tilde{u}_i}{\partial t}
+ \tilde{u}_j \frac{\partial \tilde{u}_i}{\partial x_j}
= -\frac{1}{\rho} \frac{\partial \tilde{p}}{\partial x_i}
+ \nu \frac{\partial^2 \tilde{u}_i}{\partial x_j^2}
- \frac{\partial \tau_{ij}}{\partial x_j}.
\]
Here, \( \tilde{u}_i \) and \( \tilde{p} \) represent the filtered velocity and pressure fields, respectively, and the SGS stress tensor is defined as
\[
\tau_{ij} = \widetilde{u_i u_j} - \tilde{u}_i \tilde{u}_j.
\]
This tensor accounts for the influence of unresolved small-scale motions on the resolved flow.

\subsubsection*{Subgrid-Scale Model (Smagorinsky Model)}

A commonly used SGS closure is the Smagorinsky model:
\[
\tau_{ij} - \frac{1}{3} \tau_{kk} \delta_{ij} = -2 \nu_t \tilde{S}_{ij},
\qquad
\nu_t = (C_s \Delta)^2 |\tilde{S}|.
\]
The term \( \nu_t \) is the eddy viscosity introduced to represent subgrid turbulence, \( \tilde{S}_{ij} \) is the filtered strain-rate tensor, \( C_s \) is the Smagorinsky constant, and \( \Delta \) denotes the grid filter width.

\subsubsection*{Applications}

LES is widely employed in engineering, geophysical, and environmental flows where capturing transient vortices, mixing layers, or shear-driven instabilities is essential.

\subsection{Reynolds-Averaged Navier–Stokes (RANS)}

The RANS framework \cite{ghaib2019introductionRANS} describes the mean behavior of turbulent flows by decomposing flow quantities into mean and fluctuating components, a classical method outlined in turbulence modeling literature \cite{ferziger2002computationalLESRANS}. The velocity decomposition is written as
\[
u_i = \bar{u}_i + u_i',
\]
where \( \bar{u}_i \) is the time-averaged component and \( u_i' \) is the fluctuating component.

\subsubsection*{RANS Equations}

With this decomposition, the averaged momentum equations become
\[
\frac{\partial \bar{u}_i}{\partial t}
+ \bar{u}_j \frac{\partial \bar{u}_i}{\partial x_j}
= -\frac{1}{\rho} \frac{\partial \bar{p}}{\partial x_i}
+ \nu \frac{\partial^2 \bar{u}_i}{\partial x_j^2}
- \frac{\partial \overline{u_i' u_j'}}{\partial x_j}.
\]
The term \( \overline{u_i' u_j'} \) is the Reynolds stress tensor, which must be modeled to close the system.

\subsection*{Boussinesq Approximation}

A common closure is the Boussinesq eddy viscosity hypothesis:
\[
\overline{u_i' u_j'}
= -2 \nu_t \bar{S}_{ij} + \frac{2}{3} k \delta_{ij},
\]
where \( \nu_t \) denotes the turbulent viscosity, \( \bar{S}_{ij} \) is the mean strain-rate tensor, and \( k = \frac{1}{2} \overline{u_i' u_i'} \) is the turbulent kinetic energy.

\subsubsection*{Applications}

RANS models are particularly useful for industrial, atmospheric, and design-oriented simulations due to their efficiency, especially when resolving only the mean flow is sufficient.

\subsection{Boussinesq Equation}

The Boussinesq approximation is applied to buoyancy-driven flows where density variations are sufficiently small to influence momentum but negligible in continuity.

\subsection*{Governing Equations}

The incompressible continuity equation is
\[
\nabla \cdot \mathbf{u} = 0,
\]
while the momentum equation incorporating buoyancy effects becomes
\[
\frac{\partial \mathbf{u}}{\partial t}
+ \mathbf{u} \cdot \nabla \mathbf{u}
= -\frac{1}{\rho_0} \nabla p
+ \nu \nabla^2 \mathbf{u}
+ \mathbf{g} \beta (T - T_0).
\]
Here, \( \rho_0 \) is a reference density, \( \beta \) the thermal expansion coefficient, \( T \) the temperature, and \( T_0 \) a reference temperature.

\paragraph{Reynold’s Number}

The Reynolds number is a dimensionless indicator of flow regime. Low-Re flows tend to be laminar and characterized by smooth, orderly motion, whereas high-Re flows become turbulent, exhibiting chaotic eddies and vortical structures.

\subsection{Turbulence Modeling in Practice}

In practical CFD simulations for water engineering, turbulence is one of the primary challenges because it governs energy transfer, mixing, and momentum distribution in flows over structures, around bends, and through open channels. Directly resolving all turbulent scales via Direct Numerical Simulation (DNS) is computationally prohibitive for most engineering applications, especially for field‑scale water systems. As a result, engineers and modelers routinely employ turbulence models \cite{rodriguez2019applied}that approximate the effects of unresolved turbulent motions while keeping simulations feasible in terms of computational cost.

The most widely used practical approach in industrial and environmental CFD is based on Reynolds‑Averaged Navier–Stokes (RANS) models\cite{yusuf2020short}, which provide time‑averaged representations of turbulent effects. RANS turbulence models, such as the standard $k$–$\varepsilon$ and $k$–$\omega$ families, estimate turbulent kinetic energy and its dissipation or frequency, allowing stable and efficient simulations of mean flow characteristics without resolving all turbulent fluctuations . These models are especially popular in large‑scale water engineering problems like flow over spillways, around piers, or past coastal defenses because they offer sufficient accuracy with modest computational requirements compared to higher‑fidelity methods. Validation studies using experimental data have demonstrated that well‑tuned RANS models can capture key flow features and provide reliable predictions for engineering design and assessment.

When finer resolution of flow structures is needed — for example, in regions with strong recirculation, flow separation, or vortex dynamics — Large Eddy Simulation (LES) is increasingly applied. LES resolves the larger, energy‑containing turbulent eddies while modeling only the smaller subgrid scales, delivering higher fidelity in capturing unsteady flow features at a higher computational cost. In environmental flows with complex geometry or transient behavior, LES has been shown to outperform RANS models in predicting detailed hydrodynamics, though its use is still limited by computational resources and mesh resolution constraints.

In water engineering practice, model selection is therefore a balance between computational efficiency and predictive detail. RANS models remain the workhorse for routine engineering simulations due to their robustness and lower cost, while LES is preferred in research and targeted high‑resolution studies where capturing turbulent structures is critical. In some applications, hybrid approaches that combine the strengths of RANS and LES are explored to improve prediction quality without prohibitive computational demands. Across all approaches, careful validation against experimental or field observational data is essential to ensure confidence in turbulence model predictions and to support engineering decisions.


\section{Alternative (non-PDE) Methods}
An alternative methodologically distinct approach to modeling fluid dynamics is to precisely characterize the chaotic elements as uncertainty, permitting the consideration of averaged dynamics. Conceptually, one can consider that in a casual observation of water flow, there are a number of small waves and currents that appear above the threshold of appearing on a scale of a common grid discretization choice, but below the threshold of appearing to be of consequence to the overall river flow. We can see, moreover, that precise models of fluid flow given exact initial conditions have had delicate development, yet the computational expense and fine grid discretization required for stable modeling make them a grand distance away from practical realizability in water engineering applications. Yet as far as the latter, more aggregate behavior is of interest. This presents the potential for Stochastic M0dels of Fluid Dynamics. As this has been primarily a niche line of work, we will present a brief overview, in this Section, of some of the main models and principles in the literature, without going into as much depth as in the classical hyperbolic CFD systems.

\subsection{Stochastic PDE Models}
Stochastic PDEs can refer to 1) PDEs wherein the coefficients are modeled as random fields, usually due to uncertainty as to their true values, for modeling inverse problems and uncertainty quantification (UQ) and robust control of systems governed by PDEs, 2) PDEs that include an addition stochastic process driving term, most commonly Brownian diffusion. We leave the first to the latter section on data assimilation and UQ, and consider the second in this subsection. These models are commonly used in practice to simulate turbulence, obtaining long run averages of quantities of interest, as the practical solution to the chaotic instability of turbulent flow. 

An early work to present the potential for Stochastic Modeling of PDEs governing fluid dynamics is presented in~\cite{breuer1992stochastic}. A recent monograph is given in~\cite{flandoli2023stochastic}. 

The stochastic noise-driven Navier-Stokes equation becomes
        \[
            \rho \left( \frac{\partial \mathbf{u}}{\partial t} + \mathbf{u} \cdot \nabla \mathbf{u} \right) = -\nabla p + \mu \nabla^2 \mathbf{u} + \mathbf{f}+\Sigma dW_t
        \]
wherein $\Sigma$ is some positive-semidefinite matrix and $dW_t$ is a Wiener stochastic process. Alternatively, the noise can depend on the state $\Sigma:=\Sigma(\mathbf{u})$. 

Turbulence can be modeled with~\cite{flandoli2023stochastic},
\[
dW_t=-\epsilon^{-1}W_tdt+\epsilon^{-1}d(\mathop{curl}(W_t))
\]
in which stochasticity is introduced as driving the vorticity as defined by the curl operation. This approach permits for numerical simulations of Eddy currents and other phenomena appearing in turbulence. By averaging the random drive, one can obtain statistics on the long run quantities of interest, thus assisting simulation and management. 

Recent work (e.g.~\cite{mailybaev2025rg}) suggests long run ergodicity properties of certain stochastic fluid models, thus far simplified toy constructions, however. If these properties hold for the noise-driven fluid dynamics simulations appearing in ensemble models, then this implies that the statistics of the long run average dynamics are not influenced by the typically arbitrarily chosen sampled initial conditions. 

\subsection{Peridynamic Models}

Peridynamics, while conceptually distinct from stochastics, has a significant resemblance computationally. With Peridynamics, rather than a point pass or velocity, there is a kernel integrated term,
\[
U(x,t) = \int_{K(x,\tilde{x},t)} u(\tilde{x},t)d\tilde{x}
\]
where $K(x,y,t)$ is some (positive semidefinite) Kernel that represents local interactions at a distance between quantities. This framework is amenable to turbulence and internal shear type forces, which can be understood as dynaics of influence at a magnitude faster time scale than the macro flow of the liquid. Two works presenting peridynamic models of fluids are given in~\cite{nguyen2021modelling,zhao2022construction}.

Peridynamic theory, introduced by Silling in 2000 \cite{SILLING2000}, is a nonlocal reformulation of continuum mechanics that accounts for long-range interactions. Unlike classical models that rely on partial differential equations involving spatial derivatives, peridynamic models are integral-based and depend solely on the displacement field. This characteristic enables the theory to naturally incorporate discontinuities, such as cracks and fractures, without additional criteria or special treatments.

The fundamental equation in peridynamic theory, known as the peridynamic equation of motion, is given by:
\[
\partial_{tt}\theta(x,t) = \int_{\R}C(|x-y|)[\theta(x,t)-\theta(y,t)],\mathrm{d}y, 
\] 
where $\theta(x,t)$ represents the displacement at point $x\in\R$ and time $t$, and $C(|x-y|)$ is the kernel function, often referred to as the micromodulus function. This kernel is a nonnegative, even function characterizing the material properties and interaction forces between particles at positions $x$ and $y$.

In the one-dimensional case, this equation describes the dynamics of an infinite bar composed of a linear microelastic material. The kernel function $C$ introduces a length scale parameter $\delta>0$, known as the horizon, which determines the maximum interaction distance between particles. This parameter governs the degree of nonlocality in the model and differentiates peridynamic theory from classical local approaches. As $\delta\to0$, the peridynamic model transitions to the classical wave equation:
\[
\partial_{tt}\theta(x,t) - \partial_{xx}\theta(x,t) = 0, \]
as shown in \cite{WECKNER2005,OTERKUS2014}.

Peridynamic theory is well-suited for modeling materials that exhibit long-range forces and dispersive behavior. The general initial-value problem for the peridynamic equation has been proven to be well-posed \cite{Emmrich_Puhst_2013,Emmrich_Puhst_2015}. Furthermore, the theory ensures consistency with Newton's third law by requiring the kernel to be even:
\[
C(\xi) = C(-\xi), \quad \xi \in \R, 
\]
and the dispersive effects are captured by the condition:
\[
\int_{\R} \left(1-\cos(k\xi)\right) C(\xi) ,\mathrm{d}\xi > 0, \quad \forall k \neq 0. 
\]
The long-range nature of peridynamic interactions diminishes as the distance between particles increases, which is expressed by:
\[
\lim_{\xi \to \pm\infty} C(\xi) = 0. 
\]
For materials with finite interaction ranges, the horizon restricts interactions to particles within a bounded region, $[-\delta,\delta]$, simplifying the peridynamic equation to:
\[
\partial_{tt}\theta(x,t) = \int_{B_{\delta}(x)} C(|x-y|)[\theta(x,t)-\theta(y,t)],\mathrm{d}y. 
\]
The micromodulus function $C$ plays a pivotal role in defining the material-specific constitutive model. Different choices of $C$ allow peridynamic theory to adapt to various materials and mechanical behaviors, making it a powerful tool for studying fracture mechanics, wave propagation, and other phenomena in continua with singularities.

\subsection{Empirical Models (HBV)}
The HBV hydrology model, or Hydrologiska Byr{\o a}ns Vattenbalansavdelning model, is a well known simulation software that is in a broader class of empirical models. Rather than applying first principles physics to a hydrological system, say a cascade of dams along a river, empirical models simply use piece-wise linear models to define interaction (e.g., water flow) between components, and a large volume of rich data to fit parameters to the model. The benefits include interpretability of the computational resulting state and significant reduction in computation and mathematical and coding effort. Disadvantages include a calibrated model poorly generalizing to the system under very disparate conditions. See~\cite{seibert2022retrospective,bergstrom2015interpretation,zhang1997development,kobold2006use} for some references and use cases.

\section{Lakes and Reservoirs}

Modeling lakes and reservoirs requires a hierarchy of physical equations that represent water motion, energy exchange, and mass transport across a range of spatial and temporal scales. The selection of a modeling framework depends on the study objective, whether it is water level prediction, circulation and stratification analysis, sedimentation assessment, or evaluation of hydraulic infrastructure such as spillways, outlets, and pumping stations. This section presents the governing equation systems commonly applied to lakes and reservoirs, with emphasis on what is modeled and how boundary conditions and operational controls are represented.

\subsection{Lake Models}

Lakes are generally characterized by large surface areas, relatively low mean flow velocities, and vertical density stratification driven by thermal and compositional gradients. As a result, lake modeling focuses on circulation, mixing, and transport processes, as well as interactions with inflowing and outflowing rivers.

\subsubsection{Governing Equation Systems for Lake Dynamics}


\paragraph{Navier--Stokes Equations for Lake Circulation}

The Navier--Stokes equations form the core of three-dimensional hydrodynamic lake models. They describe the conservation of momentum for an incompressible, Newtonian fluid and are coupled with the continuity equation to ensure mass conservation. These equations are essential for modeling wind-driven circulation, internal waves, vertical mixing, and density-driven flows associated with thermal stratification.

The incompressible Navier--Stokes equations are written as:
\begin{equation}
\frac{\partial \mathbf{u}}{\partial t} + (\mathbf{u} \cdot \nabla)\mathbf{u}
= -\frac{1}{\rho}\nabla p + \nu \nabla^2 \mathbf{u} + \mathbf{f},
\end{equation}
\begin{equation}
\nabla \cdot \mathbf{u} = 0,
\end{equation}
where the variables are defined in Table~\ref{tab:NS_variables}.

\begin{table}[ht]
\centering
\caption{Variables used in the Navier--Stokes equations for lake modeling.}
\label{tab:NS_variables}
\begin{tabular}{ll}
\hline
Symbol & Description \\
\hline
$\mathbf{u} = (u,v,w)$ & Velocity components in the $x$-, $y$-, and $z$-directions \\
$t$ & Time \\
$\rho$ & Water density \\
$p$ & Pressure \\
$\nu$ & Kinematic viscosity \\
$\mathbf{f}$ & Body forces (e.g., gravity, Coriolis force) \\
\hline
\end{tabular}
\end{table}

Non-dimensionalization introduces the Reynolds number, $Re = UL/\nu$, which characterizes the relative importance of inertial and viscous forces. Large lakes typically operate at high Reynolds numbers, implying turbulent flow and the need for turbulence closure models.

\paragraph{Euler and Bernoulli Equations}

Inviscid approximations are sometimes applied for idealized or preliminary analyses. Euler’s equations are obtained from the Navier--Stokes equations by neglecting viscous effects. Bernoulli’s equation represents a further simplification for steady, incompressible flow along a streamline:
\begin{equation}
p + \frac{1}{2}\rho v^2 + \rho g h = \text{constant}.
\end{equation}

In lake applications, Bernoulli’s equation is primarily used for evaluating energy relationships at outlets, spillways, or connecting channels rather than for internal circulation.

\begin{table}[ht]
\centering
\caption{Variables used in Bernoulli’s equation.}
\label{tab:Bernoulli_variables}
\begin{tabular}{ll}
\hline
Symbol & Description \\
\hline
$p$ & Pressure \\
$\rho$ & Water density \\
$v$ & Flow velocity \\
$g$ & Gravitational acceleration \\
$h$ & Elevation above a reference datum \\
\hline
\end{tabular}
\end{table}

\paragraph{Thermal Energy and Scalar Transport}

Thermal stratification strongly influences lake circulation and mixing. Heat transport is modeled using an advection--diffusion equation:
\begin{equation}
\frac{\partial T}{\partial t} + (\mathbf{u} \cdot \nabla)T
= \alpha \nabla^2 T + Q.
\end{equation}

Transport of salinity, nutrients, or pollutants is described using an analogous scalar transport equation:
\begin{equation}
\frac{\partial S}{\partial t} + (\mathbf{u} \cdot \nabla)S
= D \nabla^2 S + Q_s.
\end{equation}

\begin{table}[ht]
\centering
\caption{Variables used in thermal and scalar transport equations.}
\label{tab:scalar_variables}
\begin{tabular}{ll}
\hline
Symbol & Description \\
\hline
$T$ & Temperature \\
$S$ & Scalar concentration (e.g., salinity, pollutant) \\
$\alpha$ & Thermal diffusivity \\
$D$ & Scalar diffusivity \\
$Q, Q_s$ & Source or sink terms \\
$\mathbf{u}$ & Velocity field \\
\hline
\end{tabular}
\end{table}

Density variations link these transport processes to hydrodynamics through the equation of state:
\begin{equation}
\rho = \rho(T,S,p),
\end{equation}
which governs buoyancy forces and stratification.

\subsubsection{Inflow and Outflow Boundary Conditions: Lake--River Interaction}

Lakes exchange mass, momentum, heat, and constituents with rivers. These interactions are commonly represented through boundary fluxes rather than fully resolved internal equations. River inflows may generate density currents, while outflows control lake levels and downstream hydrology.

Unsteady river flows connected to lakes are often modeled using the Saint--Venant (shallow water) equations:
\begin{equation}
\frac{\partial h}{\partial t} + \frac{\partial (hv)}{\partial x} = 0,
\end{equation}
\begin{equation}
\frac{\partial (hv)}{\partial t}
+ \frac{\partial \left(hv^2 + \frac{1}{2}gh^2\right)}{\partial x}
= -gh\frac{\partial z}{\partial x} - f.
\end{equation}

\begin{table}[ht]
\centering
\caption{Variables used in the Saint--Venant equations.}
\label{tab:SV_variables}
\begin{tabular}{ll}
\hline
Symbol & Description \\
\hline
$h$ & Water depth \\
$v$ & Depth-averaged velocity \\
$z$ & Bed elevation \\
$g$ & Gravitational acceleration \\
$f$ & Friction or resistance term \\
\hline
\end{tabular}
\end{table}

\subsection{Reservoirs}

Reservoirs differ from natural lakes in that their hydrodynamics are strongly influenced by engineered structures and operational controls. Modeling therefore integrates natural flow processes with water management objectives and infrastructure performance.

\subsubsection{Hydrodynamic and Sediment Modeling}

Reservoir circulation can be represented using the same governing equations as lakes, but with greater emphasis on longitudinal gradients, drawdown effects, and sedimentation. Sediment transport is critical for assessing storage loss and long-term sustainability.

The conservation of sediment mass is expressed by the Exner equation:
\begin{equation}
\frac{\partial z_b}{\partial t} + \frac{\partial q_s}{\partial x} = 0.
\end{equation}

\begin{table}[ht]
\centering
\caption{Variables used in sediment transport modeling.}
\label{tab:sediment_variables}
\begin{tabular}{ll}
\hline
Symbol & Description \\
\hline
$z_b$ & Bed elevation \\
$q_s$ & Sediment flux \\
$\tau_b$ & Bed shear stress \\
$\tau_c$ & Critical shear stress \\
\hline
\end{tabular}
\end{table}

\subsubsection{Water Management and Operational Use}

Reservoirs regulate water availability for hydropower, irrigation, flood control, and domestic supply. Storage dynamics are often described using a mass balance equation:
\begin{equation}
\frac{dV}{dt} = Q_{\text{in}} - Q_{\text{out}} - Q_{\text{loss}},
\end{equation}
where inflows, controlled releases, and losses determine reservoir storage.

\subsubsection{Pumping Stations and Engineered Control Structures}

Pumping stations enable water transfer for irrigation schemes, municipal supply, and inter-basin transfers. Their operation introduces energy into the system and is commonly analyzed using Bernoulli’s equation augmented with pump head and head loss terms:
\begin{equation}
p_1 + \frac{1}{2}\rho v_1^2 + \rho g h_1 + H_p
=
p_2 + \frac{1}{2}\rho v_2^2 + \rho g h_2 + H_L.
\end{equation}

\begin{table}[ht]
\centering
\caption{Variables used in pumping system analysis.}
\label{tab:pumping_variables}
\begin{tabular}{ll}
\hline
Symbol & Description \\
\hline
$H_p$ & Pump head \\
$H_L$ & Head losses \\
$p_1, p_2$ & Inlet and outlet pressures \\
$h_1, h_2$ & Inlet and outlet elevations \\
$v_1, v_2$ & Inlet and outlet velocities \\
\hline
\end{tabular}
\end{table}

\subsection{Summary}

Modeling lakes and reservoirs relies on a structured hierarchy of physical equations, ranging from full Navier--Stokes formulations to simplified mass and energy balance relationships. Lake models emphasize circulation, stratification, and river boundary exchanges, while reservoir models additionally account for sediment dynamics, operational controls, and pumping infrastructure. The appropriate choice of equation system and dimensionality depends on the modeling objective, data availability, and management requirements.

\section{Numerical Methods - Finite Difference Methods} 

Numerical modeling of partial differential equations (PDEs) relies on replacing continuous derivatives with discrete approximations defined on a computational grid. Finite difference methods provide a systematic framework for approximating spatial and temporal derivatives using values of the dependent variable evaluated at discrete points. These approximations form the foundation for constructing numerical schemes for diffusion, advection, and wave propagation problems.

For a scalar field \( u(x,t) \), spatial derivatives are approximated using neighboring grid points separated by a uniform spacing \( \Delta x \). The choice of approximation affects the accuracy, stability, and suitability of the resulting numerical model.

The forward difference approximation estimates the first spatial derivative using information downstream of a grid point and is given by
\begin{equation}
\frac{\partial u}{\partial x} \approx \frac{u(x+\Delta x) - u(x)}{\Delta x}.
\end{equation}
This scheme is first-order accurate and is commonly used in explicit and upwind-based numerical models.

The backward difference approximation also achieves first-order accuracy and utilizes upstream information:
\begin{equation}
\frac{\partial u}{\partial x} \approx \frac{u(x) - u(x-\Delta x)}{\Delta x}.
\end{equation}
This formulation is frequently associated with implicit discretization strategies.

Higher accuracy is obtained using the central difference approximation, which symmetrically samples the solution about a grid point:
\begin{equation}
\frac{\partial u}{\partial x} \approx \frac{u(x+\Delta x) - u(x-\Delta x)}{2\Delta x}.
\end{equation}
This second-order accurate scheme is widely adopted in numerical models where truncation error control is essential.

Second-order spatial processes such as diffusion are modeled using the central difference approximation of the second derivative:
\begin{equation}
\frac{\partial^2 u}{\partial x^2} \approx
\frac{u(x+\Delta x) - 2u(x) + u(x-\Delta x)}{\Delta x^2}.
\end{equation}
This formulation forms the core spatial operator in many elliptic and parabolic PDE solvers.

\subsubsection*{Variable Definitions for Finite Difference Approximations}

\begin{table}[ht]
\centering
\begin{tabular}{ll}
\hline
\textbf{Symbol} & \textbf{Description} \\
\hline
\( u(x) \) & Dependent variable evaluated at spatial location \( x \) \\
\( x \) & Spatial coordinate \\
\( \Delta x \) & Uniform spatial grid spacing \\
\( \dfrac{\partial u}{\partial x} \) & First-order spatial derivative \\
\( \dfrac{\partial^2 u}{\partial x^2} \) & Second-order spatial derivative \\
\hline
\end{tabular}
\end{table}

\subsection{Time Integration and the Theta Scheme}

For time-dependent PDEs, spatial discretization must be coupled with a time integration strategy. A general and flexible framework is provided by the \(\theta\)-scheme, which unifies explicit and implicit time-stepping methods:
\begin{equation}
\frac{u^{n+1} - u^n}{\Delta t}
= \theta F(u^{n+1}) + (1-\theta) F(u^n).
\end{equation}

Here, the operator \( F(u) \) represents the fully discretized spatial component of the governing PDE. By adjusting the parameter \( \theta \), the numerical model can be tailored to meet stability and accuracy requirements.

The choice \( \theta = 0 \) yields the Forward Euler method, while \( \theta = 1 \) produces the Backward Euler method. The intermediate value \( \theta = \tfrac{1}{2} \) corresponds to the Crank--Nicolson scheme, which is second-order accurate in time and commonly employed in diffusion and wave propagation models.

\subsubsection*{Variable Definitions for the Theta Scheme}

\begin{table}[ht]
\centering
\begin{tabular}{ll}
\hline
\textbf{Symbol} & \textbf{Description} \\
\hline
\( u^n, u^{n+1} \) & Numerical solution at time levels \( n \) and \( n+1 \) \\
\( \Delta t \) & Time step size \\
\( F(u) \) & Spatial discretization operator \\
\( \theta \) & Time-weighting parameter \\
\hline
\end{tabular}
\end{table}

\subsection{Riemann Solvers in Conservation Law Modeling}

Many physical systems are governed by hyperbolic conservation laws. In one spatial dimension, these systems may be written in conservative form as
\begin{equation}
\frac{\partial \mathbf{U}}{\partial t}
+ \frac{\partial \mathbf{F}(\mathbf{U})}{\partial x} = 0.
\end{equation}

Accurate numerical modeling of such equations requires robust treatment of discontinuities such as shocks and contact waves. Riemann solvers address this challenge by computing numerical fluxes at cell interfaces through the solution of local initial-value problems.

Exact Riemann solvers are computationally expensive; therefore, approximate solvers such as the Roe and HLL methods are commonly employed in practical simulations to ensure stability, conservation, and efficiency.

\subsubsection*{Variable Definitions for Conservation Laws}

\begin{table}[ht]
\centering
\begin{tabular}{ll}
\hline
\textbf{Symbol} & \textbf{Description} \\
\hline
\( \mathbf{U} \) & Vector of conserved variables \\
\( \mathbf{F}(\mathbf{U}) \) & Flux vector \\
\( t \) & Time \\
\( x \) & Spatial coordinate \\
\hline
\end{tabular}
\end{table}

\subsection{Gauss' Divergence Theorem and Flux-Based Discretization}

Finite volume methods rely on integral formulations of conservation laws. A fundamental mathematical tool underpinning these methods is Gauss' Divergence Theorem, expressed as
\begin{equation}
\int_{V} \nabla \cdot \mathbf{F} \, dV
=
\int_{S} \mathbf{F} \cdot \mathbf{n} \, dS.
\end{equation}

This theorem enables the transformation of differential conservation laws into discrete balance equations over control volumes, where surface fluxes become the primary quantities of interest, ensuring conservation at the discrete level.

\subsubsection*{Variable Definitions for Gauss' Divergence Theorem}

\begin{table}[ht]
\centering
\begin{tabular}{ll}
\hline
\textbf{Symbol} & \textbf{Description} \\
\hline
\( \mathbf{F} \) & Vector field or flux \\
\( \nabla \cdot \mathbf{F} \) & Divergence of the flux \\
\( V \) & Control volume \\
\( S \) & Closed surface bounding the volume \\
\( \mathbf{n} \) & Outward-pointing unit normal vector \\
\hline
\end{tabular}
\end{table}

\subsection{ Semi-Implicit Euler-Lagrangian Discretization of 2D Shallow Water Equations}
\label{subsec:theta-semi-implicit-EL}

The two-dimensional shallow water equations (SWE) provide a reduced yet physically consistent representation of free-surface flows in which horizontal length scales are much larger than the vertical scale. In this study, the SWE are considered over a flat bottom, $b=0$, and expressed in non-conservative form to facilitate a semi-implicit treatment of gravity-wave dynamics. The governing equations describe the evolution of horizontal momentum and free-surface elevation under the influence of gravity and external forcing:
\begin{align}
\frac{\partial u}{\partial t}
+ u \frac{\partial u}{\partial x}
+ v \frac{\partial u}{\partial y}
&= - g \frac{\partial \eta}{\partial x} + F_x, \\
\frac{\partial v}{\partial t}
+ u \frac{\partial v}{\partial x}
+ v \frac{\partial v}{\partial y}
&= - g \frac{\partial \eta}{\partial y} + F_y, \\
\frac{\partial \eta}{\partial t}
+ \frac{\partial}{\partial x}\big((H+\eta)u\big)
+ \frac{\partial}{\partial y}\big((H+\eta)v\big)
&= 0.
\end{align}

These equations model the coupled evolution of mass and momentum in shallow free-surface flows, where fast-propagating gravity waves impose severe stability restrictions on explicit time-integration schemes.

\subsubsection*{Euler--Lagrangian Treatment of Advection}

Advection is handled using an Euler--Lagrangian approach, in which the flow variables are evolved along approximate trajectories of fluid parcels. For a grid point $(x,y)$ at time $t^{n+1}$, the departure point at the previous time level is estimated as
\begin{equation}
x^* = x - u^n \Delta t, \qquad
y^* = y - v^n \Delta t.
\end{equation}
The advected quantities are then interpolated from the known solution at time $t^n$:
\begin{equation}
u^* = u(x^*,y^*,t^n), \quad
v^* = v(x^*,y^*,t^n), \quad
\eta^* = \eta(x^*,y^*,t^n).
\end{equation}
This characteristic-based treatment effectively removes the nonlinear advective terms from the discrete equations, thereby enhancing numerical stability without introducing excessive numerical diffusion.

\subsubsection*{Semi-Implicit $\theta$-Time Integration}

The remaining non-advective terms, dominated by gravity-wave dynamics, are discretized in time using a generalized $\theta$-method following the semi-implicit formulation of \cite{casulli1990semi}. The discrete momentum equations are written as
\begin{align}
\frac{u^{n+1}-u^*}{\Delta t}
&= - g \left[(1-\theta)\frac{\partial \eta^n}{\partial x}
+ \theta\frac{\partial \eta^{n+1}}{\partial x}\right] + F_x^n, \\
\frac{v^{n+1}-v^*}{\Delta t}
&= - g \left[(1-\theta)\frac{\partial \eta^n}{\partial y}
+ \theta\frac{\partial \eta^{n+1}}{\partial y}\right] + F_y^n.
\end{align}

The continuity equation is discretized consistently in time as
\begin{equation}
\frac{\eta^{n+1}-\eta^*}{\Delta t}
+ (1-\theta)\nabla \cdot (H\mathbf{u}^n)
+ \theta\nabla \cdot (H\mathbf{u}^{n+1}) = 0.
\end{equation}

\subsubsection*{Reduction to a Helmholtz Problem}

Eliminating the velocity variables from the discrete system yields a Helmholtz-type elliptic equation for the free-surface elevation at the new time level:
\begin{equation}
\eta^{n+1}
- \theta \Delta t^2 g \nabla \cdot \big(H \nabla \eta^{n+1}\big)
=
\eta^*
- \Delta t \nabla \cdot (H \mathbf{u}^*)
+ (1-\theta)\Delta t^2 g \nabla \cdot \big(H \nabla \eta^n\big).
\end{equation}
This equation encapsulates the implicit treatment of gravity waves, which is the key mechanism enabling time steps that exceed the explicit gravity-wave CFL limit.

\subsubsection*{Spatial Discretization and Linear System Structure}

Spatial derivatives are approximated on a uniform Cartesian grid using second-order central differences. The resulting discrete system can be written compactly as
\begin{equation}
\mathbf{A}\,\boldsymbol{\eta}^{n+1} = \mathbf{b},
\end{equation}
where
\begin{equation}
\mathbf{A} = \mathbf{I} - \theta \Delta t^2 g \, \nabla_h \cdot (H \nabla_h).
\end{equation}
For constant water depth, the system matrix $\mathbf{A}$ is sparse, symmetric, and positive definite, allowing efficient solution using iterative solvers such as conjugate gradient or multigrid methods.

Once $\eta^{n+1}$ is obtained, the velocity field is updated explicitly from the discrete momentum equations, completing the time step.

\subsubsection*{Stability Considerations}

For $\theta \geq 0.5$, the scheme is unconditionally stable with respect to gravity waves \cite{casulli1990semi}. The only remaining stability restriction arises from the explicit Euler--Lagrangian advection step:
\begin{equation}
\frac{\max(|u|)\Delta t}{\Delta x}
+ \frac{\max(|v|)\Delta t}{\Delta y}
\leq 1.
\end{equation}

\subsection{Godunov Finite Volume Modeling of the Shallow Water Equations}

The Godunov scheme provides a fully conservative finite volume framework for modeling shallow water flows characterized by strong nonlinearities and discontinuities. In one spatial dimension, the shallow water equations are written in conservative form as
\begin{equation}
\frac{\partial \mathbf{U}}{\partial t}
+ \frac{\partial \mathbf{F}(\mathbf{U})}{\partial x}
= \mathbf{S}(\mathbf{U},x,t),
\end{equation}
where the conserved variables and corresponding fluxes are
\begin{equation}
\mathbf{U} =
\begin{bmatrix}
h \\ hu
\end{bmatrix},
\qquad
\mathbf{F}(\mathbf{U}) =
\begin{bmatrix}
hu \\
\dfrac{(hu)^2}{h} + \dfrac{1}{2} g h^2
\end{bmatrix}.
\end{equation}

The Godunov method evolves cell-averaged conserved quantities by balancing numerical fluxes across control-volume interfaces. These fluxes are obtained by solving local Riemann problems, which capture the physically correct propagation of shocks and rarefactions. In practice, approximate solvers such as HLL or HLLC are used to reduce computational cost while maintaining accuracy \cite{toro2013riemann}.

The explicit nature of the Godunov scheme imposes a Courant--Friedrichs--Lewy (CFL) stability constraint,
\begin{equation}
\Delta t
\leq \text{CFL}\,
\frac{\Delta x}{\max_i\left(|u_i| + \sqrt{g h_i}\right)},
\end{equation}
ensuring that no information propagates farther than one grid cell during a single time step. Although the first-order Godunov method is numerically diffusive, it is valued for its robustness and strict conservation properties. Higher-order accuracy can be achieved through MUSCL or WENO reconstructions combined with strong stability-preserving Runge--Kutta schemes \cite{leveque2002finite}.

\subsection{Model Variables and Parameters}

\begin{table}[h!]
\centering
\caption{Model variables and parameters used in the shallow water formulations.}
\begin{tabular}{llp{8cm}}
\hline
\textbf{Symbol} & \textbf{Description} & \textbf{Physical Meaning} \\
\hline
$u, v$ & Horizontal velocity components & Flow velocity in the $x$ and $y$ directions \\
$\eta$ & Free-surface elevation & Deviation from the mean water level \\
$h$ & Water depth & Total depth in conservative SWE \\
$H$ & Mean water depth & Reference depth for linearization \\
$g$ & Gravitational acceleration & Controls gravity-wave propagation speed \\
$F_x, F_y$ & External forcing terms & Represent friction, wind stress, or other source effects \\
$\Delta t$ & Time step & Temporal resolution of the simulation \\
$\Delta x, \Delta y$ & Grid spacing & Spatial resolution in horizontal directions \\
$\theta$ & Implicitness parameter & Controls weighting between explicit and implicit terms \\
$\mathbf{U}$ & Conserved state vector & Mass and momentum variables \\
$\mathbf{F}(\mathbf{U})$ & Flux vector & Transport of conserved quantities \\
$\mathbf{S}$ & Source term & Topography, friction, and Coriolis effects \\
\hline
\end{tabular}
\end{table}

\subsection{Concluding Perspective}

The semi-implicit Euler--Lagrangian formulation and the Godunov finite volume framework represent complementary modeling strategies for shallow water dynamics. The former emphasizes efficiency and stability for gravity-wave-dominated flows, while the latter prioritizes strict conservation and accurate resolution of nonlinear wave phenomena. Both approaches are widely used in operational hydrodynamic and geophysical flow models \cite{casulli1990semi,leveque2002finite,toro2013riemann}.

\section{Finite Element Formulations and Time–Stepping Schemes}

Finite element methods for hydrodynamic models are built by combining a
spatial discretisation—typically continuous Galerkin (CG) or
discontinuous Galerkin (DG) finite elements—with a suitable
time–integration method.  To illustrate this general strategy, consider
a generic conservation law in one space dimension,
\begin{equation}
\frac{\partial u}{\partial t} + \frac{\partial f(u)}{\partial x}
= s(u),
\label{eq:genericPDE}
\end{equation}
where $u(x,t)$ is the scalar state variable, $f(u)$ is the physical flux,
and $s(u)$ is a source or forcing term.  The shallow water equations
(SWE) fit naturally into this framework once the state vector and flux
functions are defined appropriately.

\subsection*{Spatial finite element discretisation}

In the \textbf{continuous Galerkin} (CG) formulation, the numerical
approximation is expanded as
\begin{equation}
u_h(x,t) = \sum_{i=1}^N u_i(t)\,\phi_i(x),
\end{equation}
where $\phi_i(x)$ are globally continuous basis functions and $u_i(t)$
are the nodal coefficients.  Substituting this expression into the weak
form obtained by multiplying~\eqref{eq:genericPDE} by a test function
$v(x)$ and integrating over an element gives
\begin{equation}
\int_{x_e}^{x_{e+1}} \frac{\partial u_h}{\partial t}\, v \, dx
+ \int_{x_e}^{x_{e+1}} \frac{\partial f(u_h)}{\partial x}\, v \, dx
= \int_{x_e}^{x_{e+1}} s(u_h)\, v \, dx.
\end{equation}
Applying integration by parts to the flux term yields
\begin{equation}
\int_{x_e}^{x_{e+1}} \frac{\partial u_h}{\partial t}\, v \, dx
- \int_{x_e}^{x_{e+1}} f(u_h)\,\frac{\partial v}{\partial x}\, dx
+ f(u_h)v\Big|_{x_e}^{x_{e+1}}
= \int_{x_e}^{x_{e+1}} s(u_h)\, v \, dx.
\label{eq:CGweak}
\end{equation}

Assembly over all elements produces the semi–discrete system
\begin{equation}
M \frac{d\mathbf{u}}{dt} + \mathbf{R}(\mathbf{u}) = \mathbf{0},
\label{eq:semidiscrete}
\end{equation}
where $\mathbf{u} = (u_1,\dots,u_N)^T$ contains the nodal unknowns.
Here
\[
M_{ij} = \int_{\Omega} \phi_i(x)\,\phi_j(x)\,dx
\]
is the \emph{mass matrix}, which represents the discrete $L^2$ inner
product of basis functions and encodes how the continuous field is
distributed over the mesh.  The vector $\mathbf{R}(\mathbf{u})$ collects
all contributions from the flux gradients and source terms.

A comprehensive discussion of these formulations can be found in
Hughes~\cite{hughes2003finite}.

In \textbf{discontinuous Galerkin} (DG) finite elements, the basis
functions are allowed to be discontinuous between elements.  The
semi–discrete DG formulation of~\eqref{eq:genericPDE} is
\begin{equation}
\int_{x_e}^{x_{e+1}} \frac{\partial u_h}{\partial t} v \, dx
- \int_{x_e}^{x_{e+1}} f(u_h)\,\frac{\partial v}{\partial x} \, dx
+ \hat{f}(u_L,u_R)\, v\big|_{x_e}^{x_{e+1}}
= \int_{x_e}^{x_{e+1}} s(u_h)\, v \, dx,
\label{eq:DGweak}
\end{equation}
where $u_L$ and $u_R$ are the interior and exterior traces on an element
boundary, and $\hat{f}(u_L,u_R)$ is a numerical flux.  For hydrodynamic
systems such as the SWE, approximate Riemann solvers are commonly used.
One example is the HLL flux,
\begin{equation}
F_{\mathrm{HLL}} =
\frac{\lambda^{+} f(u_L) - \lambda^{-} f(u_R)
+ \lambda^{+}\lambda^{-} (u_R - u_L)}
{\lambda^{+}-\lambda^{-}},
\end{equation}
where $\lambda^\pm$ are the extremal characteristic speeds.
.

Gaussian quadrature is used to evaluate the spatial integrals in
\eqref{eq:CGweak}–\eqref{eq:DGweak}.  For a one–dimensional element,
\[
\int_{x_e}^{x_{e+1}} g(x)\,dx
\approx \sum_{k=1}^{n} w_k\, g(x_k),
\]
with quadrature points $x_k$ and weights $w_k$.  In two dimensions, the
mapping from the reference triangle $(\xi,\eta)$ to the physical
coordinates $(x,y)$ takes the standard form
\[
x(\xi,\eta)=\sum_{i=1}^3 x_i\,\phi_i(\xi,\eta), \qquad
y(\xi,\eta)=\sum_{i=1}^3 y_i\,\phi_i(\xi,\eta),
\]
with the Jacobian determinant providing the geometric scaling.

\subsection*{Time–integration of the semi–discrete system}

The semi–discrete system \eqref{eq:semidiscrete} is a system of ordinary
differential equations in time.  Explicit Runge–Kutta (RK) methods
evaluate the residual $\mathbf{R}(\mathbf{u})$ at known intermediate
stages.  A general $s$–stage explicit RK method is written
\begin{equation}
\mathbf{u}^{(i)} = \mathbf{u}^n
+ \Delta t \sum_{j=1}^{i-1} a_{ij} \mathbf{R}(\mathbf{u}^{(j)}),
\qquad
\mathbf{u}^{n+1} = \mathbf{u}^n
+ \Delta t \sum_{i=1}^{s} b_i \mathbf{R}(\mathbf{u}^{(i)}).
\end{equation}
Explicit RK–DG \cite{cockburn2001runge} combinations are widely used for hyperbolic systems such
as the SWE but are constrained by a CFL condition tied to the mesh size
and wave speeds.

Implicit schemes remove the CFL restriction by evaluating the residual
at the new time level.  The backward Euler scheme is
\begin{equation}
M \frac{\mathbf{u}^{n+1}-\mathbf{u}^n}{\Delta t}
+ \mathbf{R}(\mathbf{u}^{n+1}) = \mathbf{0},
\end{equation}
requiring the solution of a nonlinear system at each step.  Such methods
are useful for hydrodynamic models when fast gravity waves enforce small
explicit time steps.

The $\theta$–method provides a unifying framework:
\begin{equation}
M \mathbf{u}^{n+1}
= M \mathbf{u}^n
- \Delta t \left[
(1-\theta)\, \mathbf{R}(\mathbf{u}^n)
+ \theta\, \mathbf{R}(\mathbf{u}^{n+1})
\right],
\label{eq:thetaMethod}
\end{equation}
where $\theta=0$ gives an explicit scheme, $\theta=1$ a fully implicit
scheme, and $\theta=\tfrac12$ the Crank–Nicolson method.

\subsection*{Connection to the shallow water equations}

The framework developed above applies directly to the shallow water
equations by choosing the state vector
$U=(h,\,hu,\,hv)^T$, inserting the SWE fluxes, and including the
appropriate source terms.  
The numerical flux determines how hydraulic jumps and nonlinear waves are
resolved, while the time–stepping method controls stability with respect
to the characteristic speeds $\sqrt{gh}$.  Explicit DG schemes are often
used for high–resolution modelling, whereas implicit or $\theta$–methods
are suitable for large–scale geophysical simulations dominated by fast
gravity waves.



\section{ Space--Time Finite Element Methods }
\label{sec:space_time_fe}

The shallow water equations (SWE) describe the evolution of water depth $h$ and horizontal discharge components $(hu, hv)$ over time and space. In conservative form, they read
\begin{equation}\label{eq:swe_conservative}
\partial_t U + \nabla \!\cdot\! F(U) = S(U,\mathbf{x},t), 
\qquad 
U = \begin{pmatrix} h \\ hu \\ hv \end{pmatrix},
\qquad
F(U) = 
\begin{pmatrix}
hu & hv \\
\frac{(hu)^2}{h} + \frac{1}{2} g h^2 & \frac{hu\,hv}{h} \\
\frac{hu\,hv}{h} & \frac{(hv)^2}{h} + \frac{1}{2} g h^2
\end{pmatrix},
\end{equation}
where $S(U,\mathbf{x},t)$ accounts for source terms such as bottom topography or Coriolis effects.  

Standard Galerkin finite element (FE) formulations of \eqref{eq:swe_conservative} are well known to produce spurious oscillations in advection-dominated regimes unless stabilization (e.g., SUPG, artificial viscosity) is introduced. The \emph{space--time finite element} approach overcomes this difficulty by treating space and time uniformly: instead of discretizing space and time separately, one defines a \emph{space--time slab}
\[
\mathcal{Q} = \Omega \times (t^n,t^{n+1}),
\]
and seeks a solution $U_h$ in a finite-dimensional trial space $V_h$ defined over $\mathcal{Q}$.  

The variational formulation is then posed as: find $U_h \in V_h$ such that for all test functions $W \in W_h$,
\begin{equation}\label{eq:space_time_variational}
\int_{\mathcal{Q}} \! W \cdot \partial_t U_h \, d\mathcal{Q}
+ 
\int_{\mathcal{Q}} \! \nabla W : F(U_h) \, d\mathcal{Q}
= 
\int_{\mathcal{Q}} \! W \cdot S(U_h) \, d\mathcal{Q} 
+ 
\mathcal{B}(W,U_h),
\end{equation}
where $\mathcal{B}(W,U_h)$ collects contributions from the inflow boundary $\partial\Omega_{-}$ and initial condition terms.  

The distinguishing feature of \emph{stable space--time finite element methods}, such as the automatic variationally stable finite element (AVS-FE) method introduced by Valseth and Dawson \cite{valseth2022stable}, is the \emph{choice of the test space} $W_h$. Rather than using the same space as $V_h$ (standard Galerkin), the Petrov--Galerkin approach selects $W_h$ to guarantee a uniform inf--sup condition, resulting in unconditional stability. In practice, this is done by solving local elementwise or patchwise problems that generate \emph{optimal test functions} for each basis function in $V_h$. 

Let $\{\varphi_i\}$ be a basis for $V_h$. For each $\varphi_i$, the optimal test function $\psi_i$ is defined as the solution to a local Riesz representation problem:
\begin{equation}\label{eq:optimal_test}
(\psi_i, w)_{W} = b(\varphi_i, w) \qquad \forall w \in W_h,
\end{equation}
where $(\cdot,\cdot)_W$ is the inner product in the test space and $b(\cdot,\cdot)$ is the bilinear form associated with \eqref{eq:space_time_variational}. The resulting discrete bilinear form
\[
a(U_h,V_h) = \sum_{i,j} u_j \, b(\varphi_j,\psi_i)
\]
is coercive or satisfies a uniform inf--sup condition, leading to a robust and stable method even for coarse meshes or large time steps.

For the SWE, this approach yields a fully discrete space--time system that is solved over each slab $\mathcal{Q}$. The algorithm proceeds by:
\begin{enumerate}
    \item Discretizing the solution variables $h, hu, hv$ in $V_h$ over the space--time slab.
    \item Solving \eqref{eq:optimal_test} locally to compute the optimal test functions $\psi_i$.
    \item Assembling and solving the global Petrov--Galerkin system using the computed $\psi_i$.
    \item Proceeding to the next slab $(t^{n+1}, t^{n+2})$ with the solution $U_h(t^{n+1})$ as initial data.
\end{enumerate}

Compared to traditional semi-discrete methods with explicit or implicit time-stepping, this space--time approach offers several advantages:  
(i) \emph{Unconditional stability}, i.e., no CFL restriction is required between spatial and temporal resolutions.  
(ii) \emph{Built-in error estimation}, since the residual norm in $W_h$ can be used to drive adaptive mesh refinement in both space and time.  
(iii) \emph{Robustness} for nonlinear SWE, including sharp wave fronts and wetting--drying interfaces.  

The work of Valseth and Dawson \cite{valseth2022stable} demonstrates that these methods can robustly capture transient phenomena in SWE while maintaining accuracy and stability even for coarse meshes, making them a powerful framework for modern flood and coastal wave modeling.

\section{Numerical Methods - Finite Volume Methods} 
\label{sec:app_swe}

The conservative shallow water equations may be cast into an \emph{advective} form so that the same velocity field transports the non-conservative variables.  This advective reformulation is convenient for the Finite Volume Characteristic (FVC) method, since it allows us to follow characteristics and evaluate interface states by interpolation from departure points.

We begin from the advective form used in the predictor stage of the FVC method:
\begin{equation}
\partial_t
\begin{pmatrix}
h \\[4pt]
u
\end{pmatrix}
+ u\,\partial_x
\begin{pmatrix}
h \\[4pt]
u
\end{pmatrix}
=
\begin{pmatrix}
-\,h\,\partial_x u \\[6pt]
-\,g\,\partial_x(h+Z)
\end{pmatrix},
\label{eq:advective_h_u}
\end{equation}
where $h(x,t)$ is the water depth, $u(x,t)$ the depth-averaged velocity, $Z(x)$ the bottom elevation and $g$ the gravitational acceleration.

Associated with the advective system \eqref{eq:advective_h_u} we define the characteristic curves $X_{i+1/2}(\tau)$ through the ordinary differential equation
\begin{equation}
\frac{d X_{i+1/2}(\tau)}{d\tau} \;=\; u\!\big(X_{i+1/2}(\tau),\tau\big),
\label{eq:characteristics}
\end{equation}
so that the value of the transported variables at the interface can be obtained from their values at the \emph{departure point} (the characteristic foot) at the previous time level. In practice the solution at the characteristic foot $X_{i+1/2}(t^n)$ is computed by interpolation from the grid (the control volume) that contains that departure point.

\medskip

\noindent\textbf{Alternative advective variable:} An alternative advective formulation, sometimes used when advecting conservative momentum is preferable, transports $(h,\, hu)$ with the same velocity field:
\begin{equation}
\partial_t
\begin{pmatrix}
h \\[4pt]
hu
\end{pmatrix}
+ u\,\partial_x
\begin{pmatrix}
h \\[4pt]
hu
\end{pmatrix}
=
\begin{pmatrix}
-\,h\,\partial_x u \\[6pt]
-\,hu\,\partial_x u \;-\; \tfrac{1}{2} g\,\partial_x(h^2)
\end{pmatrix}.
\label{eq:advective_h_hu}
\end{equation}

Both forms \eqref{eq:advective_h_u} and \eqref{eq:advective_h_hu} are used in the predictor stage to compute consistent interface states. The predictor uses the characteristic information to obtain provisional (predicted) values at the interfaces; these predicted interface states are then used to compute numerical fluxes for the corrector (conservative) update.

\medskip

\noindent\textbf{Corrector stage and source discretization.} The corrector stage advances the conserved variables using the numerical fluxes computed at the cell faces and a discrete representation of the source term. A central concern in discretizing the source associated with the bed slope is to preserve steady states (in particular the \emph{lake at rest}) — i.e. the well-balanced property.

Let $h^n_i$ and $Z_i$ denote the cell-averaged water depth and the bottom elevation at cell center $x_i$ at time $t^n$. For a stationary (lake-at-rest) solution we have the balance
\[
h^{n}_{i+1} - h^{n}_{i-1} = Z_{i+1} - Z_{i-1},
\]
which is the discrete manifestation of the hydrostatic equilibrium. When the discrete corrector balance is enforced consistently, the scheme preserves this equilibrium.

Consider the discrete approximation of the source term $g h \partial_x Z$. Under the above stationary relation, the discrete representation used in the corrector stage can be written in the form
\begin{equation}
\big( g h\,\partial_x Z\big)^n_i
\;=\;
g \;\frac{h^{n}_{i+\frac{1}{2}} + h^{n}_{i-\frac{1}{2}}}{2}\;
\frac{Z^{n}_{i+1} - Z^{n}_{i-1}}{2\,\Delta x},
\label{eq:discrete_gh_dZ}
\end{equation}
where $h^{n}_{i\pm 1/2}$ are suitable approximations of the depth at the half-cells (for example arithmetic averages of neighbouring cell averages or values obtained from the predictor). The form \eqref{eq:discrete_gh_dZ} shows the source written consistently as a product of an averaged depth and a centred difference of the bottom slope.

From \eqref{eq:discrete_gh_dZ} we obtain the commonly used expression (quoted in the literature) for the discrete source term at cell $i$:
\begin{equation}
\big( g h\,\partial_x Z\big)^n_i
\;=\;
g \,\frac{h^{n}_{i+\frac{1}{2}} + h^{n}_{i-\frac{1}{2}}}{2}\;
\frac{Z^{n}_{i+1} - Z^{n}_{i-1}}{2\,\Delta x}.
\label{eq:ghdxz_final}
\end{equation}

A practical choice for the source contribution used in the corrector stage (denoted here by $\widehat{h}^{\,n}_i$) which enforces the well-balanced property, is
\begin{equation}
\widehat{h}^{\,n}_i \;=\; \frac{1}{4}\,\big( h^n_{i+1} + h^n_{i-1} \big)\;
\frac{Z^n_{i+1} - Z^n_{i-1}}{\Delta x}.
\label{eq:hat_h}
\end{equation}
This discretization follows directly from the centered representation in \eqref{eq:ghdxz_final} by approximating the half-cell depths with cell averages; it is designed so that when the discrete hydrostatic relation $h^{n}_{i+1}-h^{n}_{i-1}=Z_{i+1}-Z_{i-1}$ holds, the source term precisely balances the pressure gradients and the numerical scheme remains in equilibrium.

\medskip

\noindent\textbf{Implementation notes.}
\begin{itemize}
  \item \textbf{Characteristic interpolation:} The predictor requires evaluation of the solution at departure points $X_{i+1/2}(t^n)$. These values are computed by interpolating the cell-based data from the control volume that contains the departure point. The interpolation order controls the overall spatial accuracy of the predictor.
  \item \textbf{Numerical fluxes:} Once predicted, interface states are available, a Riemann-type numerical flux or an upwind flux consistent with the advective velocity may be used in the corrector update. The choice of flux should be compatible with the predictor to avoid spurious oscillations.
  \item \textbf{Well-balanced property:} The source discretization \eqref{eq:hat_h} (or its equivalent) must be used together with the flux evaluation to guarantee that the discrete lake-at-rest is preserved exactly (up to machine precision).
  \item \textbf{Time stepping:} The predictor–corrector procedure can be embedded in a strong-stability preserving time integrator (for example SSP-RK schemes such as RK3) to obtain high-order time accuracy while maintaining stability.
\end{itemize}

\medskip

To summarize, the FVC approach for the shallow water equations proceeds by (i) rewriting the governing system in a transport (advective) form, (ii) following characteristics to compute predicted interface values by interpolation, (iii) using these predictions to assemble numerical fluxes, and (iv) performing a conservative corrector update with a carefully discretized source term such as \eqref{eq:hat_h} to preserve steady states. The combination of characteristic-based prediction and the well-balanced source discretization yields a robust finite volume scheme for problems with nontrivial bathymetry and wet/dry transitions.

 \section{Spectral Finite Volume Method}
The 2D Shallow Water Equations (SWE) in vector and conservative form are given by:
\begin{equation}\label{sfv}
    \frac{\partial U}{\partial t} + \frac{\partial E}{\partial x} + \frac{\partial G}{\partial y} = S
\end{equation}
where
\begin{align*} 
U &= \begin{bmatrix}
    h\\hu\\hv
\end{bmatrix}, &
E &=  \begin{bmatrix} hu \\ hu^2 + \frac{1}{2} g h^2\\huv\end{bmatrix}, &
G &= \begin{bmatrix} hv\\huv \\ hv^2 + \frac{1}{2} g h^2  \frac{\partial z_b}{\partial x} \end{bmatrix}, &
S &= \begin{bmatrix}0\\gh\frac{\partial h}{\partial x}\\gh\frac{\partial h}{\partial y}
\end{bmatrix}
\end{align*}
Here $h$ is the water depth, $u$ and $v$ are the depth-averaged velocity components in the $x$ and $y$ directions respectively, and $g$ is the gravitational constant. 

Applying the finite-volume formulation to \eqref{sfv} and using the Gauss theorem yields:
\begin{equation}\label{sfv1}
     \int_{\Omega} \frac{\partial U}{\partial t}\ d\Omega + \oint_{\Gamma}( E\eta_x  + G\eta_y)\ d\Gamma = \int_{\Omega}S\ d\Omega
\end{equation}
where $\eta_x$ and $\eta_y$ are the outward unit normals to the domain boundary $\Gamma$. Defining the volume-averaged state and flux terms, the equation can be rewritten as:
\begin{equation}\label{svf2}
      \frac{\partial \overline{U}}{\partial t} + \frac{1}{V}\oint_{\Gamma} \mathcal{F}\cdot\eta\ d\Gamma = \overline{S}
\end{equation}
where $E\eta_x + G\eta_y = \mathcal{F}\cdot\eta$ and 
\begin{align*}
      \overline{U} = \frac{1}{V}\int_{\Omega} U\ d\Omega.
\end{align*}
Equation \eqref{svf2} is exact and, being an ordinary differential equation in time, is integrated using a third-order Runge–Kutta (RK3) scheme to maintain a balance between accuracy and computational efficiency:
\begin{align*}
    R(\overline{U},t) &= - \frac{1}{V}\oint_{\Gamma} \mathcal{F}\cdot\eta\ d\Gamma + \overline{S},\\
    R_0 &= R(\overline{U}, t), & \overline{U}_1 &= \overline{U} + \Delta t\,R_0,\\
    R_1 &= R(\overline{U}_1, t+\Delta t), & \overline{U}_2 &= \overline{U} + \frac{\Delta t (R_0 + R_1)}{2},\\
    R_2 &= R(\overline{U}_2, t+\frac{\Delta t}{2}), & \overline{U}(t+\Delta t) &= \overline{U}(t) + \Delta t \frac{R_0 + 4R_2 + R_1}{6}.
\end{align*}

Once temporal integration is defined using the third-order Runge–Kutta scheme, the next step is to approximate the flux integrals along cell interfaces accurately. In the Spectral Finite Volume Method (SFVM), this is achieved through a local high-order reconstruction of the solution field inside each control volume. Instead of assuming piecewise-constant states as in standard finite volume schemes, SFVM expands the conserved variables using a set of local spectral basis functions. This enables the method to achieve higher accuracy without increasing the stencil size.

The reconstructed polynomial solution provides accurate left and right states at each interface. These states are then passed to an approximate Riemann solver (e.g. Roe or HLLC) to compute the numerical fluxes required in the flux integral term of \eqref{svf2}. This combination of high-order reconstruction and upwind-biased flux evaluation allows the method to resolve complex flow features such as shocks, bores, and smooth free-surface waves with reduced numerical dissipation \cite{choi2004spectral}.

Additionally, limiter functions can be applied to prevent spurious oscillations near discontinuities, ensuring stability and non-oscillatory behavior while retaining the high-order accuracy in smooth regions. This makes the SFVM particularly suitable for shallow water problems where both smooth and rapidly varying flow regions coexist.

Finally, the fully discrete update at each time step involves:
\begin{enumerate}
    \item Reconstructing the spectral solution in each control volume,
    \item Evaluating interface fluxes via Riemann solvers,
    \item Updating the cell-averaged states using the RK3 scheme.
\end{enumerate}
This procedure provides a robust and accurate framework for simulating nonlinear shallow water flows on structured grids.

\section{Other Numerical Methods}

\subsection{Artificial Viscosity Method}

In numerical treatments of the shallow water equations (SWEs), particularly in the presence of discontinuities such as hydraulic jumps or wet--dry fronts, nonphysical oscillations often appear when using standard discretizations. To suppress these oscillations and enhance stability, viscosity--based stabilization methods are introduced. A common approach is to add an artificial viscosity term into the governing equations, which regularizes the system while preserving conservation of mass and momentum \cite{stavropoulou2021residual}. 

The SWE in conservative form is written as
\begin{equation}\label{VS}
    \frac{\partial q}{\partial t} + \frac{\partial F(q)}{\partial x} + b(q) = 0 ,
\end{equation}
where
\[
q = \begin{bmatrix} h \\ hu \end{bmatrix}, 
\quad 
F(q) = \begin{bmatrix} hu \\ hu^2 + \tfrac{1}{2} g h^2 \end{bmatrix},
\quad 
b(q) = \begin{bmatrix} 0 \\ -g h \frac{\partial z_b}{\partial x} \end{bmatrix},
\]
with $h$ the water depth, $u$ the velocity, $hu$ the discharge, $g$ the gravitational acceleration, and $z_b = z_0 - \alpha x^2$ the bathymetry profile.  

To stabilize the numerical solution, we add a viscous flux $F_{\text{visc}}$ so that the equations become
\begin{equation}\label{vs1}
     \frac{\partial q}{\partial t} + \frac{\partial F(q)}{\partial x} + b(q) - \frac{\partial F_{\text{visc}}(q)}{\partial x} = 0,
\end{equation}
with
\begin{equation}\label{vs2}
    F_{\text{visc}}(q) =
    \begin{bmatrix}
        \mu \tfrac{\partial h}{\partial x} \\[6pt]
        \mu \tfrac{\partial (hu)}{\partial x}
    \end{bmatrix},
\end{equation}
where $\mu \geq 0$ denotes the artificial viscosity coefficient. In residual-based formulations, $\mu$ is not constant but depends on the local residual of the discretized equations. This means that viscosity is activated only where the residual is large, namely in regions of under-resolution or strong gradients, while remaining negligible elsewhere. In this way, sharp features are stabilized without introducing unnecessary diffusion in smooth regions. 

Let $\mathcal{T}_h$ be a subdivision of the computational domain $\Omega$ into disjoint finite elements $K$. On this mesh, we define a continuous finite element space $V_h$ and seek approximations $h_h \in V_h$ and $m_h = h_h u_h \in V_h$ such that
\[
q_h := (h_h, m_h) \in W_h = V_h \times V_h .
\]
The weak formulation of \eqref{vs1} reads: find $q_h \in W_h$ such that
\begin{equation}
    (\partial_t q_h, v_h) + (\partial_x F(q_h), v_h) + (b(q_h), v_h)
    + (F_{\text{visc}}(q_h), \partial_x v_h)
    - (\eta \cdot F_{\text{visc}}(q_h), v_h)_\Gamma = 0 ,
\end{equation}
for all test functions $v_h \in W_h$, where $(\cdot,\cdot)$ and $(\cdot,\cdot)_\Gamma$ denote element--wise volume and interior face integrals, respectively. The viscous term acts as a parabolic regularization that improves stability and ensures well-balancedness when bathymetric effects are included. To avoid singularities in dry states, the velocity can be regularized as
\[
u_h = \frac{h_h}{h_h^2 + \epsilon} m_h,
\]
with $\epsilon > 0$ preventing division by zero. We now modify the viscous component of the water equation in \eqref{vs2} to incorporate the bathymetry \( z_b \), ensuring that the numerical scheme remains well-balanced and preserves the steady-state at rest. The parabolic flux takes the form of :
        \begin{equation}  \label{vs3}
        F_{visc}(q_h) = \begin{bmatrix}
            \mu_h \frac{\partial (h + z_b)}{\partial x} \\ \mu_h \frac{\partial hu}{\partial x}
        \end{bmatrix}
             \end{equation}
             Here $\mu_h$ is the residual-based artificial viscosity coefficient.
Residual-based artificial viscosity is computed in a sequence of well-defined steps that make the added dissipation both adaptive (non-zero only where needed) and consistent with the governing equations.  The practical algorithm is:

\begin{itemize}
  \item \textbf{Element residual.}  On each element \(K\) evaluate a measure of the local PDE residual
  \[
    R_K(x) \;=\; \partial_t q_h(x) + \partial_x F(q_h(x)) + b(q_h(x)),
    \qquad x\in K,
  \]
  and form a scalar norm of this residual (commonly the \(L^2\)-norm)
  \[
    r_K \;=\; \|R_K\|_{L^2(K)} \;=\; \Big( \int_K |R_K(x)|^2 \,dx\Big)^{1/2}.
  \]
  The residual \(r_K\) quantifies how strongly the discrete solution violates the PDE on element \(K\).
  \item \textbf{Normalization / sensor.}  To make the residual dimensionless and comparable across elements introduce a sensor
  \[
    s_K \;=\; \frac{r_K}{\| q_h - \bar q_K\|_{L^2(K)} + \epsilon},
  \]
  where \(\bar q_K\) is a local average of \(q_h\) on \(K\) (or another local scale), and \(\epsilon\) is a small number to avoid division by zero in dry or very small states.  The sensor \(s_K\) is large where the residual is large compared to the local solution amplitude (e.g., near shocks or under-resolved gradients).
  \item \textbf{Viscosity coefficient.}  Use the sensor to define an element viscosity \(\mu_K\). A common (and effective) choice is
  \[
    \mu_K \;=\; C_\mu \, h_K \, s_K,
  \]
  where \(h_K\) is the element length (or characteristic mesh size) and \(C_\mu\) is a tunable constant of order unity.  This scaling ensures that the added dissipation vanishes with mesh refinement in smooth regions (because \(r_K\to 0\) there) but grows where the residual indicates under-resolution.
  \item \textbf{Capping and spectral limit.}  To avoid unphysically large dissipation a maximum value is enforced, typically linked to the local wave speed:
  \[
    \mu_K \leftarrow \min\!\big(\mu_K,\; \mu_{\max,K}\big),
    \qquad \mu_{\max,K} \;=\; C_{\max}\, h_K\, c_K,
  \]
  where \(c_K=\max_{x\in K} (|u(x)|+\sqrt{g\,h(x)})\) is the local characteristic (wave) speed and \(C_{\max}\) is a constant (often \(\le 1\)).  The cap preserves the correct physical scaling and prevents excessive smearing.
  \item \textbf{Smoothing / patch operation.}  The raw elementwise \(\mu_K\) is often smoothed (for example, by averaging over a small patch or taking a local maximum) to avoid element-to-element oscillations in the viscosity field and to improve robustness.
  \item \textbf{Viscous flux insertion.}  The computed viscosity field \(\mu(x)\) (piecewise constant or smoothed) is inserted into a viscous flux of the form
  \[
    F_{\text{visc}}(q_h) \;=\; \begin{bmatrix} \mu(x)\,\partial_x (h_h + z_b) \\[4pt] \mu(x)\,\partial_x (h_h u_h) \end{bmatrix},
  \]
  or an equivalent consistent parabolic regularization.  Using \(\partial_x(h+z_b)\) in the first component helps preserve well-balanced properties when bathymetry \(z_b\) is present.
\end{itemize}

\textbf{Rationale and properties.} The residual \(r_K\) directly measures the local imbalance of the discrete PDE, so basing the viscosity on \(r_K\) makes the dissipation adaptive: it is essentially zero in smooth, well-resolved regions and activates only near shocks, steep gradients, or under-resolved features. Normalizing by a local solution scale prevents spurious large viscosities in regions with very small solution magnitude (e.g., dry beds). The dependence on \(h_K\) yields the correct mesh-scaling so that the method is consistent (viscosity vanishes as the mesh is refined). Capping by a local wave-speed-based limit prevents excessive smearing and keeps the numerical viscosity in a physically meaningful range.

For implementation details (specific choices of the sensor, constants \(C_\mu, C_{\max}\), and smoothing operators), see Stavropoulou (2021), which presents a practical residual-based artificial viscosity method for the shallow-water equations and provides the underlying mathematical justification.
\subsection{Lattice Boltzmann Method}

The Lattice Boltzmann Method (LBM) is a mesoscopic numerical approach used to simulate fluid flows, including shallow water dynamics, by bridging kinetic theory and continuum mechanics \cite{chen1998lattice}. Unlike the Finite Difference or Finite Element Methods that discretize the macroscopic partial differential equations directly, the LBM models the evolution of particle distribution functions on a discrete lattice. Through suitable averaging, the macroscopic nonlinear two-dimensional shallow water equations (SWE) are recovered from these microscopic dynamics.

\bigskip
\noindent\textbf{Mesoscopic versus Macroscopic Description:}

In the macroscopic framework, the shallow water equations describe the evolution of the water depth $h(x,y,t)$ and the depth-averaged horizontal velocity $\mathbf{u}(x,y,t) = (u,v)$:
\begin{align}
\frac{\partial h}{\partial t} + \nabla \cdot (h \mathbf{u}) &= 0, \label{eq:contSWE}\\
\frac{\partial (h \mathbf{u})}{\partial t} + \nabla \cdot \left( h \mathbf{u} \otimes \mathbf{u} + \frac{1}{2} g h^2 \mathbf{I} \right) &= \mathbf{F}, \label{eq:momSWE}
\end{align}
where $g$ is the gravitational acceleration and $\mathbf{F}$ represents external forces such as bed friction or bottom slope effects.

In the mesoscopic picture, the LBM introduces a \textit{distribution function} $f_i(\mathbf{x}, t)$ that represents the probability of fluid ``particles’’ moving with discrete velocity $\mathbf{e}_i$ at position $\mathbf{x} = (x, y)$ and time $t$. The macroscopic quantities are obtained from velocity moments of $f_i$:
\begin{equation}
h = \sum_i f_i, \qquad h \mathbf{u} = \sum_i \mathbf{e}_i f_i.
\end{equation}

\bigskip
\noindent\textbf{Discrete Boltzmann Equation with BGK Approximation:}

The time evolution of the distribution functions follows the discrete Boltzmann equation with the Bhatnagar–Gross–Krook (BGK) collision operator:
\begin{equation}
f_i(\mathbf{x} + \mathbf{e}_i \Delta t, t + \Delta t) - f_i(\mathbf{x}, t)
= -\frac{\Delta t}{\tau} \left[ f_i(\mathbf{x}, t) - f_i^{eq}(\mathbf{x}, t) \right] + F_i,
\end{equation}
where $\tau$ is the relaxation time controlling the effective viscosity, $f_i^{eq}$ is the equilibrium distribution, and $F_i$ is the source term that accounts for external body forces.

\bigskip
\noindent\textbf{D2Q9 Lattice Configuration:}

For two-dimensional shallow water flows, the standard lattice model is D2Q9, which uses nine discrete velocities:
\[
\mathbf{e}_i =
\begin{cases}
(0, 0), & i = 0, \\
(\pm c, 0), (0, \pm c), & i = 1,2,3,4, \\
(\pm c, \pm c), & i = 5,6,7,8,
\end{cases}
\]
where $c = \Delta x / \Delta t$ is the lattice speed. Each direction is assigned a weight:
\[
w_i =
\begin{cases}
\frac{4}{9}, & i=0,\\
\frac{1}{9}, & i=1,2,3,4,\\
\frac{1}{36}, & i=5,6,7,8.
\end{cases}
\]

\bigskip
\noindent\textbf{Equilibrium Distribution Function:}

To ensure that the macroscopic equations recovered from the mesoscopic dynamics correspond to the 2D shallow water equations \eqref{eq:contSWE}–\eqref{eq:momSWE}, the equilibrium distribution function is defined as:
\begin{equation}
f_i^{eq} = w_i \left[ h + \frac{3}{c^2} (\mathbf{e}_i \cdot h \mathbf{u}) 
+ \frac{9}{2 c^4} (\mathbf{e}_i \cdot h \mathbf{u})^2 - \frac{3}{2 c^2} (h |\mathbf{u}|^2) \right].
\end{equation}
Here, the terms represent the hydrostatic pressure, advection, and kinetic energy contributions to the flow.

\bigskip
\noindent\textbf{Macroscopic Recovery of the 2D Shallow Water Equations:}

By applying a Chapman–Enskog multiscale expansion to the discrete Boltzmann equation and taking zeroth and first velocity moments, one recovers the nonlinear shallow water equations:
\begin{align}
\frac{\partial h}{\partial t} + \nabla \cdot (h \mathbf{u}) &= 0, \\
\frac{\partial (h \mathbf{u})}{\partial t} + \nabla \cdot \left( h \mathbf{u} \otimes \mathbf{u} + \frac{1}{2} g h^2 \mathbf{I} \right) &= \mathbf{F},
\end{align}
thus establishing the link between the lattice-based kinetic dynamics and the macroscopic hydrodynamic behavior.

\bigskip
\noindent\textbf{Relation to Viscosity and Stability:}

The kinematic viscosity $\nu$ of the macroscopic flow is related to the relaxation time $\tau$ by:
\begin{equation}
\nu = \frac{g h (\tau - \frac{1}{2}) \Delta t}{3}.
\end{equation}
This shows that the relaxation process on the mesoscopic scale governs the dissipative characteristics of the macroscopic flow. Numerical stability requires $\tau > \frac{1}{2}$ and a Courant–Friedrichs–Lewy (CFL) condition satisfying:
\begin{equation}
\text{CFL} = \frac{|\mathbf{u}| \Delta t}{\Delta x} \leq 1.
\end{equation}

\bigskip
\noindent\textbf{Remarks:}

\begin{itemize}
\item The Lattice Boltzmann Method offers a mesoscopic alternative to classical solvers such as FDM, FEM, and FVM for 2D shallow water flows.
\item It inherently conserves mass and momentum and is highly parallelizable.
\item The method efficiently handles complex geometries and boundary conditions.
\item The D2Q9 lattice configuration captures the essential two-dimensional hydrodynamic behavior with excellent computational efficiency.
\end{itemize}

\subsection{Virtual Element Method (VEM)}

The Virtual Element Method (VEM) is a numerical technique designed to approximate partial differential equations on general polygonal and polyhedral meshes. Unlike the classical finite element method (FEM), VEM does not require explicit basis functions inside each element. Instead, it constructs local approximation spaces that are ``virtual'' (not known in closed form) but are designed to reproduce polynomials up to degree $k$ and allow the exact computation of certain projection operators. This makes VEM particularly suited for problems with complex geometries, such as the shallow water equations (SWE) posed over irregular coastal domains or unstructured grids.

\textbf{Construction of the Method:}

Let $\Omega \subset \mathbb{R}^2$ be a polygonal domain partitioned into a mesh $\mathcal{T}_h$ consisting of generic polygons $K$. The virtual element space of order $k$ on each element $K$ is defined as
\[
V_k(K) = \left\{ v \in H^1(K) \;\middle|\; v|_{\partial K} \in \mathbb{B}_k(\partial K), \; \Delta v \in \mathbb{P}_{k-2}(K) \right\},
\]
where $\mathbb{P}_{k}(K)$ is the space of polynomials of degree $\leq k$ on $K$, and $\mathbb{B}_k(\partial K)$ denotes continuous functions on $\partial K$ whose restrictions to each edge are polynomials of degree $\leq k$.

Two fundamental requirements are built into the construction:
\begin{enumerate}
    \item \textbf{Polynomial consistency:} $\mathbb{P}_k(K) \subset V_k(K)$, ensuring that polynomials up to degree $k$ are exactly represented.
    \item \textbf{Stability:} a computable bilinear form must be constructed such that the resulting scheme is stable on arbitrary polygons.
\end{enumerate}

Since explicit basis functions are not available in $V_k(K)$, VEM relies on \emph{projection operators} onto polynomial spaces. For example, the $H^1$-projection $\Pi^\nabla_k : V_k(K) \to \mathbb{P}_k(K)$ is defined by
\[
\int_K \nabla \big( v - \Pi^\nabla_k v \big) \cdot \nabla q \, dx = 0, 
\quad \forall q \in \mathbb{P}_k(K).
\]

The discrete bilinear form for an elliptic operator (e.g., Laplace) is then defined as
\[
a^K_h(u,v) = \int_K \nabla \Pi^\nabla_k u \cdot \nabla \Pi^\nabla_k v \, dx
+ S^K \big( (I - \Pi^\nabla_k) u, (I - \Pi^\nabla_k) v \big),
\]
where $S^K(\cdot,\cdot)$ is a stabilization term computable from the degrees of freedom. This ensures both consistency (first term) and stability (second term).

\textbf{Relevance for Shallow Water Equations}

In semi-implicit formulations of the SWE, one often needs to solve elliptic or Helmholtz-type subproblems of the form
\[
- \nabla \cdot (H \nabla \eta) = f,
\]
where $\eta$ is the free-surface elevation and $H$ the depth. VEM provides a natural framework for discretizing such problems on highly irregular coastal meshes, without requiring element-shape restrictions. The use of projection operators ensures that the pressure-gradient and divergence terms are accurately captured, while stability is guaranteed by the structure of the bilinear form.

\textbf{Key Features}

The foundation of VEM can thus be summarized as:
\begin{itemize}
    \item Definition of local virtual spaces $V_k(K)$ containing polynomials.
    \item Use of projection operators (e.g., $\Pi^\nabla_k$) to achieve computability without explicit basis functions.
    \item Construction of bilinear forms combining exact polynomial consistency and stabilization.
\end{itemize}

This makes VEM a promising tool for future shallow water solvers, particularly in applications with complex bathymetry and unstructured grids, where standard finite elements or finite volumes face limitations.

\subsection{Wavelet-Based Methods for the Shallow Water Equations}

Wavelet-based methods provide a powerful multiresolution framework for efficiently solving the shallow water equations (SWE). The fundamental idea is to represent the solution on a hierarchy of spatial scales, which allows for dynamically refining the grid only in regions of interest such as shocks, bores, wet–dry fronts, or steep gradients, while keeping the grid coarse elsewhere. This adaptive strategy significantly reduces computational cost without compromising accuracy.

Mathematically, at any given time \( t \), a scalar field \( U(x,t) \) can be represented by a wavelet expansion
\[
U(x,t) = \sum_{k} c_{J,k}(t) \, \phi_{J,k}(x) 
+ \sum_{j=J}^{J_{\max}} \sum_{k} d_{j,k}(t) \, \psi_{j,k}(x),
\]
where \( \phi_{J,k} \) are the scaling functions at the coarsest level \( J \), and \( \psi_{j,k} \) are wavelet functions at finer levels \( j \). The coefficients \( c_{J,k}(t) \) represent the large-scale structure of the solution, while the detail coefficients \( d_{j,k}(t) \) encode local fluctuations. After applying the discrete wavelet transform \( \mathcal{W}[\cdot] \), small coefficients are discarded through thresholding
\[
d_{j,k}(t) = 0 \quad \text{if} \quad |d_{j,k}(t)| < \varepsilon,
\]
where \( \varepsilon \) is a user-defined tolerance. This operation generates a sparse set of significant coefficients, which directly corresponds to an adaptive grid \(\mathcal{G}(t)\) that refines around regions of strong spatial variation and coarsens elsewhere.

The two-dimensional shallow water equations can be written in conservative vector form as
\[
\frac{\partial \mathbf{U}}{\partial t} 
+ \frac{\partial \mathbf{F}(\mathbf{U})}{\partial x} 
+ \frac{\partial \mathbf{G}(\mathbf{U})}{\partial y} = 0,
\]
with
\[
\mathbf{U} =
\begin{pmatrix}
h \\ hu \\ hv
\end{pmatrix}, 
\quad
\mathbf{F}(\mathbf{U}) =
\begin{pmatrix}
hu \\ hu^2 + \tfrac{1}{2} g h^2 \\ huv
\end{pmatrix},
\quad
\mathbf{G}(\mathbf{U}) =
\begin{pmatrix}
hv \\ huv \\ hv^2 + \tfrac{1}{2} g h^2
\end{pmatrix},
\]
where \( h(x,y,t) \) is the water depth, \( \mathbf{u}=(u,v)^T \) is the velocity field, and \( g \) is the gravitational acceleration.

To apply the wavelet method, the conserved variables \( \mathbf{U}(x,y,t) \) are projected onto the wavelet basis. Multiplying the SWE by a test function \( \varphi_{j,k}(x) \) (scaling or wavelet function) and integrating over the domain gives
\[
\int_{\Omega} \left( 
\frac{\partial \mathbf{U}}{\partial t} 
+ \nabla \cdot \mathbf{F}(\mathbf{U}) 
\right) \varphi_{j,k} \, d\Omega = 0.
\]
Integration by parts yields
\[
\frac{d}{dt} \int_{\Omega} \mathbf{U} \, \varphi_{j,k} \, d\Omega 
- \int_{\Omega} \mathbf{F}(\mathbf{U}) \cdot \nabla \varphi_{j,k} \, d\Omega 
+ \int_{\partial \Omega} \mathbf{F}(\mathbf{U}) \cdot \mathbf{n} \, \varphi_{j,k} \, ds = 0.
\]
This weak formulation allows the equations to be discretized directly on the adaptive wavelet grid.

The cell averages of the conserved variables at resolution level \( j \) are approximated through scaling coefficients:
\[
\bar{\mathbf{U}}_{j,k}(t) \approx \int_{\Omega_{j,k}} \mathbf{U}(x,t) \, d\Omega 
\quad \Leftrightarrow \quad
\bar{\mathbf{U}}_{j,k}(t) \sim c_{j,k}(t),
\]
where \( \Omega_{j,k} \) denotes the control volume associated with the coefficient \( (j,k) \). A conservative finite volume update is then performed on the active grid:
\[
\bar{\mathbf{U}}_{j,k}^{n+1} 
= \bar{\mathbf{U}}_{j,k}^{n} 
- \frac{\Delta t}{\Delta x_{j,k}}
\big( \hat{\mathbf{F}}_{j+1/2,k}^{n} - \hat{\mathbf{F}}_{j-1/2,k}^{n} \big),
\]
where \( \hat{\mathbf{F}} \) denotes a numerical flux function (e.g., Roe, Rusanov, or HLLC flux) and \( \Delta x_{j,k} \) is the local grid spacing at level \( j \). Across coarse–fine interfaces, flux matching is enforced to ensure global conservation:
\[
\sum_{\text{fine}} \hat{\mathbf{F}}_{\text{fine}} = \hat{\mathbf{F}}_{\text{coarse}}.
\]

After the update, the wavelet coefficients are recomputed via the inverse transform
\[
\{ c_{J,k}^{n+1}, d_{j,k}^{n+1} \} = \mathcal{W}[\mathbf{U}^{n+1}],
\]
and the adaptive grid is adjusted for the next time step. Time integration is typically performed using explicit schemes such as second-order Runge–Kutta:
\[
\mathbf{U}^{n+1} = \mathbf{U}^n + \Delta t \, \mathcal{L}(\mathbf{U}^n),
\]
where \( \mathcal{L} \) denotes the discrete spatial operator on the adaptive grid.

This process results in a dynamically evolving mesh that refines only where sharp gradients exist, allowing accurate resolution of flow features such as dam-break waves, tidal bores, and wet–dry transitions. Haar wavelets are often used because of their compact support and simple refinement structure, making them computationally efficient.

However, several challenges arise in practice. The smallest grid spacing controls the global time step via the CFL condition, which can reduce efficiency if large refinement regions occur. Conservation of mass and momentum across refinement levels requires careful flux correction, and positivity of water depth must be preserved near wetting and drying fronts. Furthermore, efficient data structures and fast wavelet transforms are essential for high-performance implementations.

In summary, the wavelet-based method for SWE follows the adaptive cycle
\[
\text{SWE} 
\;\xrightarrow{\mathcal{W}}\;
\text{Wavelet Coefficients (Adaptive Grid)}
\;\xrightarrow{\text{FV Scheme}}\;
\text{Updated Solution}
\;\xrightarrow{\mathcal{W}^{-1}}\;
\text{Reconstructed Field},
\]
providing an accurate and efficient framework for solving shallow water flows with localized features at reduced computational cost \cite{DubosKevlahan2013, AechtnerKevlahanDubos2014}.

\subsection{Statistical Solutions}

The Shallow Water Equations (SWE) are hyperbolic partial differential equations used to model fluid flow in scenarios where the horizontal scale is much larger than the vertical depth, such as rivers, coastal areas, and atmospheric layers. The one-dimensional SWE can be written as:

\begin{equation}
\frac{\partial h}{\partial t} + \frac{\partial}{\partial x}(h u) = 0, \quad
\frac{\partial (h u)}{\partial t} + \frac{\partial}{\partial x}\left(h u^2 + \frac{g}{2}h^2\right) = - g h \frac{\partial Z}{\partial x},
\end{equation}

where $h(x,t)$ is the water depth, $u(x,t)$ is the velocity, $g$ is gravity, and $Z(x)$ represents the bed elevation. In real-world scenarios, exact solutions are generally unavailable due to uncertain topography, inflow conditions, and initial states. To address these uncertainties, the Stochastic Galerkin (SG) method represents uncertain quantities as random fields expanded using orthogonal polynomials (polynomial chaos expansion) \cite{dai2023energy}. For example, the bed elevation $Z(x,\omega)$ can be written as:

\begin{equation}
Z(x,\omega) = \sum_{k=0}^{P} Z_k(x) \Phi_k(\xi(\omega)),
\end{equation}

where $\Phi_k$ are orthogonal polynomials of the random variables $\xi(\omega)$ and $Z_k(x)$ are deterministic coefficients. Similarly, the solution variables $h(x,t,\omega)$ and $u(x,t,\omega)$ are expanded in the same basis:

\begin{equation}
h(x,t,\omega) \approx \sum_{k=0}^{P} h_k(x,t) \Phi_k(\xi(\omega)), \quad
u(x,t,\omega) \approx \sum_{k=0}^{P} u_k(x,t) \Phi_k(\xi(\omega)).
\end{equation}

Substituting these expansions into the SWE and performing a Galerkin projection onto the polynomial basis transforms the stochastic PDEs into a coupled system of deterministic PDEs for the coefficients $\{h_k, (h u)_k\}$. This step is crucial because it converts the stochastic problem into a solvable deterministic system while retaining the hyperbolic structure of the SWE. The deterministic system can be written as:

\begin{equation}
\frac{\partial h_k}{\partial t} + \frac{\partial}{\partial x}(h u)_k = 0, \quad
\frac{\partial (h u)_k}{\partial t} + \frac{\partial}{\partial x} \left( (h u^2)_k + \frac{g}{2} (h^2)_k \right) = - g (h \partial_x Z)_k,
\end{equation}

where the subscript $k$ denotes the $k$-th polynomial mode, and the terms like $(h u)_k$ are computed using convolution of coefficients in polynomial chaos expansions.

For numerical implementation, the spatial domain is discretized into a uniform grid $\{x_i\}$, and a finite volume method is applied. Denoting $h_k^i$ and $(h u)_k^i$ as the solution coefficients at grid cell $i$ for mode $k$, the semi-discrete equations become:

\begin{equation}
\frac{d h_k^i}{d t} = -\frac{F_{h,k}^{i+1/2} - F_{h,k}^{i-1/2}}{\Delta x}, \quad
\frac{d (h u)_k^i}{d t} = -\frac{F_{hu,k}^{i+1/2} - F_{hu,k}^{i-1/2}}{\Delta x} - g (h \partial_x Z)_k^i,
\end{equation}

where $F_{h,k}^{i+1/2}$ and $F_{hu,k}^{i+1/2}$ are the numerical fluxes for the $k$-th mode at cell interfaces, computed consistently with the polynomial chaos coefficients. Time integration can be carried out using explicit Runge-Kutta methods to preserve stability. Each mode is evolved simultaneously, and the coupling ensures proper propagation of uncertainty \cite{shaw2020stochastic}.

After solving for all polynomial modes, statistical quantities are reconstructed. The mean and variance of the water depth are computed as:

\begin{equation}
\mathbb{E}[h(x,t)] \approx h_0(x,t), \quad
\mathrm{Var}[h(x,t)] \approx \sum_{k=1}^{P} h_k^2(x,t) \langle \Phi_k^2 \rangle,
\end{equation}

where $\langle \Phi_k^2 \rangle$ is the norm of the polynomial basis $\Phi_k$. The mean represents the expected solution, and the variance quantifies uncertainty across realizations. This approach ensures a mathematically rigorous and computationally implementable solution for the SWE under uncertainty \cite{dai2023energy}.

In conclusion, the Stochastic Galerkin method provides a systematic and reliable framework for solving the Shallow Water Equations in the presence of uncertainty. By expanding the solution and uncertain inputs in orthogonal polynomials and projecting the equations onto this basis, stochastic PDEs are transformed into a deterministic system that can be solved using conventional numerical schemes. The method preserves the hyperbolic structure, allows accurate computation of statistical moments, and provides a precise quantification of uncertainty, making it highly suitable for practical SWE simulations where inputs or topography are uncertain.

\subsection{System Identification and Uncertainty Quantification}

The predictive ability of the shallow water equations (SWE) depends strongly on model parameters such as
bed roughness, bathymetry, and boundary forcing. These quantities are rarely known with certainty, 
which motivates the use of \emph{system identification}\cite{keesman2011system} (parameter estimation) and 
\emph{uncertainty quantification} (UQ).  
Both concepts are tightly linked: system identification reduces parameter uncertainty by assimilating 
observations, while UQ characterizes how remaining uncertainties influence SWE predictions.

\subsubsection{System Identification}

Consider the two-dimensional SWE in conservative form:
\begin{equation}
\frac{\partial \mathbf{U}}{\partial t}
+ \nabla \cdot \mathbf{F}(\mathbf{U},\boldsymbol{\theta})
= \mathbf{S}(\mathbf{U},\boldsymbol{\theta}),
\label{eq:swe_cons2}
\end{equation}
where
\[
\mathbf{U}=
\begin{bmatrix}
h \\ hu \\ hv
\end{bmatrix}, \qquad
\mathbf{F}(\mathbf{U}, \boldsymbol{\theta}) =
\begin{bmatrix}
hu & hv \\
hu^2 + \tfrac{1}{2} g h^2 & huv \\
huv & hv^2 + \tfrac{1}{2} g h^2
\end{bmatrix},
\]
and the source term
\[
\mathbf{S}=
\begin{bmatrix}
0 \\
- gh\,\partial_x b - \tau_x/\rho \\
- gh\,\partial_y b - \tau_y/\rho
\end{bmatrix}.
\]
Here, $\boldsymbol{\theta}$ denotes uncertain parameters such as Manning's roughness coefficient 
or corrections to bathymetry $b(x,y)$. Given observations 
$\mathbf{y}^{\mathrm{obs}}(x,y,t)$, which represent measurements of hydrodynamic variables 
(e.g., water depth, discharge, or velocity) collected at spatial locations $(x,y)$ and times $t$, 
parameter estimation is formulated as the following inverse problem. 
These observations provide the reference against which the model predictions are compared 
and guide the estimation of the unknown parameters $\boldsymbol{\theta}$ in the SWE model.

\begin{equation}
\hat{\boldsymbol{\theta}} 
= \arg\min_{\boldsymbol{\theta}}
J(\boldsymbol{\theta})
= \arg\min_{\boldsymbol{\theta}}
\frac{1}{2}
\sum_{k=1}^{N_t}
\left\|
\mathbf{y}^{\mathrm{obs}}_{k}
- \mathcal{H}\!\left(\mathbf{U}_{k}(\boldsymbol{\theta})\right)
\right\|_{\mathbf{R}^{-1}}^{2},
\label{eq:inverse}
\end{equation}
where $\mathcal{H}$ maps the model state to observable quantities and $\mathbf{R}$ is the measurement error covariance.
This formulation follows the standard structure of deterministic inverse problems 
as described in classical theory \cite{tarantola2005inverse}.

Gradient-based methods employ adjoint equations to compute sensitivities efficiently, while 
ensemble-based techniques approximate the parameter–observation map using multiple SWE realizations.  
Such approaches are widely used in hydrology and coastal modeling, particularly in data assimilation workflows.

\subsubsection{Uncertainty Quantification}

Even after estimating parameters, the SWE solution remains influenced by uncertainties in 
boundary conditions, observations, forcing, numerical discretization, and unresolved physical processes.
Uncertainty quantification (UQ) \cite{smith2024uncertainty}aims to describe how these uncertainties propagate through the SWE and 
affect predicted quantities of interest.

A typical mathematical representation introduces random variables 
$\boldsymbol{\xi} = (\xi_1,\dots,\xi_M)$ to model uncertain inputs such as roughness, bathymetry, or initial depth.
Spatially distributed uncertain fields may be expanded using a Karhunen--Lo\`eve (KL) representation,
\[
q(x,y,\omega)
= \bar{q}(x,y)
+ \sum_{k=1}^{K} \sqrt{\lambda_k}\,\phi_k(x,y)\,\xi_k(\omega),
\]
yielding a finite-dimensional stochastic parametrization.

For propagation of uncertainty, one may use:
\textbf{Nonintrusive methods}, such as Monte Carlo, quasi–Monte Carlo, or sparse-grid collocation, in which 
  each realization calls the deterministic SWE solver without modifying it.
\textbf{Intrusive methods}, such as stochastic Galerkin or polynomial chaos expansions, where the SWE state 
  becomes stochastic,
  \[
  \mathbf{U}(x,y,t,\boldsymbol{\xi})
  = \sum_{j=0}^{P} \widehat{\mathbf{U}}_j(x,y,t)\, \Psi_j(\boldsymbol{\xi}),
  \]
  and projecting onto the basis $\Psi_j$ yields a coupled deterministic system for 
  the coefficients $\widehat{\mathbf{U}}_j$.

After system identification, remaining parameter uncertainty can be described by a posterior distribution,
helping quantify forecast confidence intervals, flood risk, or the sensitivity of critical outputs such as
peak discharge or inundation extent.  
In digital-twin contexts, this UQ layer enables real-time updates of predictive confidence and guides the 
acquisition of new data that most effectively reduces model error.

Together, system identification and UQ provide a coherent mathematical and computational framework for 
producing reliable, data-informed SWE predictions.

\subsection{Coupled Models and Digital Twins}

Coupled modelling and digital–twin frameworks have emerged as powerful tools for
large-scale geophysical and environmental simulations. In hydraulic, coastal, and
riverine applications, a single numerical model of the Shallow Water Equations (SWE)
is often insufficient to capture multi-physics interactions, external forcings, or real-time
field conditions. Coupled models address this limitation by combining two or more
numerical components (e.g., hydrodynamics, sediment transport, rainfall–runoff,
groundwater exchange), while digital twins integrate such models with continuously
arriving observations to maintain an updated, data-consistent state of the system.

\medskip
\noindent\textbf{Coupling mechanisms and examples.}
We denote the SWE compactly as $\mathrm{SWE}(h,\mathbf{u},Q)$ to indicate the coupled
system for depth $h$, depth-averaged velocity $\mathbf{u}$, and discharge $Q= h\mathbf{u}$.
Couplings to other models may modify source terms, fluxes, or boundary/internal interface
relations; below we outline three common couplings and show how they interact with the
SWE numerically.

\paragraph{1) Local ecology (water-quality / vegetation).}
Ecological or water-quality state variables (nutrients, algae, suspended sediments,
biomass) are usually modelled as transported scalars satisfying advection–diffusion–reaction
(ADR) equations driven by the SWE velocity field:
\[
\partial_t (hC) + \nabla\!\cdot(h\mathbf{u}C) = \nabla\!\cdot\big(hD\nabla C\big) + h\,R(C,h,\mathbf{u}),
\]
where $C(x,y,t)$ is a concentration (or biomass), $D$ an effective diffusion/dispersion
coefficient, and $R(\cdot)$ chemical/biological reaction terms. This is
typically a one-way coupling (SWE $\to$ ADR) when the ecological variables are merely
transported. However, in ecohydraulic situations vegetation or high biomass can alter
momentum via a drag-like friction term, e.g.
\[
\boldsymbol{\tau}_{\mathrm{eco}} = -C_{\mathrm{veg}}(x,y)\,|\mathbf{u}|\,\mathbf{u},
\]
which enters the SWE momentum source and therefore \emph{changes the numerical stiffness
and source-term treatment} of the hydrodynamic solver . When
such feedbacks are present the coupled system must be solved either implicitly or with
careful operator splitting so that the modified friction is handled stably.

\paragraph{2) Water-engineering constructions (dams, weirs, gates).}
Hydraulic structures are represented by interface relations or rating/discharge laws that
replace or modify numerical fluxes across a computational interface. A broad-crested (or
sharp-crested) weir is commonly modelled with a discharge law such as
\[
Q = C_w\, b\, H^{3/2},
\]
where $H$ is the upstream head relative to the crest and $C_w$ the empirical coefficient
. In a finite-volume scheme the cell interface flux is replaced
by the structure flux function $\mathcal{F}_s(h_L,h_R;\theta)$ (with $\theta$ operational
parameters), so the Riemann solver or flux evaluation must be augmented to enforce the
weir/dam relation. These couplings \emph{directly enter the SWE discretisation} and
influence solver treatment of wetting/drying, well-balanced properties, and stability.

\paragraph{3) Economic and energy models (reservoir operation, hydropower).}
Economic or energy models typically use the SWE as a state constraint in optimisation
problems. For example, turbine release schedules $Q_t(t)$ may be chosen to optimise an
objective $J$ (revenue, reliability, flood-risk) subject to the SWE dynamics:
\[
\min_{Q_t(\cdot)} J(Q_t,h,\mathbf{u}) \quad\text{s.t.}\quad \mathrm{SWE}(h,\mathbf{u};Q_t)=0.
\]
Here the economic model does not itself accelerate SWE numerics, but it defines control
inputs and requires repeated forward (and sometimes adjoint) SWE solves inside an
optimization loop—hence the solver must provide sensitivities or adjoint operators if
gradient-based methods are used . Such couplings are therefore tightly
integrated into the computational workflow of operational water management.

\medskip
\noindent\textbf{Coupling strategies and numerical considerations.}
Coupling strategies include explicit exchange (loose coupling), implicit monolithic solution
(tight coupling), and operator-splitting (semi-implicit) approaches. The choice depends on
relative time scales, nonlinearity and stiffness: ecological reactions or vegetation drag may
introduce stiffness that favours implicit treatment; hydraulic structures often require
special flux treatment to preserve well-balanced states and avoid spurious oscillations;
economic optimisation typically needs efficient adjoint or reduced-order models for repeated
solves. In practice, many operational systems combine a well-balanced SWE solver with an
operator-splitting exchange of scalar transport, parameter updates for structures, and an
outer optimization loop for operational decisions .

\medskip
\noindent\textbf{Digital Twins.}
A digital twin for SWE is a live, continuously updated computational replica of a real
hydrodynamic system. It consists of:
\begin{enumerate}
    \item a physics-based SWE solver (structured or unstructured, explicit, implicit, or semi-implicit),
    \item an assimilation or inference layer (e.g., Ensemble Kalman Filter, 4D-Var, particle filtering),
    \item real-time data streams (gauges, radar rainfall, satellite altimetry, drone imagery),
    \item and a feedback loop that updates the model state, forcing, or parameters.
\end{enumerate}
The goal is to maintain a synchronised model state $U^\mathrm{DT}(t)$ that is consistent with
observations while respecting the governing PDEs. Typical updated variables include water
depth $h$, velocities $\mathbf{u}$, friction coefficients $n_m$, or boundary inflow
conditions. The digital twin thus acts as a PDE-constrained estimator: physics restricts
admissible states, while data corrects model drift, uncertainty, or mis-specified forcings.
Data-assimilation methods such as the Ensemble Kalman Filter (EnKF)\cite{evensen2009data} or variational (4D-Var)
approaches are standard choices for the assimilation/inference layer in hydrologic and
hydrodynamic digital twins . Digital-twin systems are
now applied for real-time flood forecasting, reservoir control and water-infrastructure
management; reviews and domain-specific surveys further describe practical implementation
and challenges.

\section{Measurement and data collection}


Computational fluid dynamics (CFD) models are inherently dependent on high-quality observational data to define boundary conditions, initialize flow fields, estimate model parameters, and validate numerical predictions. In water engineering applications—including rivers, reservoirs, estuaries, coastal zones, and urban drainage systems—these data are obtained through a combination of Internet of Things (IoT) sensor networks and well-established hydrometric, laboratory, and remote sensing instruments. The following discussion presents seven commonly used IoT-enabled systems and non-IoT measurement tools that provide essential physical data for CFD-based flow modeling.

\subsection{Acoustic Doppler Current Profilers (ADCPs)}

Acoustic Doppler Current Profilers (ADCPs) are among the most important field instruments used in CFD modeling of natural water bodies. ADCPs operate by emitting acoustic pulses into the water column and measuring the Doppler shift of sound waves reflected by suspended particles. From this Doppler shift, depth-resolved, three-dimensional velocity profiles are obtained across flow sections. Modern ADCP systems are frequently deployed as IoT-enabled platforms, transmitting measurements in near real time via telemetry.

The velocity profiles measured by ADCPs are fundamental for CFD model calibration and validation. They provide direct information on vertical velocity distributions, shear layers, and secondary circulation patterns, allowing numerical models to be assessed against observed momentum transport and flow structure. Such data are especially important for evaluating turbulence closures and numerical discretization schemes in water engineering CFD applications 

\subsection{Pressure Transducers and Water Level Loggers}

Pressure transducers are widely used to measure water surface elevation by converting hydrostatic pressure into water depth. These instruments are commonly installed in rivers, channels, reservoirs, and coastal environments and are often integrated into IoT monitoring networks for continuous data acquisition.

Free-surface elevation data obtained from pressure transducers are critical for CFD models that solve the shallow water equations or free-surface Navier--Stokes formulations. Water level measurements define upstream and downstream boundary conditions, support the simulation of transient wave propagation, and enable validation of modeled stage variations. Accurate water level data are therefore essential for reliable prediction of pressure gradients, flow depths, and gravity-wave dynamics 

\subsection{Meteorological Stations and Atmospheric Sensor Networks}

Meteorological stations, whether deployed as standalone instruments or IoT-connected sensor arrays, provide measurements of wind speed, wind direction, air temperature, humidity, and atmospheric pressure. In hydrodynamic and environmental CFD models, these atmospheric variables define surface forcing conditions.

Wind stress data are particularly important in simulations of lakes, reservoirs, estuaries, and coastal waters, where wind-driven circulation can dominate flow dynamics. Atmospheric heat fluxes influence thermal stratification and density-driven flows. Incorporating measured meteorological forcing improves the physical realism of CFD simulations and supports coupled air--water modeling approaches commonly used in advanced water engineering studies.

\subsection{Remote Sensing Platforms (Satellite and UAV-Based Systems)}

Remote sensing platforms, including satellite systems and unmanned aerial vehicles (UAVs), provide spatially distributed observations that complement in-situ measurements. These technologies collect data on surface water extent, surface temperature, surface roughness, and flow patterns inferred from image-based velocimetry.

For CFD modeling, remotely sensed data are particularly valuable for defining large-scale geometries, identifying dominant flow structures, and validating spatial flow patterns that cannot be captured by point sensors alone. UAV-based photogrammetry and satellite-derived bathymetry are widely used to generate computational meshes and update channel or floodplain geometries, improving the accuracy of numerical flow simulations.

\subsection{Laser Doppler Velocimetry (LDV)}

Laser Doppler Velocimetry (LDV) is a non-intrusive optical measurement technique commonly used in laboratory-scale hydraulic experiments. LDV determines flow velocity by measuring the Doppler shift of laser light scattered by tracer particles moving with the fluid.

Although LDV systems are not typically IoT-enabled, they provide highly resolved point measurements of instantaneous velocity and turbulence statistics. These data are essential for validating CFD turbulence models, particularly in controlled laboratory studies involving jets, recirculation zones, and boundary layers. LDV measurements often serve as benchmark datasets for assessing the accuracy of numerical solvers and turbulence closures .

\subsection{Particle Image Velocimetry (PIV)}

Particle Image Velocimetry (PIV) is another optical diagnostic technique widely used in experimental fluid mechanics. PIV captures pairs of images of tracer particles illuminated by a laser sheet and computes velocity fields through cross-correlation analysis.

The spatially resolved velocity fields obtained from PIV are highly valuable for validating CFD predictions of coherent structures, vortices, and flow separation. In water engineering applications, PIV data are frequently used to assess numerical accuracy in complex flow regions such as hydraulic jumps, spillways, and mixing zones, thereby strengthening confidence in CFD-based design and analysis.

\subsection{Water Quality Sensors and Multi-Parameter Probes}

Water quality sensors measure parameters such as temperature, turbidity, electrical conductivity, dissolved oxygen, and salinity. Many modern probes are IoT-enabled and deployed as part of continuous environmental monitoring networks.

In CFD models that incorporate scalar transport, density stratification, or reactive processes, water quality measurements provide essential boundary conditions and validation data. Temperature and salinity directly influence fluid density and buoyancy-driven circulation, while turbidity measurements inform sediment transport and dispersion modeling. Integrating these observations enables CFD simulations to extend beyond purely hydraulic analyses toward fully coupled hydrodynamic--environmental modeling.

\subsection{Synthesis}

Collectively, these measurement technologies form the observational foundation of computational fluid dynamics modeling in water engineering. IoT-enabled sensors supply continuous, real-time data streams for operational and forecasting models, while laboratory-scale and remote sensing tools provide high-resolution datasets for model development and validation. By anchoring numerical simulations in measured physical quantities, CFD models achieve the reliability and predictive capability required for engineering design, environmental assessment, and informed decision-making in complex water systems.


\section{Environment Coupling}




\subsection{Vessels}

Many of the river channels and lakes that water engineers are operating with have traffic. This includes commercial and recreational vesels (ships) that navigate the waterways. Classically their incorporation in any modeling and operations research of water infrastructure is eschewed because of the difficulty in accurately predicting their presence and impact. Whereas for some water channels this is largely harmless, i.e. when the traffic is light and sparse, there are many cases wherein ship traffic can displace significant quantities of water, affecting downstream river operations as well as necessary river bank protection from transient flooding. Their accurate and precise incorporation into fluid dynamics modeling and simulation for water engineering is a challenging and largely open problem at the time of writing. 

Here we present a few references on particular circumscribed investigations on the topic. The most comprehensive research work and simulation is given in~\cite{bellafiore2018modeling}. This modeled the influence of ship waves on the dynamics of the hydrological shallow water models, with an empirical validation of the canals and surrounding water channels of Venice.

Modeling the effect on the overall river flow is studied in~\cite{dempwolff2023verification}. Vessels are postulated as introducing a pressure force of either form:
\[
p(x,y)=p_0\sqrt{1-(x^2+y^2)/R^2}
\]
or
\[
p(x,y)=p_0\left[1-c_L\left(\frac{x}{L}\right)^4\right]\left[1-c_B\left(\frac{y}{B}\right)^2\right]\exp\left[-a\left(\frac{y}{B}\right)^2\right]
\]
as depending on the shape of the boat. That is, the geometric shape introduces distinct profiles of the normal force that pressure induces in the surrounding water.

The potential for, especially large and high velocity vessels, introducing waves that can travel to the river banks and possibly flood over them or sustain structural effects on their physical integrity presents an important concern for water engineers. The work~\cite{lataire2017hydrodynamic} models the bank effects of pressure in restricted waterways on the movement of the ships.

The works~\cite{fleit2019practical,fleit2021hydrodynamic} studied the impact of a ship-caused wave onto the river bed using CFD, in order to assess the possible effect on the local aquatic ecosystem. The work applied and validated it on the Hungarian Danube. 

A few other works on these and related aspects of the impact of ships on the displacement and the introduction of waves in water channels include~\cite{almaz2012simulation,he2019short,terziev2021numerical,mao2022impacts}.

\subsection{Flora and Fauna}
The consideration of flora and fauna that are present in water channels that are of interest to water engineers introduces additional considerations. Most significantly, the operation of dams and other infrastructure can induce changes in the volume, velocity and other aspects of water flow that can influence the life and health of any species that may be present. Investigation as to the acute, e.g., dam breaks, and chronic, e.g., long run changes in the average flow speed and direction, is the topic of a number of investigations. These investigations, in turn, introduce so-called ``environmental constraints'' that must be then incorporate to mitigate and prevent their ill effects. 

\paragraph{Plants and Algae}
We have identified a number of works studying the environmental impact of water engineering on water flora, with the primary biological family of interest being algae. 

Modeling and simulation of the structure and properties of flow through vegetation patches includes~\cite{anjum2020hydrodynamics,anjum2022investigation,anjum2024emergent}. In the long term, simulations suggest that the strucure of patches is influenced by water inflows, while in turn affecting the fluid dynamical properties of the outflow, in turn. The investigation of the interaction between vegetation patches and turbulent flow is done in~\cite{aydogdu2023analysis,trinci2017life}.

Environmental impacts as far as vegetation on the integrity of banks as barriers to potential flooding is given in~\cite{weber2016modelling}. Simulation of potential damage to vegetation life, and mitigation measures are given in~\cite{preuss2023cfd,kim2022harmful}. 

\paragraph{Aquatic Animals}

Consideration of acquatic animals, namely fishes, in open water channel flow is more one-directional. That is, the presence of fish does little to displace or modify the flow of water through river channels or the integrity and engineering performance of infrastructure. However, channels restricted manually by water engineering, and especially the safe and reliable navigation of fish through dams is an important environmental concern.

Studies into these considerations include~\cite{daraio2010methodological,goodwin2014fish,haro2004swimming,vcada2006efforts,willis2011modelling}. Primarily, theoretical, simulation, and numerical validation has been performed in order to analyze how the modification of water velocity vector fields influence the movement and biological health of fish. Broadly, fish are often able to adapt to changes in water flow and channel shape in identifiable patterns. Their long run ecological consequences are a matter of copious research investigation. 

\section{Case Descriptions}

\subsection{USA: CALSIM by California Department of Water Resources}

The California Simulation Model (CalSim) is a \textbf{hydrological modeling platform} designed to simulate and analyze water resource systems in California \cite{california_dwr_models}. It serves as a critical tool for planning, management, and decision-making in water resource allocation, particularly in complex systems involving multiple reservoirs, rivers, and water users. CalSim is widely used by water authorities and agencies to address the challenges of water distribution, drought management, and climate change adaptation in California.

CalSim primarily focuses on \textbf{California's water resources}, covering major river basins such as the Sacramento River, San Joaquin River, and their tributaries. It also includes the Sacramento-San Joaquin Delta, a critical hub for water distribution in the state. The platform models key reservoirs like Shasta, Oroville, and Folsom, as well as aqueducts such as the California Aqueduct and the Central Valley Project. Additionally, CalSim integrates both \textbf{surface water} and \textbf{groundwater systems}, addressing the water demands of agricultural, urban, and environmental users across California. This comprehensive coverage allows the platform to provide a unified framework for understanding and managing the state's water resources.

Several key water authorities and agencies in California rely on CalSim for their operations. The \textbf{California Department of Water Resources (DWR)}, the primary agency responsible for managing the state's water resources, is the lead user of the platform. The \textbf{U.S. Bureau of Reclamation (USBR)}, which manages federal water projects like the Central Valley Project, also utilizes CalSim. Additionally, the \textbf{Delta Stewardship Council}, which oversees the management of the Sacramento-San Joaquin Delta, and numerous local water districts employ CalSim for regional water planning and management. These collaborations ensure that the platform remains a vital tool for addressing California's diverse water management needs.

The development of CalSim was a collaborative effort involving multiple stakeholders. The \textbf{California Department of Water Resources (DWR)} led the initiative, with significant contributions from the \textbf{U.S. Bureau of Reclamation (USBR)} for federal water projects. Academic institutions and consulting firms also played a role in refining the model over the years. CalSim is continuously updated to reflect changes in water infrastructure, regulations, and climate conditions, ensuring its relevance and accuracy in addressing contemporary water management challenges.

CalSim functions as a \textbf{decision-support tool}, simulating the operation of California's water resource systems. One of its primary functionalities is \textbf{water allocation and management}, where it simulates the distribution of water resources among competing users, including agriculture, urban areas, and environmental needs. The platform models the operation of reservoirs, aqueducts, and pumping stations to optimize water delivery. Additionally, CalSim supports \textbf{scenario analysis}, enabling users to evaluate the impacts of different water management strategies, such as changes in reservoir operations, water transfers, or climate change scenarios. It also assesses the effects of regulatory policies, such as the Sustainable Groundwater Management Act (SGMA).

The platform's \textbf{hydrological simulation} capabilities allow it to model the flow of water through rivers, reservoirs, and aqueducts. It incorporates climate data, such as precipitation and temperature, to simulate hydrological processes accurately. CalSim can also be integrated with other models, such as \textbf{CalSim II} (an updated version) and \textbf{CALSIM 3}, to provide more detailed and accurate simulations. Furthermore, it interfaces with groundwater models to account for interactions between surface water and groundwater. This integration is particularly important for addressing the state's complex water systems.

CalSim also plays a crucial role in \textbf{environmental and ecosystem support}. It models the impacts of water management decisions on ecosystems, such as fish habitats in the Delta, and supports the implementation of environmental regulations, such as the Endangered Species Act. The platform's \textbf{modular structure} allows users to customize the model for specific regions or water systems, while its \textbf{user-friendly interface} provides graphical outputs and visualization tools for easy interpretation of results. With \textbf{high-resolution modeling}, CalSim captures the complexity of California's water systems, offering detailed spatial and temporal resolution. Its \textbf{scenario testing} feature enables users to evaluate "what-if" scenarios, assessing the impacts of different policies or climate conditions.

The applications of CalSim are diverse and far-reaching. It is used for \textbf{water supply planning}, helping water agencies plan for future water needs under different scenarios. During droughts, CalSim assesses the impacts of water shortages and identifies strategies to mitigate them. It also supports \textbf{climate change adaptation} by evaluating the effects of changing climate conditions on water availability and infrastructure. Additionally, CalSim aids in \textbf{regulatory compliance}, ensuring that water management practices align with state and federal regulations.

A typical use case for CalSim involves simulating the operation of the \textbf{State Water Project (SWP)} and \textbf{Central Valley Project (CVP)} under different climate scenarios. For example, in a dry-year scenario with reduced precipitation and increased temperatures, CalSim predicts water shortages, impacts on agriculture, and potential mitigation strategies, such as increased groundwater pumping or water transfers. These simulations provide valuable insights for decision-makers, enabling them to develop effective water management strategies.

Despite its strengths, CalSim has certain limitations. The platform requires extensive input data, which can be challenging to obtain and validate. Its \textbf{computational complexity} can make large-scale simulations resource-intensive. Additionally, like all hydrological models, CalSim's predictions are subject to uncertainty due to variability in climate and hydrological processes. These limitations highlight the need for continuous refinement and validation of the model to ensure its accuracy and reliability.

In conclusion, CalSim is a powerful tool for managing California's water resources, offering comprehensive simulations and scenario analyses to support decision-making. Its development and application involve collaboration among multiple agencies, including the \textbf{California Department of Water Resources (DWR)} and the \textbf{U.S. Bureau of Reclamation (USBR)}. While the platform has limitations, its ability to integrate surface water and groundwater systems, support environmental regulations, and address complex water management challenges makes it an indispensable resource for California's water authorities.

\subsection{AUSTRALIA : Source by eWater and Murray - Darling Basin Authority }

\begin{figure}
    \centering
    \includegraphics[width=0.5\linewidth]{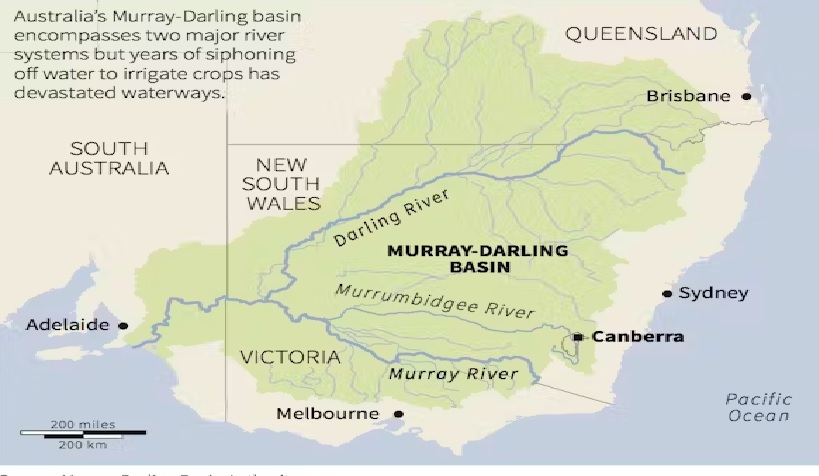}
    \caption{Murray Darling Basin in Australia}
    \label{fig: MDBA}
\end{figure}

The \textbf{Source Water Model}, developed by eWater in Australia \cite{woldemeskel2020watertools}, is a sophisticated hydrological modeling tool designed to support water resource management and planning. It is widely used by water authorities, researchers, and policymakers to simulate and analyze water systems, ensuring sustainable water use and allocation. The model is particularly significant in Australia, where it has been applied across diverse hydrological regions, including the \textbf{Murray-Darling Basin}, which is the largest and most complex river system in the country. The Murray-Darling Basin spans multiple states and territories and provides water for five of Australia's six largest cities: Brisbane, Sydney, Canberra, Melbourne, and Adelaide. Additionally, the basin supports the majority of Australia's agricultural output, making it a critical area for water management. Beyond the Murray-Darling Basin, the Source Water Model has been used in coastal catchments across Queensland, New South Wales, Victoria, and Western Australia, as well as in urban and rural water systems. While its primary focus is Australia, the model's flexible framework allows it to be adapted to other regions globally, particularly in areas with similar water management challenges.

The Source Water Model is utilized by a wide range of water authorities and organizations in Australia. Key users include the \textbf{Murray-Darling Basin Authority (MDBA)}, which relies on the model to manage water resources in the Murray-Darling Basin. State water agencies, such as the \textbf{Queensland Department of Natural Resources, Mines and Energy} and the \textbf{New South Wales Department of Planning, Industry and Environment}, also use Source for water planning and management. Local water utilities employ the model for urban water supply planning, while research institutions like the \textbf{Commonwealth Scientific and Industrial Research Organisation (CSIRO)} and various universities use it for hydrological research and analysis. These organizations depend on Source to simulate water systems, assess water availability, and develop strategies for sustainable water use. The model's ability to integrate data on system inflows, flow routing and losses, water demand for irrigation, stock, domestic use, town water supply, and environmental needs, as well as interstate water sharing and allocation, makes it a versatile tool for comprehensive water management.

The development of the Source Water Model was a collaborative effort involving the \textbf{Murray-Darling Basin Authority}, state governments, and the \textbf{eWater Cooperative Research Centre}, which is managed by eWater Limited. The model builds on earlier tools like \textbf{REALM} and \textbf{IQQM}, incorporating their strengths while addressing their limitations. Its development process included input from water agencies, researchers, and policymakers to ensure it meets real-world needs. Source is designed as an open-source framework, allowing for modularity and customization to adapt to specific regional or operational requirements. The model undergoes continuous improvement, with regular updates and enhancements based on user feedback and advancements in hydrological science. This collaborative and iterative approach has made Source a robust and reliable tool for water resource management.

The Source Water Model provides a wide range of functionalities to support hydrological modeling and water resource management. It simulates rainfall-runoff processes, river flows, and groundwater interactions, supporting both lumped and distributed modeling approaches. The model also enables water allocation, storage, and distribution system modeling, allowing users to evaluate the impacts of climate change, land use changes, and water policy decisions. One of its key features is scenario analysis, which allows users to test "what-if" scenarios, such as changes in water demand, infrastructure, or environmental flows. Source can be integrated with other tools, such as \textbf{Geographic Information Systems (GIS)}, to enhance its analytical capabilities. Additionally, the model offers a user-friendly graphical interface for model setup, calibration, and visualization of results, making it accessible to a broad range of users.

Despite its strengths, the Source Water Model has some limitations. Its flexibility and comprehensive features can make it challenging for new users to learn and operate. High-quality, detailed input data, such as rainfall, streamflow, and land use data, are required for accurate simulations, and these data may not always be available. Large-scale or highly detailed models can be computationally intensive, requiring significant processing power and time. While Source can be adapted to other regions, its default settings and assumptions are tailored to Australian conditions, which may limit its applicability elsewhere without customization. Additionally, the model is primarily designed for long-term planning and scenario analysis, with limited support for real-time water management. These limitations highlight the need for ongoing development and user training to maximize the model's potential.

In conclusion, the Source Water Model is a powerful tool for hydrological modeling and water resource management, particularly in Australia. Its development through collaboration and continuous improvement has made it a versatile and reliable platform for simulating water systems and supporting sustainable water use. While it has some limitations, its ability to integrate diverse data and provide comprehensive analysis makes it an invaluable resource for water authorities, researchers, and policymakers. By addressing its challenges and leveraging its strengths, the Source Water Model can continue to play a critical role in water management efforts both in Australia and globally.

\subsection{CANADA : Green Kenue Water Resource Model by the National Research Council Canada (NRC)}

\begin{figure}
    \centering
    \includegraphics[width=0.5\linewidth]{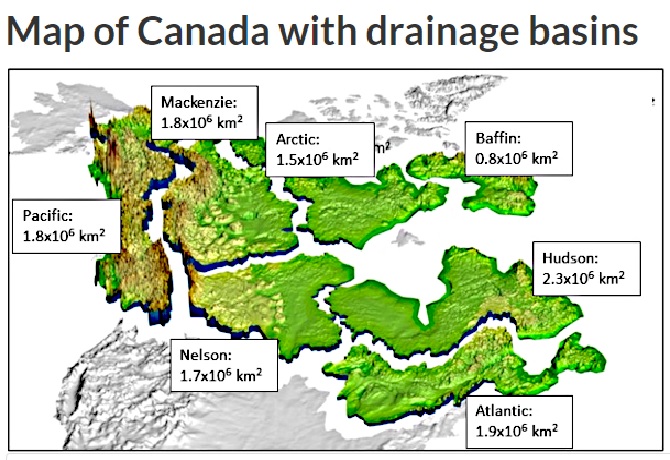}
    \caption{Canada's Drainage Basins}
    \label{fig:CDB}
\end{figure}

Figure \ref{fig:CDB} is an image illustrating the seven drainage basins (Pacific Basin, Mackenzie Basin, Nelson Basin, Artic Basin, Baffin Basin, Hudson Basin and Atlantic Basin).

The \textbf{Green Kenue Water Resource Model}, developed by the \textbf{National Research Council Canada (NRC)}, is an advanced hydrological modeling tool designed for water resource management and analysis. It integrates various physical processes and hydrodynamic components to simulate surface water flow, rainfall-runoff relationships, ice formation, sediment transport, and water quality. The platform includes sub-models such as \textbf{HYDROTEL}, which focuses on hydrological modeling of catchments in Canada, and \textbf{TANH}, which is used for hydraulic river modeling. Green Kenue is widely recognized for its ability to support flood forecasting, urban stormwater management, and water resource planning, making it a valuable tool for addressing complex hydrological challenges. \cite{greenkenue2022}

Geographically, Green Kenue has been applied extensively across \textbf{Canada}, including major watersheds such as the Great Lakes Basin, the Fraser River Basin, and the Prairie Provinces. Its versatility has also led to international adoption, with applications in countries like the United States, Australia, and parts of Europe. The platform is designed to function in diverse environments, from urban areas requiring stormwater management to rural regions needing hydrological modeling for agricultural and ecological purposes. This adaptability makes Green Kenue a globally relevant tool for addressing water resource challenges in varying climatic and hydrological conditions.

The development of Green Kenue was a collaborative effort led by the \textbf{National Research Council Canada (NRC)}, with input from federal and provincial water agencies, researchers, and industry stakeholders. The platform was designed with a user-centered approach, ensuring that it is accessible to both experts and non-experts. It incorporates advanced technologies such as Geographic Information Systems (GIS) and hydrological modeling techniques, enabling users to analyze and visualize complex water systems effectively. Regular updates and improvements are made to the platform based on user feedback and advancements in hydrological science, ensuring that it remains at the forefront of water resource modeling.

Green Kenue is utilized by a wide range of water authorities and organizations. In Canada, \textbf{Environment and Climate Change Canada (ECCC)} employs the platform for flood forecasting and water resource management. Provincial agencies such as \textbf{Alberta Environment and Parks} and the \textbf{Ontario Ministry of Natural Resources and Forestry} also rely on Green Kenue for their hydrological modeling needs. Municipalities use the platform for urban stormwater management and infrastructure planning, while research institutions and consulting firms leverage its capabilities for academic and professional projects. The platform’s advanced features and user-friendly interface make it a preferred choice for organizations involved in water resource management.

The functionality of Green Kenue is comprehensive, offering tools for hydrological modeling, flood forecasting, urban stormwater management, and data visualization. For hydrological modeling, the platform simulates rainfall-runoff processes, river flows, and watershed dynamics, supporting both lumped and distributed modeling approaches. In flood forecasting, Green Kenue models floodplains and predicts flood events using real-time data, aiding in emergency preparedness and response planning. For urban stormwater management, the platform designs and evaluates systems such as detention ponds and drainage networks. Additionally, Green Kenue provides advanced visualization tools for analyzing and presenting hydrological data, and it integrates seamlessly with GIS software and other hydrological models for enhanced analysis.

Despite its robust capabilities, Green Kenue has some limitations. The platform’s advanced features can present a steep learning curve for new users, requiring time and training to master. Accurate simulations depend on high-quality, detailed input data, such as rainfall, topography, and land use data, which may not always be readily available. Large-scale or highly detailed models can be computationally intensive, demanding significant processing power and time. While Green Kenue can be adapted to various regions, its default settings and assumptions are tailored to Canadian conditions, which may limit its applicability elsewhere without customization. Additionally, the platform is primarily designed for planning and analysis, with limited support for real-time water management.

Green Kenue incorporates several key equations to model hydrological processes. For example, the \textbf{Saint-Venant equations}, which describe one-dimensional unsteady flow in open channels, are used for river and floodplain modeling see ~\eqref{subsec:swe}. These equations enable the simulation of water flow dynamics in rivers and floodplains, providing critical insights for flood forecasting and water resource management.

In conclusion, the Green Kenue Water Resource Model is a powerful and versatile tool for hydrological modeling and water resource management. Its development by the National Research Council Canada, combined with its wide adoption by water authorities and organizations, underscores its importance in addressing water-related challenges. While the platform has some limitations, its advanced capabilities, user-friendly design, and continuous updates make it an invaluable resource for researchers, planners, and decision-makers in the field of water resources (see, e.g, \cite{greenkenue2022}).

\subsection{India, Central Water Commission : Google Floodhub}

In 2020, the Indian Central Water Commission partnered with Google Research Lab to create a water resource modeling tool called Google Flood Hub.

\begin{figure}
    \centering
    \includegraphics[width=0.5\linewidth]{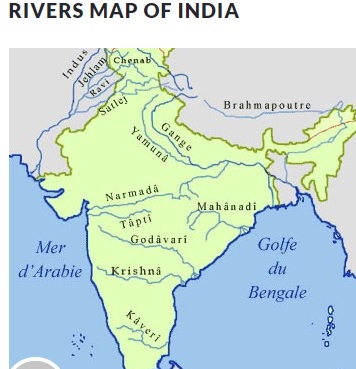}
    \caption{Map of India's Major Rivers}
    \label{fig:MIMR}
\end{figure}

\begin{figure}
    \centering
    \includegraphics[width=0.5\linewidth]{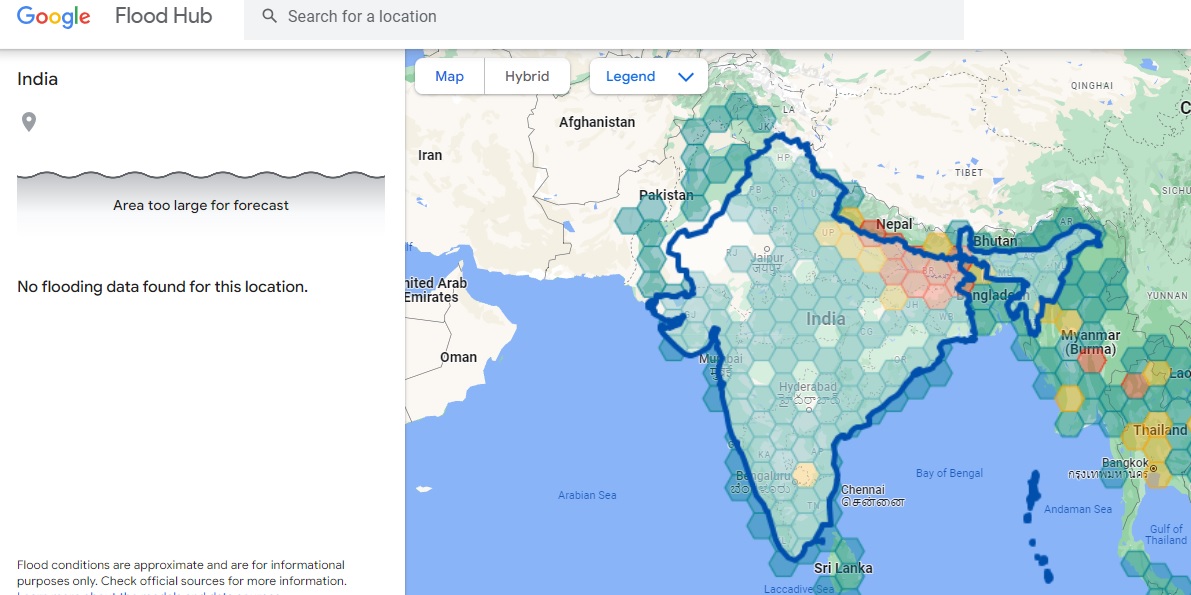}
    \caption{Google Flood Hub's Coverage of India's River Network (Every Coloured dot represents a river gauge that is remotely monitored}
    \label{fig:GFCI}
\end{figure}

\begin{figure}
    \centering
    \includegraphics[width=0.5\linewidth]{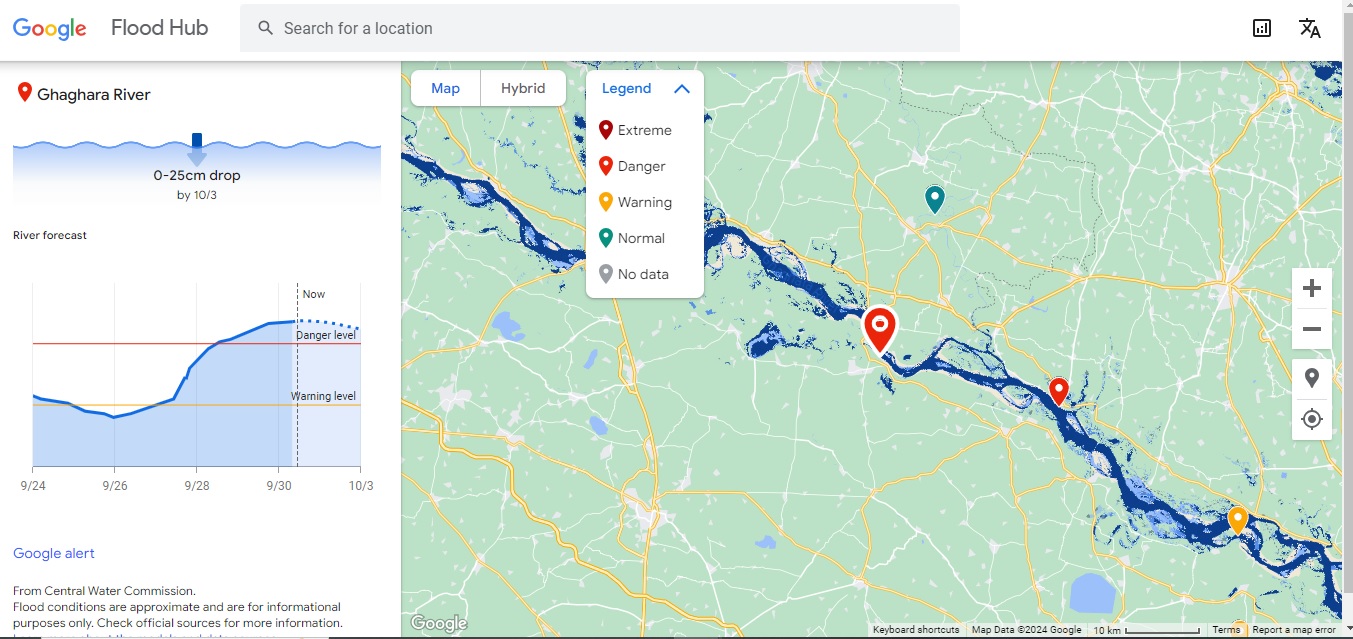}
    \caption{The Ghaghara River, a Tributary to the Gange River Flooding on September 30th 2024. Inundation of the River is captured}
    \label{fig:GRG}
\end{figure}

Google Flood Hub relies on a variety of water resource and hydrological equations to forecast floods, utilizing advanced machine learning models, satellite imagery, and traditional hydrological models. The equations are rooted in hydrology, hydraulics, and meteorology. Below is a comprehensive breakdown of key equations and concepts likely involved in their flood prediction framework:

\subsubsection{Hydrological Modeling}


\paragraph{Precipitation and Runoff}
Floods are primarily driven by precipitation and the conversion of that precipitation into runoff. For the most common equations used in hydrological modeling include, we refer to, e.g., \cite{cea2010experimental,sun2024assessing,turan2024high}.

\paragraph{Rainfall-Runoff Relationship (SCS Curve Number Method)}
The SCS Curve Number method estimates the amount of runoff from rainfall. The equation is:

\[
Q = \frac{(P - I_a)^2}{(P - I_a) + S},
\]
where:
\begin{itemize}
    \item $Q$: Runoff (depth),
    \item $ P $: Precipitation (depth),
    \item $ I_a $: Initial abstraction (loss before runoff starts, e.g., infiltration),
    \item $ S $: Maximum potential retention after runoff begins, which depends on land use, soil type, and moisture;
\end{itemize}
moreover,
\[
S = \frac{1000}{CN} - 10
\]
$ CN $ is the curve number based on land cover, soil type, and antecedent moisture conditions.

\subsubsection{Continuity Equation (Water Balance)}
The water balance equation is fundamental in tracking water in the system over time:
\[
P = R + ET + \Delta S,
\]
where
\begin{itemize}
    \item $ P $: Precipitation,
    \item $ R $: Runoff,
    \item $ ET $: Evapotranspiration,
    \item $ \Delta S $: Change in storage (e.g., soil, aquifers).
\end{itemize}

This equation forms the basis of calculating the accumulation of water in catchments and floodplains.

\subsubsection{River Hydraulics}
For predicting how water flows through rivers and channels, hydraulic models calculate the movement of water using principles from fluid dynamics.


In recent years, machine learning (ML) has begun to play an increasingly significant role in augmenting traditional CFD workflows by providing data-driven enhancements that can improve both accuracy and computational efficiency. Rather than replacing the core numerical solvers, ML techniques are typically used in hybrid frameworks where key components of the CFD process — such as surrogate modeling, solution acceleration, or data assimilation — are enhanced through learned representations. This approach is especially useful in water engineering applications for assimilating diverse observational datasets, such as satellite measurements and sensor networks, into hydraulic models like Google Flood Hub, where ML can help refine flood extent and timing predictions.

A major area of ML–CFD integration is the development of \textbf{data-driven surrogate models} that approximate the behavior of expensive fluid simulations. These surrogates, often based on neural networks or other regression frameworks, are trained on high-fidelity CFD data and can rapidly predict flow fields or system responses across a range of conditions, enabling real-time or near-real-time forecasting \cite{wang2024recent}. \textbf{Physics-informed machine learning} methodologies go a step further by embedding governing physical principles directly into the training process, ensuring that learned models respect conservation laws and minimizing unphysical artifacts in predictions.

Another practical direction is \textbf{ML-assisted numerical solutions}, where machine learning augments traditional solvers. For example, trained models can provide educated initial guesses for iterative pressure–velocity coupling or turbulence closure terms, significantly reducing the number of solver iterations required. In water engineering contexts with complex boundary conditions and heterogeneous datasets, such hybrid strategies help balance the need for accuracy with practical computational resource limits.

Despite the promise of ML in accelerating and enhancing CFD, challenges remain. High-quality training data, robust generalization outside training regimes, and interpretability of learned models are ongoing research focuses. Nevertheless, the integration of machine learning with CFD represents a transformative direction, enabling more responsive, informed, and efficient modeling workflows that complement traditional numerical analysis in environmental and hydraulic engineering.

\subsubsection{Rainfall Forecasting Models}
Rainfall forecasts from NWP are integrated into flood models. Precipitation values are processed to estimate the potential volume of water contributing to floods.

\paragraph{Remote Sensing and Satellite Data}
Google Flood Hub likely incorporates satellite data for real-time monitoring of water bodies. This data can help calculate:

\begin{itemize}
    \item \textbf{Water surface area and height} from radar and optical sensors
    \item \textbf{Soil moisture content}, which affects runoff potential
    \item \textbf{Vegetation and land cover} that modify floodplain behavior
\end{itemize}

\section{Conclusion and Contemporary Research Challenges}

This survey has reviewed the mathematical foundations and numerical methodologies that underpin computational fluid dynamics in water engineering, spanning depth-averaged models, fully three-dimensional formulations, and selected alternative approaches. By situating governing equations, discretization strategies, and computational practices within their historical and practical contexts, the discussion highlights both the maturity of classical CFD techniques and their continued relevance in modern hydraulic applications. Across a wide range of spatial and temporal scales, physics-based models remain indispensable for understanding and managing water systems, particularly when supported by robust numerical schemes and careful treatment of boundary conditions, source terms, and multiscale interactions.

Despite significant progress, several contemporary research challenges remain open. One persistent issue concerns the trade-off between model fidelity and computational efficiency, especially for large-scale or real-time applications such as flood forecasting and basin-wide water management. Advances in adaptive mesh refinement, high-order discretizations, and parallel computing continue to mitigate these limitations, but their integration into operational models is often constrained by robustness and implementation complexity. In addition, the faithful representation of multiphysics processes—including sediment transport, ecohydraulics, and surface–subsurface coupling—poses ongoing challenges for both model formulation and numerical stability.

A further emerging direction lies in the systematic integration of data with CFD models. While data-driven and machine-learning-assisted techniques offer promising avenues for accelerating simulations, improving parameter estimation, and quantifying uncertainty, their reliable coupling with conservation-law-based solvers remains an active area of research. Ensuring physical consistency, interpretability, and generalization across flow regimes is essential if such methods are to complement rather than undermine established modeling practices. Addressing these challenges will require continued collaboration between applied mathematicians, numerical analysts, and water engineering practitioners, with the ultimate goal of developing CFD tools that are not only accurate and efficient, but also transparent and decision-relevant for complex water management problems.

\paragraph*{Acknowledgements}
This work received funding from the National Centre for Energy II (TN02000025). FVD acknowledges INdAM Research Group GNCS.

\bibliographystyle{plain}
\bibliography{refs, mybib}

@article{BrianSpalding,
  author    = {Brian Launder and Andrew Pollard},
  title     = {Memoirs Dudley Brian Spalding. 9 January 1923—27 November 2016},
  journal   = {Biographical Memoirs of Fellows of the Royal Society},
  year      = {2019},
  doi       = {10.1098/rsbm.2018.0024},
  url       = {https://royalsocietypublishing.org/doi/10.1098/rsbm.2018.0024},
}

@book{anderson1997history,
  title = {A History of Aerodynamics and Its Impact on Flying Machines},
  author = {Anderson, John D., Jr.},
  year = {1997},
  publisher = {Cambridge University Press},
  address = {Cambridge}
}

@book{ferziger2002computational,
  title = {Computational Methods for Fluid Dynamics},
  author = {Ferziger, Joel H. and Perić, Milovan},
  edition = {3rd},
  year = {2002},
  publisher = {Springer},
  address = {Berlin}
}

@book{ferziger2002computationalFDM,
  title = {Computational Methods for Fluid Dynamics},
  author = {Ferziger, Joel H. and Perić, Milovan},
  edition = {3rd},
  year = {2002},
  publisher = {Springer},
  address = {Berlin},
  pages = {292-295}
}

@book{ferziger2002computationalFEM,
  title = {Computational Methods for Fluid Dynamics},
  author = {Ferziger, Joel H. and Perić, Milovan},
  edition = {3rd},
  year = {2002},
  publisher = {Springer},
  address = {Berlin},
  page = {40}
}

@book{ferziger2002computationalFVM,
  title = {Computational Methods for Fluid Dynamics},
  author = {Ferziger, Joel H. and Perić, Milovan},
  edition = {3rd},
  year = {2002},
  publisher = {Springer},
  address = {Berlin},
  pages = {81-107}
}

@book{ferziger2002computationalLESRANS,
  title = {Computational Methods for Fluid Dynamics},
  author = {Ferziger, Joel H. and Perić, Milovan},
  edition = {3rd},
  year = {2002},
  publisher = {Springer},
  address = {Berlin},
  pages = {366-380,395-414}
}

@article{chen1998lattice,
  title = {Lattice Boltzmann Method for Fluid Flows},
  author = {Chen, Shiyi and Doolen, Gary D.},
  journal = {Annual Review of Fluid Mechanics},
  volume = {30},
  number = {1},
  pages = {329--364},
  year = {1998},
  publisher = {Annual Reviews}
}

@article{moin1998direct,
  title = {Direct Numerical Simulation: A Tool in Turbulence Research},
  author = {Moin, Parviz and Mahesh, Krishnan},
  journal = {Annual Review of Fluid Mechanics},
  volume = {30},
  number = {1},
  pages = {539--578},
  year = {1998},
  publisher = {Annual Reviews}
}

@article{brenner2019perspective,
  title = {Perspective on Machine Learning for Advancing Fluid Mechanics},
  author = {Brenner, Michael P. and Eldredge, John D.},
  journal = {Physical Review Fluids},
  volume = {4},
  number = {10},
  pages = {100501},
  year = {2019},
  publisher = {American Physical Society}
}

@book{castro2019shallow,
  title={Shallow water hydraulics},
  author={Castro-Orgaz, Oscar and Hager, Willi H and others},
  year={2019},
  publisher={Springer}
}

@book{ghaib2019introductionLES,
  title = {Introduction to Computational Fluid Dynamics},
  author = {Ghaib, Karim},
  year = {2019},
  publisher = {Springer},
  page = {65-68}
}

@book{ghaib2019introductionRANS,
  title = {Introduction to Computational Fluid Dynamics},
  author = {Ghaib, Karim},
  year = {2019},
  publisher = {Springer},
  page = {47-64}
}

@book{maliska2023fundamentals,
  title = {Fundamentals of Computational Fluid Dynamics: The Finite Volume Method},
  author = {Maliska, Clovis R.},
  year = {2023},
  publisher = {Springer},
  page = {41-78}
}

@book{khan2014modeling2DGQ,
  title = {Modeling Shallow Water Flows Using the Discontinuous Galerkin Method},
  author = {Khan, Abdul A. and Lai, Wencong},
  year = {2014},
  publisher = {CRC Press},
  page = {22}
}

@book{shivamoggi2023introduction3DHIF,
  title = {Introduction to Theoretical and Mathematical Fluid Dynamics},
  author = {Shivamoggi, Bhimsen K.},
  edition = {Third},
  year = {2023},
  publisher = {University of Central Florida},
  page = {32-35}
}

@book{shivamoggi2023introduction3DIF,
  title = {Introduction to Theoretical and Mathematical Fluid Dynamics},
  author = {Shivamoggi, Bhimsen K.},
  edition = {Third},
  year = {2023},
  publisher = {University of Central Florida},
  page = {99-114}
}

@book{HCManning,
  author    = {M. Hanif Chaudhry},
  title     = {Open-Channel Flow},
  edition   = {2nd},
  publisher = {Springer},
  year      = {2008},
  page = {94-100}
}

@book{GKBReynolds,
  author    = {G. K. Batchelor},
  title     = {An Introduction to Fluid Dynamics},
  publisher = {Cambridge University Press},
  year      = {2000},
  page      = {211-229},
}

@book{SPReynoldsNumber,
  author    = {Stephen B. Pope},
  title     = {Turbulent Flows},
  publisher = {Cambridge University Press},
  year      = {2000},
  page      = {83-92},
}

@article{delestre2013swashes,
  title={SWASHES: a compilation of shallow water analytic solutions for hydraulic and environmental studies},
  author={Delestre, Olivier and Lucas, Carine and Ksinant, Philippe-Antoine and Darboux, Fran{\c{c}}ois and Laguerre, Cl{\'e}ment and Vo, Th{\`e}~Minh and James, Fr{\'e}d{\'e}ric and Cordier, St{\'e}phane},
  journal={International Journal for Numerical Methods in Fluids},
  volume={72},
  number={3},
  pages={269--300},
  year={2013},
  publisher={Wiley Online Library}
}

@misc{stavropoulou2021residual,
  title={A residual based artificialviscosity method for the stabilizationof the shallow water equations},
  author={Stavropoulou, Charitini},
  year={2021}
}

@article{casulli1990semi,
  title={Semi-implicit finite difference methods for the two-dimensional shallow water equations},
  author={Casulli, Vincenzo},
  journal={Journal of Computational Physics},
  volume={86},
  number={1},
  pages={56--74},
  year={1990},
  publisher={Elsevier}
}

@book{boyd2001chebyshev,
  title={Chebyshev and Fourier spectral methods},
  author={Boyd, John P},
  year={2001},
  publisher={Courier Corporation}
}

@book{canuto2006spectral,
  title={Spectral methods: fundamentals in single domains},
  author={Canuto, Claudio and Hussaini, M Youssuff and Quarteroni, Alfio and Zang, Thomas A},
  year={2006},
  publisher={Springer}
}

@article{crandall1983viscosity,
  title={Viscosity solutions of Hamilton-Jacobi equations},
  author={Crandall, Michael G and Lions, Pierre-Louis},
  journal={Transactions of the American mathematical society},
  volume={277},
  number={1},
  pages={1--42},
  year={1983}
}

@mastersthesis{turan2024high,
  title={{High Resolution Flash Flood Forecasting by Combining a Hydrometeorological Modeling System with a Computational Fluid Dynamics Model}},
  author={Turan, Azim},
  year={2024},
  school={Middle East Technical University (Turkey)}
}

@article{sun2024assessing,
  title={Assessing the impact of rainfall on water quality in a coastal urban river utilizing the environmental fluid dynamics code},
  author={Sun, Qingqing and Chang, Suyun and Wang, Jingfu and Chen, Jing-An and Qin, Caiqing and Shi, Weiwei and Zhang, Wen-Xi and Zhang, Yutong and Yue, Fu-Jun and Li, Si-Liang},
  journal={Urban Climate},
  volume={56},
  pages={102082},
  year={2024},
  publisher={Elsevier}
}

@article{cea2010experimental,
  title={Experimental validation of two-dimensional depth-averaged models for forecasting rainfall--runoff from precipitation data in urban areas},
  author={Cea, L and Garrido, M and Puertas, J},
  journal={Journal of Hydrology},
  volume={382},
  number={1-4},
  pages={88--102},
  year={2010},
  publisher={Elsevier}
}

@article{fleit2019practical,
  title={A practical framework to assess the hydrodynamic impact of ship waves on river banks},
  author={Fleit, G{\'a}bor and Baranya, S{\'a}ndor and Kr{\'a}mer, Tam{\'a}s and Bihs, Hans and J{\'o}zsa, J{\'a}nos},
  journal={River Research and Applications},
  volume={35},
  number={9},
  pages={1428--1442},
  year={2019},
  publisher={Wiley Online Library}
}

@article{almaz2012simulation,
  title={Simulation modeling of the vessel traffic in Delaware River: Impact of deepening on port performance},
  author={Almaz, Ozhan Alper and Altiok, Tayfur},
  journal={Simulation modelling practice and Theory},
  volume={22},
  pages={146--165},
  year={2012},
  publisher={Elsevier}
}

@article{lataire2017hydrodynamic,
  title={Hydrodynamic interaction between ships and restricted waterways},
  author={Lataire, Evert and Vantorre, Marc},
  journal={International journal of maritime engineering},
  volume={159},
  number={A1},
  year={2017}
}

@article{terziev2021numerical,
  title={A numerical assessment of the scale effects of a ship advancing through restricted waters},
  author={Terziev, Momchil and Tezdogan, Tahsin and Incecik, Atilla},
  journal={Ocean Engineering},
  volume={229},
  pages={108972},
  year={2021},
  publisher={Elsevier}
}

@phdthesis{fleit2021hydrodynamic,
  title={Hydrodynamic Impact Analysis of Ship-Induced Waves},
  author={Fleit, G{\'a}bor},
  year={2021},
  school={Budapest University of Technology and Economics (Hungary)}
}

@article{bellafiore2018modeling,
  title={Modeling ship-induced waves in shallow water systems: The Venice experiment},
  author={Bellafiore, D and Zaggia, L and Broglia, R and Ferrarin, C and Barbariol, F and Zaghi, S and Lorenzetti, G and Manf{\`e}, G and De Pascalis, F and Benetazzo, A},
  journal={Ocean Engineering},
  volume={155},
  pages={227--239},
  year={2018},
  publisher={Elsevier}
}

@article{dempwolff2023verification,
  title={Verification of a free-surface pressure term extension to represent ships in a nonhydrostatic shallow-water-equations solver},
  author={Dempwolff, Le{\'o}n-Carlos and Windt, Christian and Goseberg, Nils and Martin, Tobias and Bihs, Hans and Melling, Gregor},
  journal={Journal of Offshore Mechanics and Arctic Engineering},
  volume={145},
  number={2},
  pages={021202},
  year={2023},
  publisher={American Society of Mechanical Engineers}
}

@article{he2019short,
  title={Short-term vessel traffic flow forecasting by using an improved Kalman model},
  author={He, Wei and Zhong, Cheng and Sotelo, Miguel Angel and Chu, Xiumin and Liu, Xinglong and Li, Zhixiong},
  journal={Cluster Computing},
  volume={22},
  pages={7907--7916},
  year={2019},
  publisher={Springer}
}

@article{mao2022impacts,
  title={Impacts of ship waves on bed morphology of a trapezoidal cross-sectional channel},
  author={Mao, Lilei and Li, Xin and Chen, Yimei},
  journal={Journal of Offshore Mechanics and Arctic Engineering},
  volume={144},
  number={5},
  pages={051201},
  year={2022},
  publisher={American Society of Mechanical Engineers}
}

@article{trinci2017life,
  title={Life in turbulent flows: interactions between hydrodynamics and aquatic organisms in rivers},
  author={Trinci, Giuditta and Harvey, Gemma L and Henshaw, Alexander J and Bertoldi, Walter and H{\"o}lker, Franz},
  journal={Wiley interdisciplinary reviews: Water},
  volume={4},
  number={3},
  pages={e1213},
  year={2017},
  publisher={Wiley Online Library}
}

@article{kim2022harmful,
  title={Harmful algal bloom dynamics in a tidal river influenced by hydraulic control structures},
  author={Kim, Jaeyoung and Seo, Dongil and Jones, John R},
  journal={Ecological Modelling},
  volume={467},
  pages={109931},
  year={2022},
  publisher={Elsevier}
}

@article{daraio2010methodological,
  title={A methodological framework for integrating computational fluid dynamics and ecological models applied to juvenile freshwater mussel dispersal in the Upper Mississippi River},
  author={Daraio, Joseph A and Weber, Larry J and Newton, Teresa J and Nestler, John M},
  journal={Ecological Modelling},
  volume={221},
  number={2},
  pages={201--214},
  year={2010},
  publisher={Elsevier}
}

@article{willis2011modelling,
  title={Modelling swimming aquatic animals in hydrodynamic models},
  author={Willis, Jay},
  journal={Ecological Modelling},
  volume={222},
  number={23-24},
  pages={3869--3887},
  year={2011},
  publisher={Elsevier}
}

@article{goodwin2014fish,
  title={Fish navigation of large dams emerges from their modulation of flow field experience},
  author={Goodwin, R Andrew and Politano, Marcela and Garvin, Justin W and Nestler, John M and Hay, Duncan and Anderson, James J and Weber, Larry J and Dimperio, Eric and Smith, David L and Timko, Mark},
  journal={Proceedings of the National Academy of Sciences},
  volume={111},
  number={14},
  pages={5277--5282},
  year={2014},
  publisher={National Acad Sciences}
}

@article{vcada2006efforts,
  title={Efforts to reduce mortality to hydroelectric turbine-passed fish: locating and quantifying damaging shear stresses},
  author={{\v{C}}ada, Glenn and Loar, James and Garrison, Laura and Fisher, Richard and Neitzel, Duane},
  journal={Environmental Management},
  volume={37},
  pages={898--906},
  year={2006},
  publisher={Springer}
}

@article{haro2004swimming,
  title={Swimming performance of upstream migrant fishes in open-channel flow: a new approach to predicting passage through velocity barriers},
  author={Haro, Alex and Castro-Santos, Theodore and Noreika, John and Odeh, Mufeed},
  journal={Canadian journal of fisheries and aquatic sciences},
  volume={61},
  number={9},
  pages={1590--1601},
  year={2004},
  publisher={NRC Research Press Ottawa, Canada}
}

@article{weber2016modelling,
  title={Modelling the influence of aquatic vegetation on the hydrodynamics of an alternative bank protection measure in a navigable waterway},
  author={Weber, Arnd and Zhang, Jingxin and Nardin, Alejandro and Sukhodolov, Alexander and Wolter, Christian},
  journal={River Research and Applications},
  volume={32},
  number={10},
  pages={2071--2080},
  year={2016},
  publisher={Wiley Online Library}
}

@article{OTERKUS2014,
title = "Peridynamic thermal diffusion",
journal = "Journal of Computational Physics",
volume = "265",
pages = "71 - 96",
year = "2014",
issn = "0021-9991",
doi = "https://doi.org/10.1016/j.jcp.2014.01.027",
author = "S. Oterkus and E. Madenci and A. Agwai"
}

@ARTICLE{Emmrich_Puhst_2015,
AUTHOR = "Emmrich, E. and Puhst, D.",
TITLE = "Survey of existence results in nonlinear peridynamics in comparison with local elastodynamics",
JOURNAL = "Comput. Methods Appl. Math. ",
VOLUME = "15",
NUMBER = "4",
PAGES = "483--496",
YEAR = "2015",
DOI = "https://doi.org/10.1515/cmam-2015-0020"}

@ARTICLE{Emmrich_Puhst_2013,
AUTHOR = "Emmrich, E. and Puhst, D.",
TITLE = "Well-posedness of the peridynamic model with  {L}ipschitz continuous pairwise force function",
JOURNAL = "Commun. Math. Sci.",
VOLUME = "11",
NUMBER = "4",
PAGES = "1039--1049",
YEAR = "2013",
DOI="https://doi.org/10.4310/CMS.2013.v11.n4.a7"
}

@book{anderson1995,
  author    = {Anderson, John D.},
  title     = {Computational Fluid Dynamics: The Basics with Applications},
  publisher = {McGraw--Hill},
  year      = {1995}
}

@article{WECKNER2005,
title = "The effect of long-range forces on the dynamics of a bar",
journal = "Journal of the Mechanics and Physics of Solids",
volume = "53",
number = "3",
pages = "705 - 728",
year = "2005",
issn = "0022-5096",
doi = "https://doi.org/10.1016/j.jmps.2004.08.006",
url = "",
author = "Weckner, O. and Abeyaratne, R."
}

@article{SILLING2000,
title = {Reformulation of elasticity theory for discontinuities and long-range forces},
journal = {Journal of the Mechanics and Physics of Solids},
volume = {48},
number = {1},
pages = {175-209},
year = {2000},
issn = {0022-5096},
doi = {https://doi.org/10.1016/S0022-5096(99)00029-0},
author = {S.A. Silling}
}

@article{aydogdu2023analysis,
  title={Analysis of the effect of rigid vegetation patches on the hydraulics of an open channel flow with Realizable k-$\varepsilon$ and Reynolds stress turbulence models},
  author={Aydogdu, Mahmut},
  journal={Flow Measurement and Instrumentation},
  volume={94},
  pages={102477},
  year={2023},
  publisher={Elsevier}
}

@article{anjum2020hydrodynamics,
  title={Hydrodynamics of longitudinally discontinuous, vertically double layered and partially covered rigid vegetation patches in open channel flow},
  author={Anjum, Naveed and Tanaka, Norio},
  journal={River Research and Applications},
  volume={36},
  number={1},
  pages={115--127},
  year={2020},
  publisher={Wiley Online Library}
}

@article{anjum2022investigation,
  title={Investigation of the flow structures through heterogeneous vegetation of varying patch configurations in an open channel},
  author={Anjum, Naveed and Ali, Mustajab},
  journal={Environmental Fluid Mechanics},
  volume={22},
  number={6},
  pages={1333--1354},
  year={2022},
  publisher={Springer}
}

@article{preuss2023cfd,
  title={CFD analysis of environmentally friendly wave mitigation measures in river waterways},
  author={Preu{\ss}, Julia and Fleit, G{\'a}bor and Baranya, S{\'a}ndor},
  journal={River Research and Applications},
  volume={39},
  number={5},
  pages={847--860},
  year={2023},
  publisher={Wiley Online Library}
}

@article{anjum2024emergent,
  title={How Emergent Vegetation Patch Shapes Affect Flow Structure in Open Channel Environments},
  author={Anjum, Naveed and Pasha, Ghufran Ahmed and Ashraf, Jawad and Ghani, Usman and Abbas, Fakhar Muhammad and Rahman, Md Abedur},
  journal={Ecohydrology},
  pages={e2740},
  year={2024},
  publisher={Wiley Online Library}
}

@article{valseth2022stable,
  title={A stable space-time FE method for the shallow water equations},
  author={Valseth, Eirik and Dawson, Clint},
  journal={Computational Geosciences},
  pages={1--18},
  year={2022},
  publisher={Springer}
}

@article{zhao2022construction,
  title={Construction of a peridynamic model for viscous flow},
  author={Zhao, Jiangming and Larios, Adam and Bobaru, Florin},
  journal={Journal of Computational Physics},
  volume={468},
  pages={111509},
  year={2022},
  publisher={Elsevier}
}

@book{flandoli2023stochastic,
  title={Stochastic partial differential equations in fluid mechanics},
  author={Flandoli, Franco and Luongo, Eliseo and others},
  volume={2330},
  year={2023},
  publisher={Springer}
}

@article{nguyen2021modelling,
  title={Modelling of Eulerian incompressible fluid flows by using peridynamic differential operator},
  author={Nguyen, Cong Tien and Oterkus, Selda and Oterkus, Erkan and Amin, Islam and Ozdemir, Murat and El-Aassar, Abdel-Hameed and Shawky, Hosam},
  journal={Ocean Engineering},
  volume={239},
  pages={109815},
  year={2021},
  publisher={Elsevier}
}

@article{breuer1992stochastic,
  title={A stochastic approach to computational fluid dynamics},
  author={Breuer, HP and Petruccione, F},
  journal={Continuum Mechanics and Thermodynamics},
  volume={4},
  pages={247--267},
  year={1992},
  publisher={Springer}
}

@book{toro2013riemann,
  title={Riemann solvers and numerical methods for fluid dynamics: a practical introduction},
  author={Toro, Eleuterio F},
  year={2013},
  publisher={Springer Science \& Business Media}
}

@article{zhang1997development,
  title={Development of an automatic calibration scheme for the HBV hydrological model},
  author={Zhang, Xingnan and Lindstr{\"o}m, G{\"o}ran},
  journal={Hydrological Processes},
  volume={11},
  number={12},
  pages={1671--1682},
  year={1997},
  publisher={Wiley Online Library}
}

@article{kobold2006use,
  title={The use of HBV model for flash flood forecasting},
  author={Kobold, Mira and Brilly, Mitja},
  journal={Natural Hazards and Earth System Sciences},
  volume={6},
  number={3},
  pages={407--417},
  year={2006},
  publisher={Copernicus GmbH}
}

@article{bergstrom2015interpretation,
  title={Interpretation of runoff processes in hydrological modelling—experience from the HBV approach},
  author={Bergstr{\"o}m, Sten and Lindstr{\"o}m, G{\"o}ran},
  journal={Hydrological Processes},
  volume={29},
  number={16},
  pages={3535--3545},
  year={2015},
  publisher={Wiley Online Library}
}

@article{seibert2022retrospective,
  title={A retrospective on hydrological catchment modelling based on half a century with the HBV model},
  author={Seibert, Jan and Bergstr{\"o}m, Sten},
  journal={Hydrology and Earth System Sciences},
  volume={26},
  number={5},
  pages={1371--1388},
  year={2022},
  publisher={Copernicus Publications G{\"o}ttingen, Germany}
}

@article{mailybaev2025rg,
  title={RG approach to the inviscid limit for shell models of turbulence},
  author={Mailybaev, Alexei A},
  journal={Nonlinearity},
  volume={38},
  number={8},
  pages={085010},
  year={2025},
  publisher={IOP Publishing}
}

@misc{greenkenue2022,
  title={GreenKenue: Software Tool for Hydrologic Modellers},
  author={National Research Council Canada (NRC)},
  year={2022},
  url={https://nrc.canada.ca/en/research-development/products-services/software-applications/green-kenuetm-software-tool-hydrologic-modellers},
  note={Accessed: October 2023},
  howpublished={National Research Council Canada},
  date={9 August 2022}
}

@article{woldemeskel2020watertools,
  title={WaterTools: A Guide to Three National Level Platforms that Support the Management of Australia's Scarce Water Resources},
  author={Woldemeskel, Fitsum},
  year={2020},
  journal={ResearchGate},
  url={https://www.researchgate.net/publication/346889523_WaterTools_A_Guide_to_three_national_level_platforms_that_support_the_management_of_Australia's_scarce_water_resources},
  note={Accessed: October 2023}
}

@article{california_dwr_models,
  title={Central Valley Models and Tools},
  author={California Department of Water Resources},
  year={2023},
  journal = {California Department of Water Resources},
  url={https://water.ca.gov/Library/Modeling-and-Analysis/Central-Valley-models-and-tools},
}

@book{evans2022partial,
  title={Partial differential equations},
  author={Evans, Lawrence C},
  volume={19},
  year={2022},
  publisher={American Mathematical Society}
}

@article{choi2004spectral,
  title={A spectral finite-volume method for the shallow water equations},
  author={Choi, Byoung-Ju and Iskandarani, Mohamed and Levin, Julia and Haidvogel, Dale B},
  journal={Monthly weather review},
  volume={132},
  number={7},
  pages={1777--1791},
  year={2004}
}

@article{DubosKevlahan2013,
  author    = {Dubos, Thomas and Kevlahan, Nicholas K.-R.},
  title     = {A Conservative Adaptive Wavelet Method for the Shallow-Water Equations on the Sphere},
  journal   = {Quarterly Journal of the Royal Meteorological Society},
  volume    = {139},
  number    = {671},
  pages     = {1997--2020},
  year      = {2013},
  doi       = {10.1002/qj.2063}
}

@article{AechtnerKevlahanDubos2014,
  author    = {Aechtner, Matthias and Kevlahan, Nicholas K.-R. and Dubos, Thomas},
  title     = {A Conservative Adaptive Wavelet Method for the Shallow-Water Equations on the Sphere},
  journal   = {Journal of Computational Physics},
  volume    = {268},
  pages     = {80--122},
  year      = {2014},
  doi       = {10.1016/j.jcp.2014.02.005}
}

@article{dai2023energy,
  title={Energy stable and structure-preserving schemes for the stochastic Galerkin shallow water equations},
  author={Dai, Dihan and Epshteyn, Yekaterina and Narayan, Akil},
  journal={ESAIM: Mathematical Modelling and Numerical Analysis},
  volume={58},
  number={1},
  pages={1--31},
  year={2024},
  publisher={EDP Sciences},
  doi={10.1051/m2an/2023031}
}

@article{shaw2020stochastic,
  title={Stochastic Galerkin finite volume shallow flow model: well-balanced treatment over uncertain topography},
  author={Shaw, James and Kesserwani, Georges},
  journal={Journal of Hydraulic Engineering},
  volume={146},
  number={10},
  pages={04020072},
  year={2020},
  publisher={American Society of Civil Engineers},
  doi={10.1061/(ASCE)HY.1943-7900.0001705}
}

@book{leveque2002finite,
  title={Finite volume methods for hyperbolic problems},
  author={LeVeque, Randall J},
  volume={31},
  year={2002},
  publisher={Cambridge university press}
}

@book{tarantola2005inverse,
  title={Inverse Problem Theory and Methods for Model Parameter Estimation},
  author={Tarantola, Albert},
  year={2005},
  publisher={SIAM},
  address={Philadelphia}
}

@book{hughes2003finite,
  title={The finite element method: linear static and dynamic finite element analysis},
  author={Hughes, Thomas JR},
  year={2003},
  publisher={Courier Corporation}
}

@article{cockburn2001runge,
  title={Runge--Kutta discontinuous Galerkin methods for convection-dominated problems},
  author={Cockburn, Bernardo and Shu, Chi-Wang},
  journal={Journal of scientific computing},
  volume={16},
  number={3},
  pages={173--261},
  year={2001},
  publisher={Springer}
}

@article{wang2024recent,
  title={Recent advances on machine learning for computational fluid dynamics: A survey},
  author={Wang, Haixin and Cao, Yadi and Huang, Zijie and Liu, Yuxuan and Hu, Peiyan and Luo, Xiao and Song, Zezheng and Zhao, Wanjia and Liu, Jilin and Sun, Jinan and others},
  journal={arXiv preprint arXiv:2408.12171},
  year={2024}
}

@misc{delis2021shallow,
  title={Shallow water equations in hydraulics: Modeling, numerics and applications},
  author={Delis, Anargiros I and Nikolos, Ioannis K},
  journal={Water},
  volume={13},
  number={24},
  pages={3598},
  year={2021},
  publisher={MDPI}
}

@article{sarker2021short,
  title={A short review on computational hydraulics in the context of water resources engineering},
  author={Sarker, Shiblu},
  journal={Open Journal of Modelling and Simulation},
  volume={10},
  number={1},
  pages={1--31},
  year={2021},
  publisher={Scientific Research Publishing}
}

@book{ortloff2024hydraulic,
  title={Hydraulic Engineering and Modelling of Water Flow by Use of Computational Fluid Dynamics (CFD) and Modern Hydraulic Analysis Methods},
  author={Ortloff, Charles R},
  year={2024},
  publisher={MDPI-Multidisciplinary Digital Publishing Institute}
}

@book{versteeg2007introduction,
  title={An introduction to computational fluid dynamics the finite volume method, 2/E},
  author={Versteeg, Henk Kaarle},
  year={2007},
  publisher={Pearson Education India}
}

@book{toro2024computational,
  title={Computational Algorithms for Shallow Water Equations},
  author={Toro, Eleuterio F},
  year={2024},
  publisher={Springer}
}

@book{smith2024uncertainty,
  title={Uncertainty quantification: theory, implementation, and applications},
  author={Smith, Ralph C},
  year={2024},
  publisher={SIAM}
}

@book{keesman2011system,
  title={System identification: an introduction},
  author={Keesman, Karel J},
  year={2011},
  publisher={Springer Science \& Business Media}
}

@book{evensen2009data,
  title={Data assimilation: the ensemble Kalman filter},
  author={Evensen, Geir},
  year={2009},
  publisher={Springer}
}

@incollection{rodriguez2019applied,
  title={Applied Computational Fluid Dynamics and Turbulence Modeling: Practical Tools},
  author={Rodriguez, S},
  booktitle={Tips and Techniques},
  year={2019},
  publisher={Springer International Publishing Cham}
}

@article{yusuf2020short,
  title={A short review on rans turbulence models},
  author={Yusuf, Siti Nurul Akmal and Asako, Yutaka and Sidik, Nor Azwadi Che and Mohamed, Saiful Bahri and Japar, Wan Mohd Arif Aziz},
  journal={CFD letters},
  volume={12},
  number={11},
  pages={83--96},
  year={2020}
}

@book{chorin1990mathematical,
  title={A mathematical introduction to fluid mechanics},
  author={Chorin, Alexandre Joel and Marsden, Jerrold E and Marsden, Jerrold E},
  volume={168},
  year={1990},
  publisher={Springer}
}

@book{bates2005computational,
  title={Computational fluid dynamics: applications in environmental hydraulics},
  author={Bates, Paul D and Lane, Stuart N and Ferguson, Robert I},
  year={2005},
  publisher={John Wiley \& Sons}
}
\end{document}